\documentclass[showpacs,epsfig,preprintnumbers,amsmath,amssymb,nofootinbib,20pt]{revtex4-1}
\usepackage{graphicx}% Include figure files
\usepackage{dcolumn}% Align table columns on decimal point
\usepackage{bm}% bold math

\usepackage{subfigure}
\usepackage{epstopdf}

\begin{document}
\def\d{{\rm d}}
%%%%%%%%%%%%%%%%%%%%%%%
\def\Epos{E_{\rm pos}}
\def\ap{\approx}
\def\eff{{\rm eft}}
\def\L{{\cal L}}
\newcommand{\vev}[1]{\langle {#1}\rangle}
\newcommand{\CL}   {C.L.}
\newcommand{\dof}  {d.o.f.}
\newcommand{\eVq}  {\text{eV}^2}
\newcommand{\Sol}  {\textsc{sol}}
\newcommand{\SlKm} {\textsc{sol+kam}}
\newcommand{\Atm}  {\textsc{atm}}
\newcommand{\Chooz}{\textsc{chooz}}
\newcommand{\Dms}  {\Delta m^2_\Sol}
\newcommand{\Dma}  {\Delta m^2_\Atm}
\newcommand{\Dcq}  {\Delta\chi^2}
\newcommand{\nbb}{$\beta\beta_{0\nu}$ }
\newcommand {\be}{\begin{equation}}
\newcommand {\ee}{\end{equation}}
\newcommand {\ba}{\begin{eqnarray}}
\newcommand {\ea}{\end{eqnarray}}
\def\VEV#1{\left\langle #1\right\rangle}
\let\vev\VEV
\def\e6{E(6)}
\def\10{SO(10)}
\def\21{SA(2) $\otimes$ U(1) }
\def\321{$\mathrm{SU(3) \otimes SU(2) \otimes U(1)}$ }
\def\lr{SA(2)$_L \otimes$ SA(2)$_R \otimes$ U(1)}
\def\422{SA(4) $\otimes$ SA(2) $\otimes$ SA(2)}

\newcommand{\AHEP}{AHEP Group, Institut de F\'{i}sica Corpuscular --
	Universitat de Val\`{e}ncia/CSIC, Parc Cientific de Paterna.\\
  C/Catedratico Jos\'e Beltr\'an, 2 E-46980 Paterna (Val\`{e}ncia) - SPAIN}

\newcommand{\Tehran}{%
School of physics, Institute for Research in Fundamental Sciences (IPM)
\\
P.O.Box 19395-5531, Tehran, Iran}
\def\roughly#1{\mathrel{\raise.3ex\hbox{$#1$\kern-.75em
      \lower1ex\hbox{$\sim$}}}} \def\lsim{\roughly<}
\def\gsim{\roughly>}
\def\ltap{\raisebox{-.4ex}{\rlap{$\sim$}} \raisebox{.4ex}{$<$}}
\def\gtap{\raisebox{-.4ex}{\rlap{$\sim$}} \raisebox{.4ex}{$>$}}
\def\lsim{\raise0.3ex\hbox{$\;<$\kern-0.75em\raise-1.1ex\hbox{$\sim\;$}}}
\def\gsim{\raise0.3ex\hbox{$\;>$\kern-0.75em\raise-1.1ex\hbox{$\sim\;$}}}

%\title{Non-Standard Neutrino Interactions: theory and observation}
%\title{Reviewing the status of Neutrino oscillations and Non-Standard Neutrino Interactions}
\title{Neutrino oscillations and Non-Standard Interactions}

%\preprint{ IPM/P-2012}
\date{\today}

\author{Y. Farzan}\email{yasaman@theory.ipm.ac.ir}
\affiliation{\Tehran}

\author{M. T\' ortola}\email{mariam@ific.uv.es}
\affiliation{\AHEP}
%--------------------------------------------------
\begin{abstract}
%% Standard oscillations part %%
Current neutrino experiments are measuring the neutrino mixing parameters
with an unprecedented accuracy. The upcoming generation of neutrino experiments
will be sensitive to subdominant neutrino oscillation effects that can in principle give information on the yet-unknown neutrino parameters:
the Dirac CP-violating phase in the  PMNS mixing matrix, the neutrino mass ordering and the octant of $\theta_{23}$.  Determining  the exact values of neutrino  mass and mixing parameters is crucial to test various neutrino models and flavor symmetries  that are designed to predict these neutrino parameters.
% added
In the first part of this review, we summarize the current status of the neutrino oscillation parameter determination.
We  consider the most recent data from all solar neutrino experiments and  the atmospheric neutrino data from Super-Kamiokande, IceCube and ANTARES.   We also implement the data   from the reactor neutrino experiments KamLAND, Daya Bay, RENO and Double Chooz as well as  the long baseline neutrino data from MINOS, T2K and NO$\nu$A.
%
%% NSI part %%
%

If in addition to the standard interactions, neutrinos have subdominant yet-unknown Non-Standard Interactions (NSI)  with matter fields, extracting the values of these parameters will suffer from new degeneracies and ambiguities. We review such effects and formulate the conditions on the NSI parameters under which   the precision measurement of neutrino oscillation parameters can be distorted.
Like standard weak interactions, the non-standard interaction  can be categorized into two groups: Charged Current (CC) NSI and Neutral Current (NC) NSI. Our focus will be mainly on neutral current NSI because it is possible to build a class of models that give rise to sizeable NC NSI with discernible effects on neutrino oscillation.
These models are based on new  $U(1)$ gauge symmetry with a gauge boson of mass $\lesssim
 10$~MeV.
The UV complete model should be of course electroweak invariant which in general implies that along with neutrinos, charged fermions also acquire new interactions on which there are strong bounds.
We enumerate the bounds that already exist on the electroweak symmetric models and demonstrate that it is possible to build viable models  avoiding all these bounds.
In the end, we review methods to test these models and suggest approaches to break the degeneracies in deriving neutrino mass parameters caused by NSI.
\end{abstract}
%--------------------------------------------------------
 %\pacs{12.60.-i,14.80.-j,95.35.+d,98.70.Sa}
{}
%--------------------------------------------------------
\date{\today}
\maketitle
%---------------------------------------------------

\section{Introduction\label{intro}}

In the framework of ``old" electroweak theory, formulated by Glashow, Weinberg and Salam, lepton flavor is conserved and neutrinos are massless. As a result, a neutrino of flavor $\alpha$ ($\alpha \in \{ e , \mu , \tau \}$) created in charged current weak interactions in association with a charged lepton of flavor $\alpha$ will maintain its flavor. Various observations have however shown that the flavor of neutrinos change upon propagating long distances. Historically, solar neutrino anomaly (deficit of the solar neutrino flux relative to standard solar model predictions) \cite{Davis:1968cp} and atmospheric neutrino anomaly (deviation of the ratio of muon neutrino flux to the electron neutrino flux from 2 for atmospheric neutrinos that cross the Earth before reaching the detector) \cite{Fukuda:1998mi} were two main observations that showed the lepton flavor was violated in nature. This conclusion was further confirmed by observation of flavor violation of man-made neutrinos after propagating sizable distances in various reactor~\cite{Abe:2011fz,An:2012eh,Ahn:2012nd} and long baseline experiments~\cite{Aliu:2004sq,Michael:2006rx}. The established paradigm for flavor violation which impressively explain all these anomalies is the three neutrino mass and mixing scheme. According to this scheme, each neutrino flavor is a mixture of different mass eigenstates. As neutrinos propagate, each component mass eigenstate acquires a different phase so neutrino of definite flavor will convert to a mixture of different flavors; hence, lepton flavor violation takes place.

Within this scheme, the probability of conversion of $\nu_\alpha$ to $\nu_\beta$ (as well as that of  $\bar{\nu}_\alpha$ to $\bar{\nu}_\beta$) in vacuum or in matter with constant density has a
oscillatory dependence on time or equivalently on  the distance traveled by neutrinos \footnote{One should bear in mind that in a medium with varying density, such as the Sun interior, the conversion may not have an oscillatory behavior for a certain  energy range. Likewise, the presence of strong matter effects may suppress the oscillatory behavior even in the case of constant density \cite{Smirnov:2016xzf}.}.
For this reason, the phenomenon of flavor conversion in neutrino sector is generally known as neutrino oscillation.
Neutrino flavor eigenstates are usually denoted by $\nu_\alpha$. That is $\nu_\alpha$ is defined as a state which appears in $W$ boson vertex along with charged lepton $l_\alpha$ ($\alpha \in \{ e, \mu , \tau \}$). The latter  corresponds to charged lepton mass eigenstates. Neutrino mass eigenstates are denoted by $\nu_i$ with mass $m_i$ where $i \in \{ 1,2,3 \}$. The flavor eigenstates are related to mass eigenstate by a $3 \times 3$ unitary matrix, $U$, known as PMNS after Pontecorvo-Maki-Nakagawa-Sakata: $\nu_\alpha= \sum_i U_{\alpha i} \nu_i$.
The unitary mixing matrix can be decomposed as follows
\begin{eqnarray} U \equiv \left( \begin{matrix}   1 & 0 & 0 \cr
0 & \cos \theta_{23} & \sin \theta_{23} \cr 0 & -\sin \theta_{23} & \cos \theta_{23}\end{matrix} \right)\left( \begin{matrix}   \cos \theta_{13} & 0 & \sin\theta_{13} e^{-i \delta} \cr
0 & 1 & \cr -\sin\theta_{13} e^{i \delta} & 0 & \cos \theta_{13}\end{matrix} \right)\left( \begin{matrix}   \cos \theta_{12}& \sin \theta_{12} & 0 \cr
-\sin\theta_{12} & \cos \theta_{12} & 0 \cr 0 & 0 & 1\end{matrix} \right)  ,
\end{eqnarray}
where the mixing angles $\theta_{ij}$ are defined to be in the $[0,\pi/2]$ range while the phase $\delta$ can vary in $[0,2\pi)$. In this way, the whole physical parameter space is covered.
 Historically,  $\nu_1$, $\nu_2$ and $\nu_3$ have been defined  according to their contribution to $\nu_e$. In other words, they are ordered such that $|U_{e 1}|> |U_{e2}|>|U_{e3}|$ so $\nu_1$ ($\nu_3$) provides largest (smallest) contribution to $\nu_e$. Notice that with this definition, $\theta_{12}, \theta_{13} \leq \pi/4$. It is then of course a meaningful question to ask which $\nu_i$ is the lightest and which one is the heaviest; or equivalently, what is the sign of $\Delta m_{ij}^2=m_i^2-m_j^2$. The answer to this question comes from observation.
Time evolution of ultra-relativistic neutrino state is governed by the following Hamiltonian: $H_{vac}+H_m$, where the effective Hamiltonian in vacuum is given by
\be H_{vac}= U \cdot {\rm Diag}\left(\frac{m_1^2}{2 E} , \frac{m_2^2}{2 E}, \frac{m_3^2}{2 E}\right) \cdot U^\dagger. \ee
Within the Standard Model (SM) of particles, the effective Hamiltonian in matter $H_m$ in the framework of the medium in which neutrinos are propagating can be written as
 \be
 \label{eq:Hm}
 H_m=\left( \begin{matrix}
 \sqrt{2} G_F N_e- \frac{ \sqrt{2}}{2} G_F N_n &0 &0 \cr
 0& -\frac{ \sqrt{2}}{2} G_F N_n &0 \cr
 0 & 0 & -\frac{ \sqrt{2}}{2} G_F N_n\end{matrix} \right) ,
\ee
where it is assumed that the medium is electrically neutral ($N_e=N_p$), unpolarized and composed of non-relativistic particles.
In vacuum, $H_m=0$ and we can write
 \be \label{Pab} P(\nu_\alpha \to \nu_\beta)=\left|\sum_{ij} U_{\alpha i} U_{\beta j}^* e^{i \Delta m_{ij}^2 L/(2E)}\right|^2. \ee
By adding or subtracting a matrix proportional to the identity $I_{3 \times 3}$ to the Hamiltonian, neutrinos obtain an overall phase with no observable physical consequences. That is why neutrino oscillation probabilities (both in vacuum and in matter) are sensitive only to $\Delta m_{ij}^2$ rather than to $m_i^2$. As a result, it is impossible to derive the mass of the lightest neutrino from oscillation data alone. Similarly, neutrino oscillation pattern within the SM only depends on $N_e$ not on $N_n$.
Similar arguments can be repeated for antineutrinos by replacing $U$ with $U^*$ (or equivalently $\delta \to -\delta$) and replacing $H_m \to -H_m$. The phase $\delta$, similarly to its counterpart in the CKM mixing matrix of quark sector, violates CP. Just like in the quark sector,  CP violation in neutrino sector is given by the Jarlskog invariant: $\mathcal{J}= \sin \theta_{13} \cos^2 \theta_{13} \sin \theta_{12} \cos \theta_{12} \cos\theta_{23} \sin \theta_{23} \sin \delta$.

 As we will see in detail in  section \ref{oscillation}, the mixing angles $\theta_{12}$, $\theta_{13}$ and $\theta_{23}$ are derived from observations with remarkable precision. The mixing angle $\theta_{23}$ has turned out to be close $45^\circ$ but it is not clear within present uncertainties whether $\theta_{23} <\pi/4$ or $\theta_{23}>\pi/4$. This uncertainty is known as the octant degeneracy. The value of $\delta$ is also unknown for the time being, although experimental data start indicating a preferred value close to $3\pi/2$. The absolute value and the sign of $\Delta m_{21}^2$ are however determined.  While $|\Delta m_{31}^2|$ is measured, the sign of $\Delta m_{31}^2$ is not yet determined. If $\Delta m_{31}^2 >0$ ($\Delta m_{31}^2<0$), the scheme is called normal (inverted) ordering or normal (inverted) mass spectrum. The main goals of current and upcoming neutrino oscillation experiments are determining $sgn(\cos 2 \theta_{23})$, $sgn (\Delta m_{31}^2)$ and the value of the CP--violating phase $\delta$.

The neutrino oscillation program is entering a precision era, where the known parameters are being measured with an ever increasing accuracy. Next generation of long--baseline neutrino experiments will resolve the subdominant effects in oscillation data sensitive to the yet unknown oscillation parameters ({\it e.g.} $\delta$). Of course, all these derivations are within $3 \times 3$ neutrino mass and mixing scheme under the assumption that neutrinos interact with matter only through the SM weak interactions (plus gravity which is too weak
to be relevant).  Allowing for Non-Standard Interaction (NSI) can change the whole picture.
Non-Standard Interaction of neutrinos can be divided into two groups: Neutral Current (NC) NSI and Charged Current  (CC) NSI.
 While the CC NSI of neutrinos with the matter fields ($e, u,d$) affects in general the production and detection of neutrinos, the NC NSI  may  affect the neutrino propagation in matter.
As a result, both types of interaction may show up  at various neutrino experiments.
In recent years, the effects of both types of NSI on neutrino experiments have been extensively studied in the literature, formulating the lower limit on the values of couplings in order to have a resolvable impact on the oscillation pattern in upcoming experiments.
On the other hand,   non-standard interaction of neutrinos can  crucially affect the interpretation of the experimental data in terms of the relevant neutrino mass parameters. Indeed, as it will be discussed in this work, the presence of NSI in the neutrino propagation may give rise, among other effects,  to a degeneracy in the measurement of the solar mixing angle $\theta_{12}$~\cite{Miranda:2004nb,Escrihuela:2009up,Gonzalez-Garcia:2013usa}.
Likewise, CC NSI at the production and detection of reactor antineutrinos can affect the very precise measurement of the  mixing angle $\theta_{13}$ in Daya Bay~\cite{Leitner:2011aa,Agarwalla:2014bsa}.
Moreover, it has been shown that NSI can cause degeneracies in deriving the CP--violating phase $\delta$~\cite{Forero:2016cmb,Masud:2015xva,Masud:2016bvp,Liao:2016hsa}, as well as the correct octant of the atmospheric mixing angle $\theta_{23}$~\cite{Agarwalla:2016fkh} at current and future long--baseline neutrino experiments.
Along this review, we will discuss possible ways to resolve the parameter degeneracies due to NSI, by exploiting the capabilities of some of the planned experiments such as the intermediate baseline reactor neutrino experiments JUNO and RENO50~\cite{Bakhti:2014pva}.

Most of the analyses involving NSI in neutrino experiments parameterize such interactions in terms of  effective four-Fermi couplings. However, one may ask whether it is possible to build viable renormalizable electroweak symmetric UV complete models that underlay this effective interaction with coupling large enough to be discernible at neutrino oscillation experiments. Generally speaking if the effective coupling comes from integrating out a new state ($X$) of mass $m_X$ and of coupling $g_X$, we expect the strength of the effective four-Fermi interaction to be given by $g_X^2/m_X^2$. We should then justify why $X$ has not been so far directly produced at labs. As far as NC NSI is concerned, two solutions exist: (i) $X$ is too heavy; {\it i.e.,} $m_X\gg m_{EW}$.
Recent bounds from the LHC imply $m_X > 4-5$ TeV \cite{Aaboud:2017buh} which for even $g_X \sim 1$ implies $g_X^2/m_X^2 \ll G_F$ \cite{Franzosi:2015wha,Berezhiani:2001rs}.  Moreover, as shown in \cite{Friedland:2011za}, in the range $10~{\rm GeV}<m_{Z^\prime}<{\rm TeV}$, the monojet searches at the LHC constrain this ratio to values much smaller than 1. (ii) Second approach is to take $m_X \ll  m_{EW}$ and $g_X \ll 1$ such that $g_X^2/m_X^2\sim G_F$.  In this approach, the null result for direct production of $X$ is justified with its very small
coupling. In \cite{Farzan:2015doa,Farzan:2016wym,Farzan:2015hkd}, this approach has been evoked to build viable models for NC NSI with large effective couplings. For CC NSI, the intermediate state, being charged, cannot be light. That is, although its Yukawa couplings to neutrinos and matter fields can be set to arbitrarily small values, the gauge coupling to the photon is  set by its charge so the production at LEP and other experiments cannot be avoided. We are not aware of any viable model that can lead to a sizable CC NSI. Interested reader may see Refs.~\cite{VanegasForero:2017ezf,Bakhti:2016gic,Agarwalla:2014bsa,Khan:2011zr,Kopp:2010qt,Bellazzini:2010gn,Akhmedov:2010vy,Biggio:2009nt}.
Notice that throughout this review, we focus on the interaction of neutrinos with matter fields. Refs. \cite{Das:2017iuj,Dighe:2017sur} study the effects of non-standard self-interaction of neutrinos in supernova. Refs  \cite{deSalas:2016svi,Brdar:2017kbt} discuss propagation of neutrino in presence of interaction with dark matter. The effect of NSI on the decoupling of neutrinos in the early Universe has been considered in Ref.~\cite{Mangano:2006ar}.

This review is organized as follows. In section \ref{oscillation}, we review the different neutrino oscillation experiments and discuss how neutrino oscillation parameters within the standard three neutrino scheme can be derived. We then discuss the prospect of measuring  yet unknown parameters: $\delta$, $sgn (\Delta m_{31}^2)$ and $sgn (\cos 2 \theta_{23})$.
In section \ref{non-stan}, we discuss how NSI can affect this picture and review the bounds that the present neutrino data sets on the effective $\epsilon$ parameters. We then discuss the potential effects of NSI on future neutrino experiments and suggest strategies to solve the degeneracies.
In section \ref{viableMODELS}, we introduce models that can lead to effective NSI of interest and briefly discuss their potential effects on various observables.
In section \ref{upcoming}, we review methods  suggested to test these models.
Results will be summarized in section \ref{summ}.

%%%%%%%%%%%%%%%%%%%%%%%%%%%%%%%%%%%

\section{Neutrino oscillations\label{oscillation}}

In this section, we will present the current status of neutrino oscillation data in the standard three--neutrino framework.
Most recent global neutrino fits to neutrino oscillations can be found in Refs.~\cite{deSalas:2017kay,Capozzi:2017ipn,Esteban:2016qun}. Here we will focus on the results of Ref.~\cite{deSalas:2017kay}, commenting also on the comparison with the other two analysis.
First, we will describe the different experiments entering in the global neutrino analysis, grouped in the solar, reactor, atmospheric and long--baseline sectors.
For each of them we will also discuss their main contribution to the determination of the oscillation parameters.

%%%%  solar %%%%

\subsection{The solar neutrino sector: ($\sin^2\theta_{12}$, $\Delta m^2_{21}$)}

Under the  denomination of \textit{solar neutrino sector}, one finds traditionally not only all the solar neutrino experiments, but also the reactor KamLAND experiment, sensitive to the same oscillation channel, under the assumption of CPT conservation.
Solar neutrino analysis include the historical radiochemical experiments Homestake~\cite{Cleveland:1998nv}, Gallex/GNO~\cite{Kaether:2010ag}, SAGE~\cite{Abdurashitov:2009tn},
sensitive only to the interaction rate of electron neutrinos, but not to their energy or arrival time to the detector.
This more detailed information became available with the start-up of the real--time solar neutrino experiment Kamiokande~\cite{Hirata:1991ub}, that confirmed the solar neutrino deficit already observed
by the previous experiments. Its successor Super--Kamiokande, with a volume 10 times larger, has provided very precise observations in almost 20 years of operation. Super--Kamiokande is a
Cherenkov detector that uses 50 kton of ultra pure water as target for solar neutrino interactions, that are detected  through elastic neutrino-electron scattering. This process  is sensitive to all
neutrino flavors, with a larger cross section for $\nu_e$ due to the extra contribution from the charged--current neutrino--electron interaction. The correlation between the incident
neutrino and the recoil electron in the observed  elastic scattering makes possible the reconstruction of the incoming neutrino energy and arrival direction.
After its first three solar phases~\cite{hosaka:2005um,Cravens:2008aa,Abe:2010hy}, Super-Kamiokande is already in its fourth phase, where a very low energy detection threshold of 3.5 MeV
 has been achieved~\cite{Nakano:PhD}. Moreover, during this last period, Super--Kamiokande has reported a 3$\sigma$ indication of Earth matter effects in the solar neutrino flux, with the following measured value of the day--night asymmetry~\cite{Abe:2016nxk,SK-ICHEP:16}
 \be
    A_{DN} = \frac{\Phi_D-\Phi_N}{(\Phi_D + \Phi_N)/2} = (-3.3 \pm 1.0 \, \rm{(stat)} \pm 0.5 \, \rm{(syst)} )~ \% \ .
\ee
Likewise, they have reconstructed a neutrino survival probability consistent with the MSW  prediction at 1$\sigma$~\cite{Wolfenstein:1977ue,Mikheev:1986gs}.

The Sudbury Neutrino Observatory (SNO) used a similar detection technique with 1 kton of pure heavy water as neutrino target.
The use of the heavy water allowed the neutrino detection through three different processes: charged--current  $\nu_e$ interactions with the deuterons
in the heavy water (CC), neutral--current $\nu_\alpha$ with the deuterons (NC), and as well as elastic scattering of all neutrino flavors with electrons (ES).
The measurement of the neutrino rate for each of the three reactions allows the  determination of the $\nu_e$ flux and the total active $\nu_\alpha$ flux
of $^8$B neutrinos from the Sun.
SNO took data during three phases, each of them characterized by a  different way of detecting the neutrinos produced in the neutrino NC interaction
 with the heavy water~\cite{Aharmim:2008kc,Aharmim:2009gd}.

Apart from Super--Kamiokande, the only solar neutrino detector at work nowadays is the Borexino experiment.
Borexino is a liquid--scintillator experiment sensitive to solar neutrinos through the elastic neutrino--electron scattering, with a design optimized  to  measure
the lower energy part of the spectrum. During its first detection phase, Borexino has reported precise observations of the  $^7$Be solar neutrino flux, as well as
the first direct observation of the mono-energetic $pep$ solar neutrinos and the strongest upper bound on the $CNO$ component of the solar neutrino flux~ \cite{Bellini:2013lnn}.
Moreover, Borexino has also measured the solar $^8$B rate with a very low energy threshold of 3 MeV~\cite{Bellini:2008mr} and it has also
provided the first real--time observation of the very low energy $pp$ neutrinos~\cite{Bellini:2014uqa}.\\

The simulation of the production and propagation of solar neutrinos requires the knowledge of the neutrino fluxes produced in the Sun's interior.
This information is provided by the Standard Solar Model (SSM), originally built by John Bahcall~\cite{Bahcallwww}.
The more recent versions of the SSM offer at least two different versions according to the solar metallicity assumed~\cite{Serenelli:2009yc,Vinyoles:2016djt}.
Ref.~\cite{deSalas:2017kay} uses the low metallicity model while \cite{Esteban:2016qun} reports its main results for the high--metallicity model.
For a discussion on the impact of the choice of a particular SSM  over the neutrino oscillation analysis see for instance Refs.~\cite{Schwetz:2008er,Esteban:2016qun}.\\

%% KamLAND %%

KamLAND is a reactor neutrino experiment designed to probe the existence of  neutrino oscillations in the so-called LMA region,  with $\Delta m^2_{21}
\sim 10^{-5} \eVq$. KamLAND detected reactor antineutrinos produced at an average distance of  180 kilometers, providing the first evidence
for the disappearance of neutrinos traveling to a detector from a power reactor~\cite{eguchi:2002dm}. In KamLAND, neutrinos are observed through  the inverse beta
decay process $\bar\nu_e + p \to e^+ + n$, with a delayed coincidence between the positron annihilation and the neutron capture in the medium that allows
the efficient reduction of the background.
The final data sample released by KamLAND contains a total live time of
 2135 days, with a total of 2106 reactor antineutrino events observed to be compared with
2879$\pm$118 reactor antineutrino events plus  325.9$\pm$26.1 background
events  expected in  absence of neutrino oscillations~\cite{Gando:2010aa}.\\

\begin{figure}[t!]
 \centering
\includegraphics[width=0.4\textwidth]{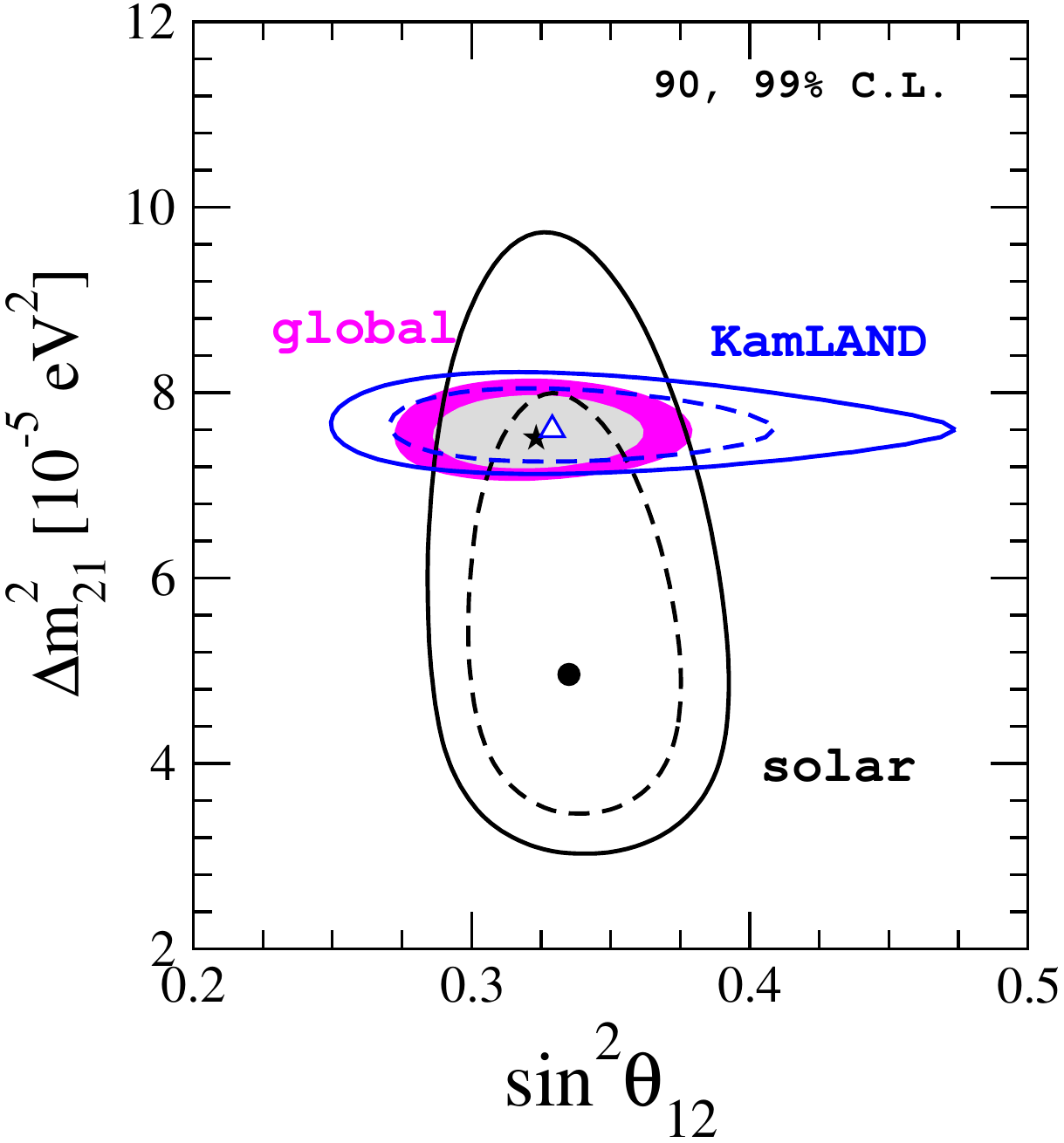}
\caption{\label{fig:sol2017} Allowed regions at 90\% and 99\% C.L. from the analysis of solar data (black lines), KamLAND (blue lines) and the global fit (colored regions).
$\theta_{13}$ has been marginalized according to the latest reactor measurements~\cite{deSalas:2017kay}.}
\end{figure}

Fig.~\ref{fig:sol2017} reports the allowed region in the $\sin^2\theta_{12}$-$\Delta m^2_{21}$ plane from the analysis of all solar neutrino data (black lines),
from the analysis of the KamLAND reactor experiment (blue lines) and from the combined analysis of solar + KamLAND data (colored regions). Here the value of the $\theta_{13}$ has been marginalized following the most recent short--baseline reactor experiments which will be described in the next subsection.
From the figure, one can see that the determination of $\theta_{12}$ is mostly due to solar neutrino experiments, while the
very accurate measurement of $\Delta m^2_{21}$ is obtained thanks to the spectral information from KamLAND.
There is also a mild but noticeable tension between the preferred values of $\Delta m^2_{21}$ by KamLAND and by solar experiments.
While the first one shows a preference for $\Delta m^2_{21} =  4.96\times 10^{-5}$ eV$^2$, the combination of all solar experiments prefer a lower value:
$\Delta m^2_{21} =  7.6\times 10^{-5}$ eV$^2$.
This discrepancy appears at the  2$\sigma$ level. As we will see in the next section, non--standard neutrino interactions have been proposed as a way to solve the tension between solar and KamLAND data.

The best fit point for the global analysis corresponds to:
\be
\sin^2\theta_{12} = 0.321^{+0.018}_{-0.016}\, , \, \Delta m^2_{21} = 7.56 \pm 0.19 \times 10^{-5} \rm{eV}^2\,.
\ee
Maximal mixing is excluded at more than 7$\sigma$.

%%%%  reactor %%%%

\subsection{Short--baseline reactor neutrino experiments and $\theta_{13}$}

Until  recently,  the mixing angle $\theta_{13}$ was pretty much unknown.
Indeed, the only available information about the reactor  angle was an upper-bound obtained from the
non-observation of  antineutrino disappearance at the CHOOZ and Palo Verde reactor
 experiments~\cite{Apollonio:2002gd,Piepke:2002ju}: $\sin^2\theta_{13} < 0.039$ at 90\%~C.L.\ for
 $\Delta m^2_{31} = 2.5\times 10^{-3} \eVq$.
Later on, the interplay between different data samples in the global neutrino
oscillation analyses started showing some sensitivity to the reactor mixing angle $\theta_{13}$.
In particular, from the combined analysis of solar and KamLAND
neutrino data, a non-zero $\theta_{13}$ value was preferred~\cite{Schwetz:2008er,Fogli:2008jx,Maltoni:2008ka}.
The  non-trivial constraint on $\theta_{13}$ mainly appeared as a result of the
different correlation between $\sin^2\theta_{12}$ and
$\sin^2\theta_{13}$ present in the solar and KamLAND neutrino data
samples~\cite{Maltoni:2004ei,Goswami:2004cn}. Moreover,
 a value of $\theta_{13}$ different from zero helped to reconcile the tension between
the  $\Delta m^2_{21}$ best fit points for  solar and KamLAND separately.
Another  piece of evidence for a non-zero value of $\theta_{13}$ was obtained from the combination of atmospheric and long-baseline
neutrino data~\cite{Fogli:2005cq, Escamilla:2008vq}.
Due to a small tension between the preferred values of $|\Delta m^3_{31}|$ at $\theta_{13}=0$  by MINOS experiment and Super-Kamiokande atmospheric neutrino data,
the combined analysis of both experiments showed a preference  for $\theta_{13} > 0~$\cite{GonzalezGarcia:2010er,Schwetz:2011qt,Fogli:2011qn}. \\

%% new generation reactor experiments %%

Nevertheless, the precise determination of $\theta_{13}$ was possible thanks to the new generation of reactor neutrino experiments,
Daya Bay, RENO and Double Chooz.
The main features of these new reactor experiments are, on one side,
 their increased reactor power compared to their predecessors and, on the other side,
 the use of several identical antineutrino detectors located at different distances from the reactor cores.
Combining these two features results in an impressive increase on the number of detected events.
 Moreover, the observed event rate at the closest detectors is used to predict
the expected  number of events at the more distant detectors,  without
relying on the theoretical predictions of the  antineutrino flux at reactors.
Several years ago, in the period between 2011 and 2012, the three experiments
 found  evidence for the disappearance of reactor antineutrinos over short distances, providing the first
measurement of the angle $\theta_{13}$~\cite{Abe:2011fz,An:2012eh,Ahn:2012nd}. We will now briefly
discuss the main details of each experiment  as well as their latest results.\\

%%%% Daya Bay %%%%

The Daya Bay reactor experiment~\cite{An:2012eh} in China  is a multi-detector
and multi-core reactor experiment. Electron antineutrinos produced at six reactor cores
with $2.9\,\text{GW}$ thermal power are observed  at eight antineutrino $20~\text{ton}$ Gadolinium-doped
liquid scintillator detectors, located at distances between $350$ and $2000$~m from the cores.
The latest data release from  Daya Bay has reported the detection of more than 2.5 millions of reactor antineutrino events, after
$1230$~days of data taking~\cite{An:2016ses}. This enormous sample of antineutrino events, together with a significant reduction of systematical errors
has made possible the most precise determination of the reactor mixing angle to date \cite{An:2016ses}
\be
\sin^2 2\theta_{13} = 0.0841 \pm 0.0027 \, \rm{(stat.)} \, \pm 0.0019 \, \rm{(syst.)}\ .
\ee
Likewise, the sensitivity to the effective mass splitting $\Delta m^2_{ee}$ has been substantially improved,\footnote{Ref. \cite{Parke:2016joa} discusses the correct form of the definition of $\Delta m_{ee}^2$.}
\be
|\Delta m_{ee}^2| = 2.50 \pm 0.06\, \rm{(stat.)} \, \pm 0.06 \, \rm{(syst.)}   \times 10^{-3} \,\rm{eV}^2\, ,
\ee
 reaching the accuracy of the long--baseline accelerator experiments, originally designed to measure this parameter.\\

%%% RENO %%%

The RENO experiment~\cite{Ahn:2012nd} in South Korea consists of six aligned reactor cores, equally distributed over a distance of 1.3 km.
Reactor antineutrinos are observed by two identical 16 ton  Gadolinium-doped Liquid Scintillator detectors, located at
approximately  300 (near) and 1400 m (far detector) from the reactor array center.
The RENO Collaboration has recently reported their 500~live days observation of the reactor neutrino spectrum~\cite{RENO:2015ksa,Seo:2016uom},
showing an improved sensitivity to the atmospheric mass splitting, $|\Delta m_{ee}^2| = 2.62^{+0.24}_{-0.26}\, \times 10^{-3} \,\rm{eV}^2$.
Their determination for $\theta_{13}$ is consistent with the results of Daya Bay:
\be
\sin^2 2\theta_{13} = 0.082 \pm 0.009\,\rm{(stat.)}\,\pm0.006\,\rm{(syst.)}
\ee

%%% DChooz %%

The Double Chooz experiment in France detects antineutrinos
produced at two reactor cores in a near and far detectors located at distances of $0.4$~km and $1$~km
 from the neutrino source, respectively~\cite{Abe:2011fz}.
The latest results presented by the Double Chooz collaboration correspond to a period of $818$~days
of data at far detector plus $258$~days of observations with the near detector.
From the spectral analysis of the multi-detector neutrino data, the following best fit value for $\theta_{13}$ is obtained~\cite{dc-MORIOND:17}
\be
 \sin^2 2 \theta_{13} = 0.119 \pm 0.016 \, (\text{stat. + syst.}).
 \ee

\begin{figure}[t!]
 \centering
\includegraphics[width=0.7\textwidth]{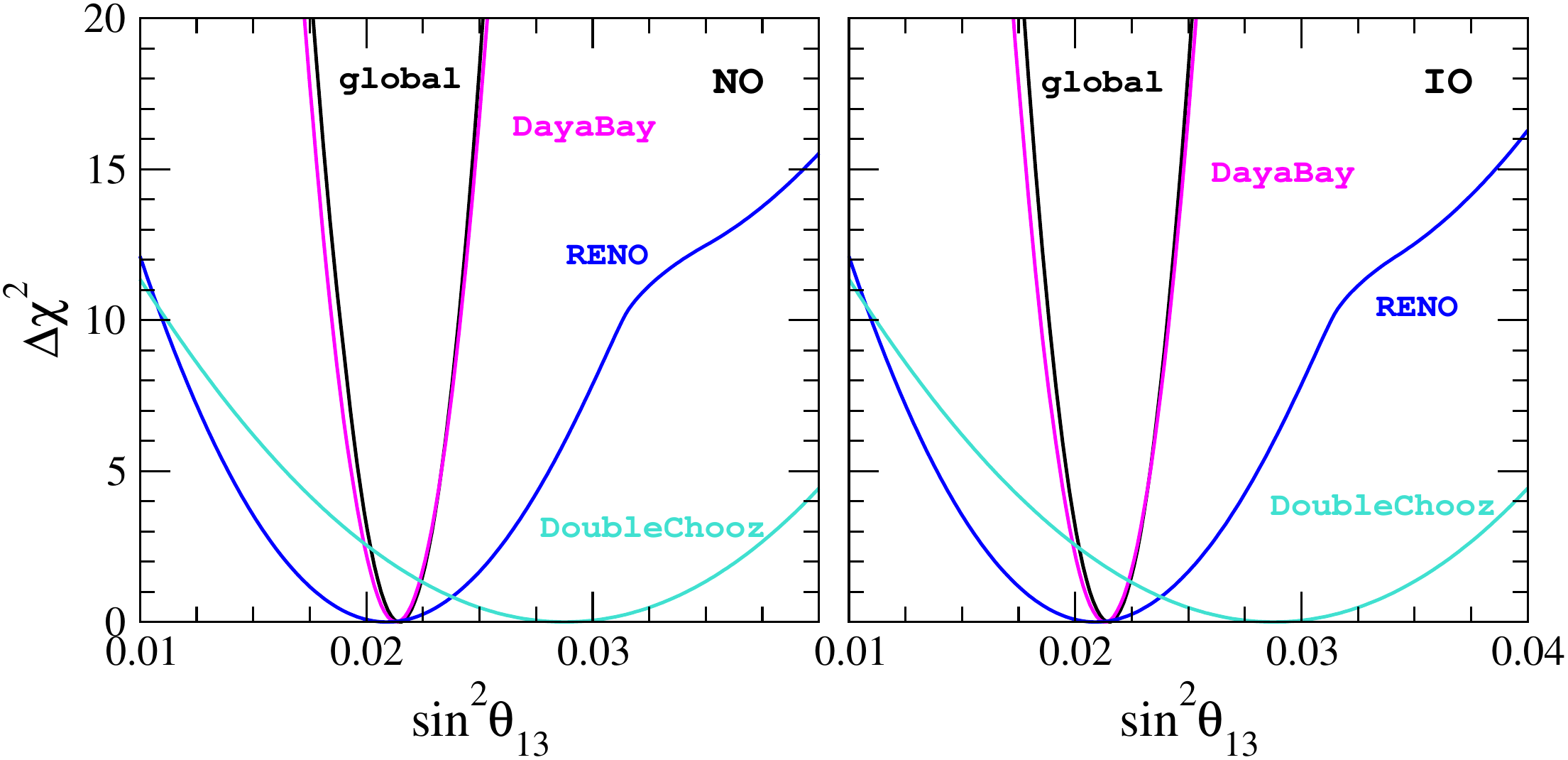}
\caption{\label{fig:sq13}
$\Delta\chi^2$ profile as a function of
    $\sin^2\theta_{13}$ from the analysis of global neutrino data (black line), as well as from the separate
     analysis of the reactor experiments: Daya Bay (magenta), RENO (blue) and Double Chooz (turquoise).
     Left (right) panel corresponds to normal (inverted) mass ordering.
     Figure adapted from Ref.~\cite{deSalas:2017kay}.}
\end{figure}

Fig.~\ref{fig:sq13} illustrates the sensitivity to  $\sin^2\theta_{13}$  obtained from the analysis of  reactor and global neutrino data
for normal and inverted mass ordering. The black line corresponds to the result obtained from the combination of all the reactor neutrino data samples
while the others correspond to the individual reactor data samples, as indicated. One can see from the figure that the more constraining
results come from Daya Bay and RENO experiments, while Double Chooz  shows a more limited sensitivity  to $\theta_{13}$.
Moreover, the global constraint on $\theta_{13}$ is totally dominated by the Daya Bay measurements, with some contributions from RENO
to its lower bound. Notice also that global analyses of neutrino data do not show relevant differences between the preferred value of $\theta_{13}$ for normal
or inverted mass ordering, as we will discuss later. For more details on the analysis of reactor data presented in Fig.~\ref{fig:sq13}, see Ref.~\cite{deSalas:2017kay}.

%%%%  atmospheric %%%%

\subsection{The atmospheric neutrino sector: ($\sin^2\theta_{23}$, $\Delta m^2_{31}$)}

The atmospheric neutrino flux was originally studied as the main source of background for the
nucleon-decay experiments~\cite{Berger:1990rd,BeckerSzendy:1992hr,Hirata:1992ku}.
For several years, most of the dedicated experiments observed  a deficit in the detected number of
atmospheric neutrinos with respect to the predictions.
The solution to this puzzling situation arrived in 1998, when the observation of the zenith angle dependence of the $\mu$-like atmospheric neutrino
data in Super-Kamiokande indicated an  evidence for neutrino oscillations~\cite{Fukuda:1998mi}.
 Some years later, the Super-Kamiokande Collaboration reported a $L/E$ distribution of the atmospheric $\nu_\mu$ data sample characteristic of  neutrino
 oscillations~\cite{Ashie:2004mr}.
Super-Kamiokande has been taking data almost continuously since 1996, being now in its fourth phase. Super-Kamiokande is
sensitive to the atmospheric neutrino flux in the range from 100 MeV to TeV.
The observed neutrino events are classified in three types, fully contained, partially contained
and upward-going muons, based on the topology of the event.
The subsequent data releases by the Super-Kamiokande Collaboration have  increased in complexity. Currently it is
 very complicated to analyze the latest results by independent groups~\cite{Esteban:2016qun,deSalas:2017kay}.
From the analysis of the latest Super-Kamiokande atmospheric data, the following best fit values have been obtained for
the oscillation parameters~\cite{Koshio:2017ngh}:
\be
\sin^2\theta_{23} = 0.587\, , \quad \Delta m^2_{32} = 2.5 \times 10^{-3} \rm{eV}^2 \,.
\ee
Thus, a slight preference for $\theta_{23}>\pi/4$ is reported. Likewise, the normal mass ordering  ({\it i.e.,} $\Delta m_{31}^2>0$) is preferred over the inverted one ({\it i.e.,} $\Delta m_{31}^2<0$) .\\

%%   neutrino telescopes  %%

 In recent years, atmospheric neutrinos are also  being detected by neutrino telescope experiments. IceCube and ANTARES,
originally designed to detect higher energy neutrino fluxes, have reduced their energy threshold  in such a way that they can
measure the most energetic part of the atmospheric neutrino flux.\\

The ANTARES telescope~\cite{Collaboration:2011nsa}, located under the Mediterranean Sea, observes atmospheric neutrinos with energies as low as 20 GeV.
Neutrinos are detected via the Cherenkov light emitted after the neutrino interaction with the medium in the vicinity of the detector.
In Ref.~\cite{AdrianMartinez:2012ph}, the ANTARES Collaboration has analyzed the atmospheric neutrino data collected  during a period of 863 days. Their results for the oscillation parameters are in good agreement with current world data. For the first time, ANTARES results have also been included in a global neutrino oscillation fit~\cite{deSalas:2017kay}.\\

The IceCube DeepCore detector is a sub-array of the IceCube neutrino observatory, operating at the South Pole~\cite{Collaboration:2011ym}.
DeepCore was designed with  a denser instrumentation compared to  IceCube, with the goal  of lowering the energy threshold for the detection of atmospheric
muon neutrino events below 10 GeV. Neutrinos are identified trough the Cherenkov radiation emitted by the secondary particles produced after their interaction in the ice.
The most recent data published by DeepCore  correspond to a live time of three years~\cite{Aartsen:2014yll}.  A total of 5174 atmospheric neutrino events were observed,
compared to a total 6830 events  expected in absence of neutrino oscillations. The obtained best fit values for the atmospheric neutrino parameters
$\sin^2\theta_{23} = 0.53^{+0.09}_{-0.12}$ and $\Delta m^2_{32} = 2.72^{+0.19}_{-0.20} \times 10^{-3}$ eV$^2$ are also compatible with the atmospheric results of the
Super--Kamiokande experiment.\\

The left panel of Fig.~{\ref{fig:all-atm} shows  the allowed regions at 90\% and 99\% C.L. in the
atmospheric neutrino oscillation parameters $\sin^2\theta_{23}$ and $\Delta m^2_{31}$ obtained from ANTARES, DeepCore and
Super-Kamiokande phases I to III~\cite{deSalas:2017kay}.
From the combination one sees that DeepCore results start being competitive with  the determination of the
 atmospheric  oscillation parameters by long-baseline experiments. Indeed, a recent reanalysis of DeepCore
atmospheric data~\cite{Aartsen:2017nmd} shows an improved sensitivity with respect to the region plotted in Fig.~{\ref{fig:all-atm}.
The sensitivity of ANTARES shown in Fig.~{\ref{fig:all-atm} is not yet competitive with the other experiments. However, it is expected that the
ANTARES collaboration will publish an updated analysis that will certainly  improve their sensitivity to the atmospheric neutrino parameters.

\begin{figure}
\centering
\includegraphics[width=0.4\textwidth]{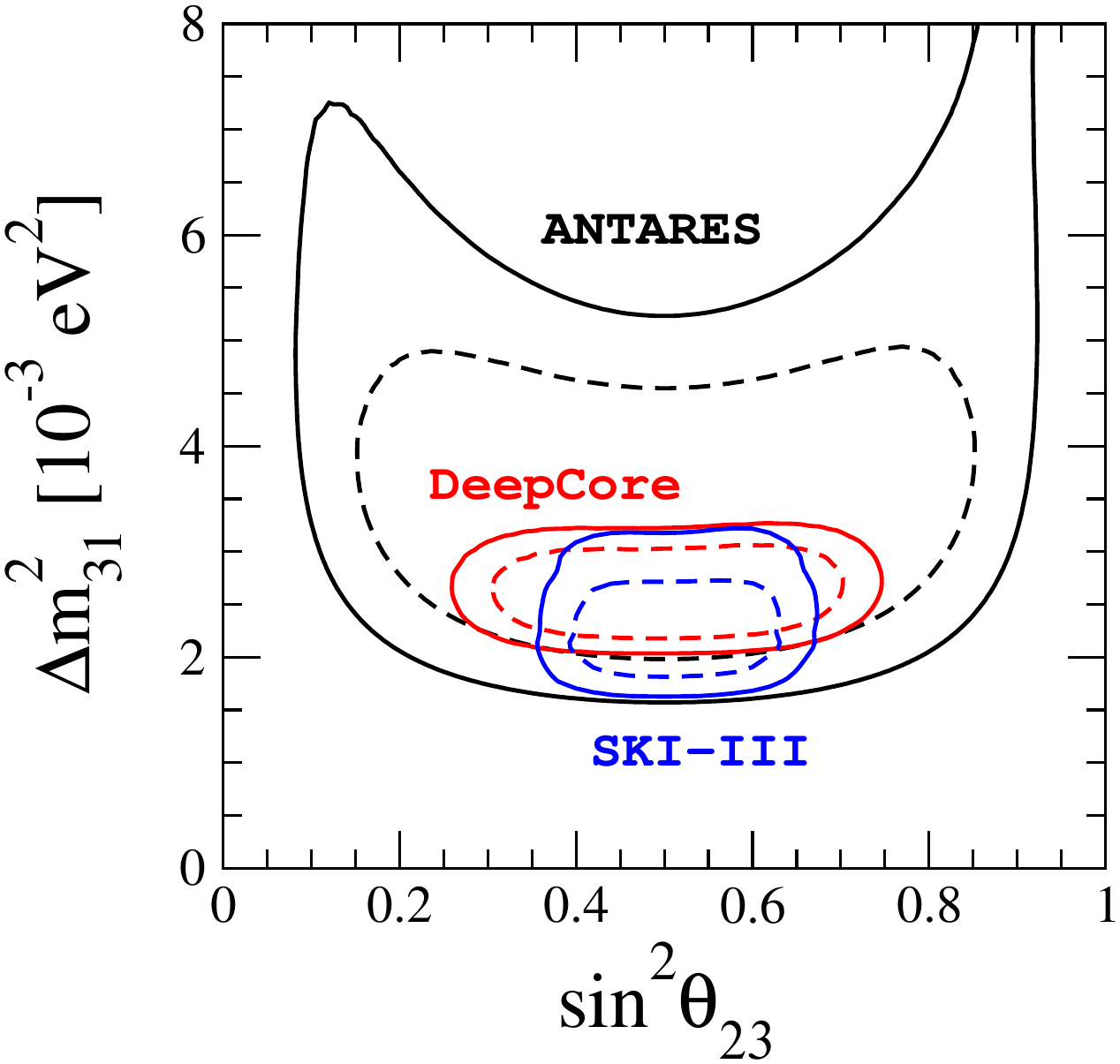}
\includegraphics[width=0.4\textwidth]{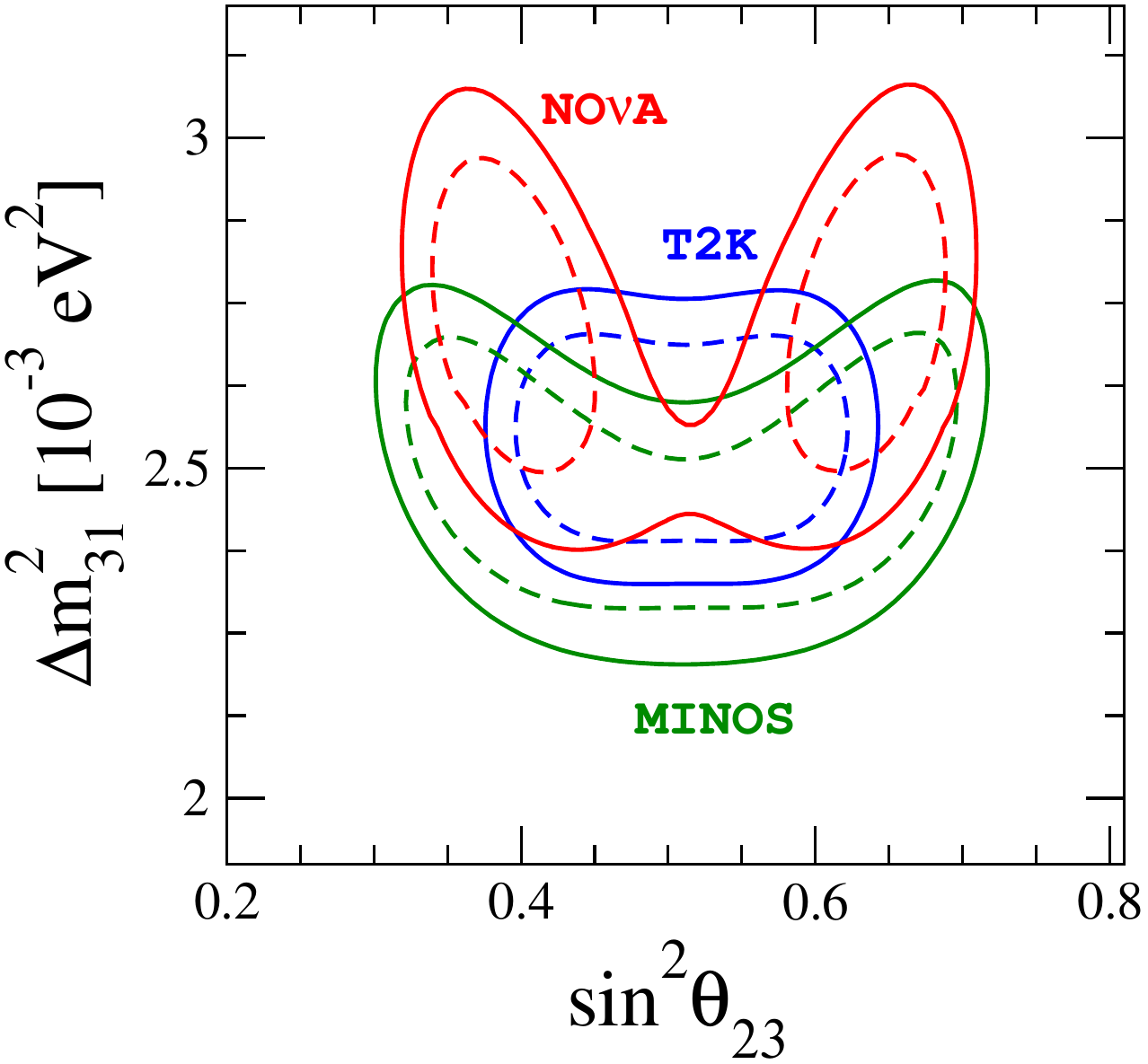}
\caption{
90\% and 99\% C.L. allowed regions at the $\sin^2\theta_{23}$--$\Delta m^2_{31}$ plane obtained from the atmospheric (left panel) and
   long--baseline accelerator experiments (right panel), see the text for details. Notice the different scale in the $\Delta m^2_{31}$ parameter.
   Both plots correspond to the normal ordered neutrino mass spectrum. Figures adapted from Ref.~\cite{deSalas:2017kay}.}
\label{fig:all-atm}
\end{figure}

%%%%  accelerator %%%%

\subsection{Long--baseline accelerator experiments}

After the discovery of neutrino oscillations in the atmospheric neutrino flux, several long-baseline accelerator
experiments were planned to confirm the oscillation phenomenon with a man--made neutrino source.
The first two experiments trying to probe the $\nu_\mu$ disappearance oscillation channel in
the same region of $\Delta m^2$ as explored by atmospheric neutrinos were K2K and MINOS.
Their successors, T2K and NO$\nu$A are still  at work today.\\

%% K2K %%

The KEK to Kamioka (K2K) experiment used a  neutrino beam  produced by a 12~GeV proton beam from the
KEK proton synchrotron.  The neutrino beam was detected by  a near detector 300~m away from the proton target and by the Super--Kamiokande detector, at a distance of 250~km.
The number of detected neutrino events, as well as the spectral distortion of the neutrino flux observed by K2K was fully consistent with
the hypothesis of neutrino oscillation~\cite{Ahn:2006zza}.\\

 %% MINOS %%

The Main Injector Neutrino Oscillation Search (MINOS) experiment observed neutrino oscillations from a beam produced
by the NuMI (Neutrinos at Main Injector) beamline at Fermilab in an underground detector located at the Soudan Mine, in Minnesota, 735 km away.
MINOS searched for  oscillations in the disappearance ($\nu_\mu \to \nu_\mu$) and appearance channels ($\nu_\mu \to \nu_e$), for neutrinos and antineutrinos as well.
After a period of 9 years, the  MINOS experiment collected a data sample corresponding to an exposure of $10.71\times 10^{20}$ protons on
target (POT) in the neutrino mode,  and $3.36\times 10^{20}$ POT in the antineutrino mode ~\cite{Adamson:2013whj,Adamson:2013ue}.
The combined analysis of all MINOS data shows a slight preference for inverted mass ordering and $\theta_{23}$ below maximal as well as a disfavored status for maximal mixing with $\Delta\chi^2 = 1.54$~\cite{Adamson:2014vgd}.
The allowed ranges for the atmospheric parameters  from the joint analysis of all MINOS data are the following
\begin{eqnarray}
& \sin^2\theta_{23} \in  [0.35, 0.65] \, \, (90\% \,\rm{C.L.})\, , & |\Delta m^2_{32}| \in [2.28, 2.46]\times 10^{-3} \, \rm{eV}^2 \, \, (1\sigma) \, \, \rm{for} \, \rm{normal}\,\rm{ordering}\\
& \sin^2\theta_{23} \in  [0.34, 0.67] \, \, (90\% \,\rm{C.L.})\, , & |\Delta m^2_{32}| \in [2.32, 2.53]\times 10^{-3} \, \rm{eV}^2 \, \, (1\sigma) \, \, \rm{for} \, \rm{inverted}\,\rm{ordering}\,.
\end{eqnarray}
\\

%% T2K %%

The Tokai to Kamioka (T2K) experiment uses a neutrino beam consisting
mainly of muon neutrinos, produced at the J-PARC accelerator facility
and observed at a distance of 295~km and an off-axis angle of
$2.5^\circ$ by the Super-Kamiokande detector.
The most recent results of the T2K collaboration for the neutrino and antineutrino channel have been published
  in Refs.~\cite{PhysRevD.96.011102,Abe:2017uxa}.
  A separate analysis of the disappearance data in the neutrino and antineutrino channels has provided the determination of the best fit oscillation parameters  for neutrinos and
  antineutrinos~\cite{PhysRevD.96.011102}. The obtained results are consistent so, no hint for CPT violation in the neutrino sector has been obtained~\cite{Barenboim:2002tz,Barenboim:2002rv}~\footnote{See Ref.~\cite{Barenboim:2017ewj} for updated bounds on CPT violation from neutrino oscillation data.}. In both cases, the preferred value of the atmospheric  angle is compatible with  maximal mixing.
The combined analysis of the neutrino and antineutrino appearance and disappearance searches in T2K, that corresponds to a total sample of  $7.482\times 10^{20}$
POT in the neutrino mode, and  $7.471\times 10^{20}$ POT in the antineutrino mode, results in the best determination of the atmospheric oscillation
parameters to date~\cite{Abe:2017uxa}
\be
  \sin^2\theta_{23} = 0.532 \, (0.534) \, , \, |\Delta m^2_{32}| =2.545 \, (2.510)\times10^{-3} \, \rm{eV}^2\, ,
\ee
 for  normal  (inverted) mass ordering  spectrum. Furthermore, thanks to the combination of neutrino and antineutrino data, T2K
 has already achieved a mild  sensitivity to the CP violating phase, reducing the allowed  90\% C.L. range of $\delta$ in radians to $[-3.13, -0.39]$ for normal
 and $[-2.09, -0.74]$ for inverted mass ordering~\cite{Abe:2017uxa}.\\

%% NOvA %%

In the NO$\nu$A (NuMI Off-Axis $\nu_e$ Appearance) experiment,  neutrinos produced at the Fermilab's NuMI beam are detected in
Ash River, Minnesota, after traveling 810 km  through the Earth. In the same way as the T2K experiment, the NO$\nu$A far
 detector is located slightly off the centerline of the neutrino beam coming from Fermilab. Thanks to this configuration, a large neutrino flux  is
 obtained at energies close to 2 GeV, where the maximum of the muon to electron neutrino oscillations is expected.
The most recent data release from the NO$\nu$A collaboration corresponds to an accumulated statistics of
6.05$\times 10^{20}$ POT in the neutrino run~\cite{Adamson:2017qqn,Adamson:2017gxd}.
For the muon antineutrino disappearance channel,  78 events have been observed,
to be compared with 82 events expected for oscillation and  $473\pm 30$ events predicted under the
no-oscillation hypothesis. The searches for $\nu_\mu \to \nu_e$ transitions in the accelerator neutrino flux have reported the
observation of 33 electron neutrino events, with an expected background of 8.2$\pm$0.8 $\nu_e$ events.
The analysis of the NO$\nu$A Collaboration  disfavors maximal values of $\theta_{23}$ at the $2.6\sigma$ level~\cite{Adamson:2017qqn}.
On the other hand, from the analysis of the appearance channel it is found that
the inverted mass ordering is disfavored at 0.46$\sigma$, due to the small number of event predicted for this ordering in comparison to the observed results ~\cite{Adamson:2017gxd}.
Furthermore,  the combination of appearance and disappearance NO$\nu$A data  with the $\theta_{13}$ measurement at the reactor experiments results disfavors
 the scenario with inverted neutrino mass ordering and
 $\theta_{23} < \pi/2$ at 93\% C.L., regardless of the value of $\delta$~\cite{Adamson:2017gxd}}.\\

%%%% Figure

The right panel of Fig.~{\ref{fig:all-atm} shows  the 90\% and 99\% C.L. allowed region in the
atmospheric neutrino oscillation parameters $\sin^2\theta_{23}$ and $\Delta m^2_{31}$ according to the MINOS, T2K and NO$\nu$A data for normal mass ordering~\cite{deSalas:2017kay}.
Note the different scale for $\Delta m^2_{31}$ with respect to the left panel. The three long--baseline experiments  provide similar constraints on this parameter,
while the constraint on $\theta_{23}$ obtained from T2K is a bit stronger. One can also see some small differences between the preferred values of   $\theta_{23}$ by the three experiments.
While T2K prefers maximal mixing, MINOS and NO$\nu$A show a slight preference for non-maximal $\theta_{23}$. In any case, these differences are not significant and the agreement among the three experiments is quite good. Although not shown here, the agreement for inverted mass ordering is a bit worse, since in that case the rejection of NO$\nu$A against maximal mixing is stronger, whereas the preference for $\theta_{23} \sim \pi/4$ in T2K remains the same as for normal ordering.

%%%%  global %%%%

\subsection{Global fit to neutrino oscillations}

In the previous subsections, we have reviewed the different experimental neutrino data samples, discussing their dominant sensitivity to one or two oscillation parameters.
However, every data sample offers subleading sensitivities to other parameters as well.  Although the information they can provide about such parameters may be limited, in combination with the rest of data samples, relevant information can emerge.
This constitutes the main philosophy behind global analyses of neutrino oscillation data: joint analyses trying to exploit the complementarity of the different experiments to
 improve our knowledge on the neutrino oscillation parameters. Here, we will show the results of a combined analysis of neutrino oscillation data in the framework of the
 three-flavor neutrino oscillation scheme.\\

Fig.~\ref{fig:2dimfit} reports the 90\%, 95\% and 99\% C.L. allowed regions in the parameters
$\sin^2\theta_{23}$, $\sin^2\theta_{13}$, $|\Delta m^2_{31}|$ and  $\delta$  from the global fit in Ref.~\cite{deSalas:2017kay} for normal and inverted mass ordering. For the allowed regions in the solar plane $\sin^2\theta_{12}$--$\Delta m^2_{21}$, see Fig.~\ref{fig:sol2017}. The best fit points, along with the corresponding 1$\sigma$ uncertainties and 90\% C.L. ranges for each parameter, are quoted
 in Table.~\ref{tab:fit2017}.
\begin{figure}[h!]
 \centering
 \includegraphics[width=0.6\textwidth]{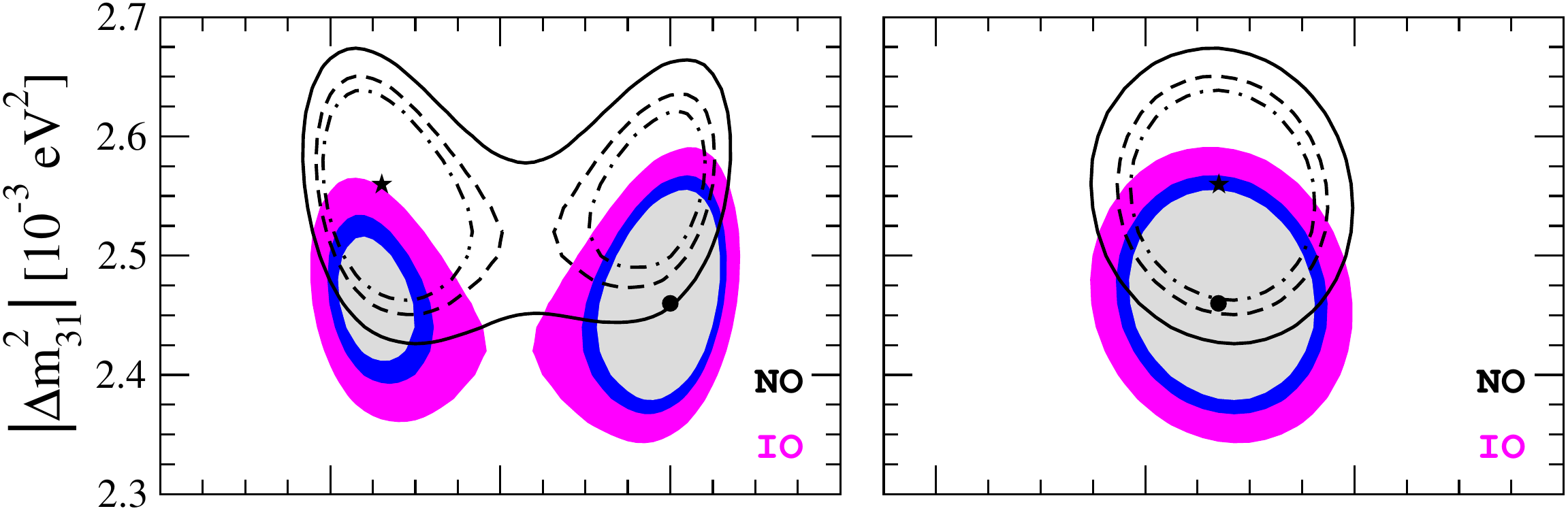}\\
\hspace{5.5mm}\includegraphics[width=0.6\textwidth]{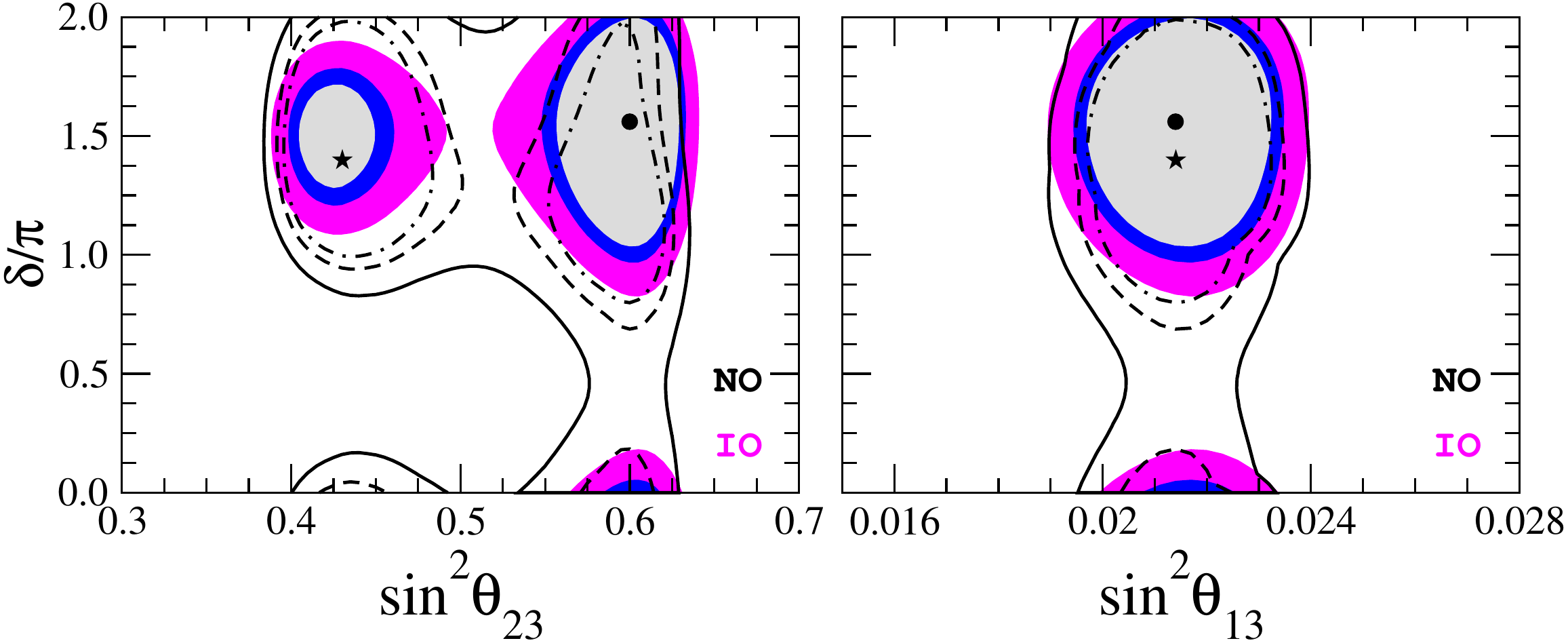}
\caption{\label{fig:2dimfit} Allowed regions at 90\%, 95\%  and 99\% C.L. in the planes $\sin^2\theta_{23}$--$|\Delta m^2_{31}|$,  $\sin^2\theta_{13}$--$|\Delta m^2_{31}|$, $\sin^2\theta_{23}$--$\delta$ and   $\sin^2\theta_{13}$--$\delta$ for normal (lines) and inverted mass ordering (colored regions). The star indicates the global best fit point, corresponding to normal ordering, while the circle indicates the local minimum in inverted ordering. Adapted from Ref.~\cite{deSalas:2017kay}.}
\end{figure}
\begin{table}[h!]\centering
  \catcode`?=\active \def?{\hphantom{0}}
   \begin{tabular}{lcc}
 \hline\hline\\[-2.5mm]
    parameter & best fit $\pm$  $1\sigma$  &  90\% C.L. range \\
    \hline\hline\\[-2.5mm]
        $\Delta m^2_{21}\: [10^{-5}\eVq]$
    & 7.56$\pm$0.19  &  7.26--7.87 \\[3mm]
 \, 	$\Delta m^2_{31}\: [10^{-3}\eVq]$ (NO)
    &  2.55$\pm$0.04 &  2.48--2.62 \\
     $|\Delta m^2_{31}|\: [10^{-3}\eVq]$ (IO)
    &  2.47$^{+0.04}_{-0.05}$ & 2.40--2.53  \\[3mm]
    $\sin^2\theta_{12} / 10^{-1}$
    & 3.21$^{+0.18}_{-0.16}$ & 0.294--0.352\\  [3mm]
     $\sin^2\theta_{23} / 10^{-1}$ (NO)
              &	4.30$^{+0.20}_{-0.18}$$^a$
                                      & 0.403--0.466  \&  0.577--0.608   \\
          $\sin^2\theta_{23} / 10^{-1}$ (IO)
              & 5.98$^{+0.17}_{-0.15}$$^b$
                                         & 0.569--0.623  \\[3mm]
    $\sin^2\theta_{13} / 10^{-2}$ (NO)
    & 2.155$^{+0.090}_{-0.075}$ & 0.0201--0.0228  \\
        $\sin^2\theta_{13} / 10^{-2}$ (IO)
    & 2.155$^{+0.076}_{-0.092}$ & 0.0201--0.0228 \\[3mm]
   $\delta/\pi$ (NO)
   	& 1.40$^{+0.31}_{-0.20}$ & 0.98--2.00  \\
       $\delta/\pi$ (IO)	
   	& 1.56$^{+0.22}_{-0.26}$ &  1.15--1.90\\[1mm]
                \hline\hline \\[-1mm]
                  \multicolumn{3}{l}{ \footnotesize{$^a$ Local min. at
                $\sin^2\theta_{23}$=0.596 with $\Delta\chi^2 = 2.1$ w.r.t. the
                global min.}}\\
       \multicolumn{3}{l}{ \footnotesize{$^b$ Local min.  at
                $\sin^2\theta_{23}$=0.426 with $\Delta\chi^2 = 3.0$ w.r.t. the
                global min. for IO.}}
     \end{tabular}
     \caption{ \label{tab:fit2017}
        Neutrino oscillation parameters summary determined from the
        global analysis in Ref.~\cite{deSalas:2017kay}. The ranges for inverted ordering refer to the
        local minimum of this neutrino mass ordering.
     }
\end{table}
The relative uncertainties on the oscillation  parameters at 1$\sigma$ range from around 2\% for the mass splittings to 7-10\% (depending on the mass ordering) for $\sin^2\theta_{23}$.
 In case of the CP phase, the 1$\sigma$ uncertainties are of the order of 15-20\%. Note also that, at the 3$\sigma$ level, the full range of $\delta$ is still allowed for normal ordering.
 For the case of inverted ordering, a third of the total range is now excluded at  the  3$\sigma$ level.
These results are in good agreement with Refs.~\cite{Capozzi:2017ipn,Esteban:2016qun}.\\

Despite the remarkable sensitivity reached in the determination of most of the neutrino oscillation parameters,
there are still three unknown parameters in the oscillation of standard three neutrino scheme: the octant of  $\theta_{23}$, the value of the CP phase $\delta$ and the neutrino mass ordering. The current status of these still unknown parameters will be discussed next.

%% theta_23 octant %%
Let us now comment on the maximality/non-maximality and octant preference for the atmospheric mixing angle.
So far, experimental neutrino data have not shown a conclusive preference for values of $\theta_{23}$ smaller, equal or larger than $\pi/4$.
Different experiments may show a limited preference for one of the choices, but for the moment all the results are consistent at the 3$\sigma$ level.
On the other hand, one finds that the available  global analyses of neutrino data~\cite{Esteban:2016qun,Capozzi:2017ipn,deSalas:2017kay}, using very similar data samples
 show slightly different results for the octant preference. For this particular case, one can find the origin of the possible discrepancies in the different treatment of the
 Super--Kamiokande atmospheric  data. See the previous references for more details on the chosen approach at each work.
The results in Fig.~\ref{fig:2dimfit} and Table \ref{tab:fit2017}, corresponding to the analysis in Ref.~\cite{deSalas:2017kay}, show a preference for $\theta_{23}$ in the first octant.
This global best fit point corresponds to normal mass ordering, but a local minimum can also be found with $\theta_{23} > \pi/4$ and inverted mass ordering with a $\Delta\chi^2 = 4.3$.
In the same way, additional local minima can be found with $\theta_{23}$ in the second octant and inverted mass spectrum and the other way around. All these possibilities are allowed
 at 90\% C.L. as can be seen in the right panels of Fig.~\ref{fig:2dimfit}.
With current data, the status of the maximal atmospheric mixing is a bit delicate, being allowed only at 99\% C.L. However, this result may change after the implementation of the
partially published data release of T2K~\cite{t2k-hartz} in the global fit.\\

%% neutrino mass ordering %%%

In the same way, the current neutrino oscillation data do not offer a definitive determination for the neutrino mass ordering.
Individual neutrino experiments show in general a  limited sensitivity to the mass ordering, with the exception of the latest atmospheric data from Super-Kamiokande, that prefer normal mass
ordering with a significance of $\Delta\chi^2 = 4.3$. Note however that this data sample is not included in some of the global analyses of neutrino
oscillations~\cite{Esteban:2016qun,deSalas:2017kay}.
The sensitivity to the mass ordering in the global analysis arises instead from the interplay of the different neutrino data, as a result of the existing correlations and tensions among the other neutrino parameters. Indeed, the three global analysis discussed in this review show a preference for  normal mass ordering, although the significance may be different in each case, depending on the particular details of the specific global fit. In the work in Ref.~\cite{deSalas:2017kay}, discussed in a bit more details here, a preference for  normal ordering over inverted is obtained, with a significance of
$\Delta\chi^2 = 4.3$.
In any case, the results reported are not conclusive yet, and we will have to wait for the next generation of experiments devoted to this purpose (among others), such as
DUNE~\cite{Acciarri:2015uup}, PINGU~\cite{Aartsen:2014oha}, ORCA~\cite{Brunner:2015ltd}, JUNO~\cite{An:2015jdp} or RENO-50~\cite{Kim:2014rfa}.\\

%% delta_CP %%

Finally, we comment on the sensitivity to the CP-violating phase $\delta$. Prior to the publication of the antineutrino run data from T2K, combined analyses were already showing a
weak preference for $\delta = 3\pi/2$, while $\delta = \pi/2$ was disfavored above the 2$\sigma$ level~\cite{Forero:2014bxa,Capozzi:2013csa,Gonzalez-Garcia:2014bfa}. This sensitivity,
absent in all the individual data samples, emerged from the tension between the  value of $\theta_{13}$ measured at the reactor experiments and the preferred value of $\theta_{13}$ for
$\delta = \pi/2$ in T2K.
This scenario has changed after the release of T2K results from its antineutrino run and now the sensitivity to $\delta$ comes mainly from the combined analysis of the neutrino and antineutrino channel in T2K. The remaining experiments contribute only marginally to the determination of the CP--violating phase.

%%%%%%%%%%%%%%%%%%%%%%%%%%

\section{Current bounds on non--standard interactions\label{non-stan}}

New neutrino interactions beyond the Standard Model are  natural
features in most neutrino mass models~\cite{Ohlsson:2012kf,Miranda:2015dra}.
As commented in the introduction, these Non--Standard Interactions (NSI) may be of Charged-Current (CC) or of Neutral-Current (NC)
type.
In the low energy regime, neutrino NSI with matter fields can be formulated in terms of
the  effective four-fermion Lagrangian terms as follows:
%
%--------------------------------------------------------------------------------------------------------------------------------------
\begin{eqnarray}
  \mathcal{L}_{\rm CC-NSI} \, & = & \,
 -2\sqrt{2} G_{F} \, {\epsilon}^{ff^\prime X}_{\alpha\beta}
     \left(\bar{\nu}_\alpha \gamma^{\mu} P_L \ell_\beta \right)
     \left(\bar{f}^\prime \gamma_{\mu} P_X f \right)\, ,
  \label{eq:CC-NSI-Lagrangian}\\
%\end{equation}
%--------------------------------------------------------------------------------------------------------------------------------------
%
%--------------------------------------------------------------------------------------------------------------------------------------
%\begin{equation}
  \mathcal{L}_{\rm NC-NSI} \, & = &  \,
  - 2\sqrt{2} G_{F}  \, {\epsilon}^{f X}_{\alpha\beta}
      \left(\bar{\nu}_\alpha \gamma^{\mu} P_L \nu_\beta \right)
      \left(\bar{f}\gamma_{\mu} P_X f \right) \, .
  \label{eq:NC-NSI-Lagrangian}
\end{eqnarray}
%--------------------------------------------------------------------------------------------------------------------------------------
%
where  $G_F$ is the Fermi constant and $P_X$ denote the left and right chirality projection operators $P_{R,L} = (1\pm \gamma_5)/2$. The dimensionless
coefficients $\epsilon_{\alpha\beta}^{ff' X}$ and  $\epsilon_{\alpha\beta}^{f X}$ quantify the strength of the NSI between
 leptons of  $\alpha$ and $\beta$ flavour and the matter field $f\in\{ e, u,d \}$ (for NC-NSI) and $f\ne f'\in \{u,d \}$ (for CC-NSI).   At the limit $\epsilon_{\alpha \beta}^{f X}\to 0$, we recover the standard interactions, while $\epsilon_{\alpha \beta}\sim 1$ corresponds to new interactions with strength comparable to that of SM weak interactions. If  $\epsilon_{\alpha \beta}$ is non-zero for $\alpha \ne \beta$, the NSIs violate lepton flavor. If $\epsilon_{\alpha \alpha}-\epsilon_{\beta \beta} \ne 0$, the lepton flavor universality is violated by NSI.

The presence of  neutrino NSI may affect the
neutrino production and detection at experiments as well as their propagation in a medium through modified matter effects
~\cite{Wolfenstein:1977ue,Mikheev:1986gs}.
In the literature, it is common denoting the CC-NSI couplings as
$\epsilon^{s}_{\alpha\beta}$  or $\epsilon^{d}_{\alpha\beta}$ since they may often affect the
source ($s$) and detector ($d$) interactions at neutrino experiments.
On the other hand,  $\epsilon^{m f}_{\alpha\beta}$ is used to refer to the NC-NSI
couplings with the fundamental fermion $f$ generally affecting the neutrino propagation in matter ($m$).
In this case, what is relevant for neutrino propagation in a medium is the vector part of interaction
$\epsilon_{\alpha\beta}^{f V} = \epsilon_{\alpha \beta}^{f L} + \epsilon_{\alpha \beta}^{f R}$.
In fact, the neutrino propagation in a medium is sensitive to the following combinations~\footnote{Note that some  references use the definition $\epsilon_{\alpha\beta} = \sum_f \frac{N_f}{N_d}\epsilon_{\alpha\beta}^{fV}$. It is very relevant to distinguish between both notations,
since the reported bounds will be different by a factor of 3. For this reason, we prefer to quote directly the results in terms of the effective lagrangian coefficients $\epsilon_{\alpha\beta}^{fV}$. In any case, for the analysis including Earth matter effects, the relation between the corresponding NSI couplings
is straightforward.}
\be
\epsilon_{\alpha \beta} \equiv \epsilon_{\alpha \beta}^{eV}+\frac{N_u}{N_e} \epsilon_{\alpha \beta}^{uV}+\frac{N_d}{N_e} \epsilon_{\alpha \beta}^{dV}\label{combination}
\ee
so most of the bounds from oscillation experiments are presented in the literature in terms of $\epsilon_{\alpha\beta}$ rather than in terms of $\epsilon_{\alpha\beta}^{fV}$. Inside the Sun, $N_u/N_e \simeq 2 N_d/N_e \simeq 1$~\cite{Serenelli:2009yc} and inside the Earth, $N_u/N_e \simeq  N_d/N_e \simeq 3$~\cite{Lisi:1997yc}.
When studying the effect of NSI at neutrino detection, there will be independent sensitivity for the left and right chirality coefficients
$\epsilon_{\alpha \beta}^{f L}$  and $\epsilon_{\alpha \beta}^{f R}$.

Although this kind of interactions has not been confirmed experimentally, their potential  effects have been extensively
studied in a large variety of physical scenarios. As a result, stringent bounds on their strength have been derived \cite{Miranda:2015dra, Ohlsson:2012kf,Gonzalez-Garcia:2013usa}. Moreover,
it has been shown that NSI may interfere with neutrino oscillations in different contexts, giving rise to parameter degeneracies
that can affect the robustness of the neutrino parameter determination.
In this section, we will review these results.

%%%% solar %%%%%

\subsection{NSI in solar experiments}

NSI may affect the propagation of solar neutrinos within the Sun and the Earth as well as the detection, depending on the type of NSI considered.
Before the confirmation of neutrino oscillation as the phenomenon responsible for the solar neutrino anomaly by the KamLAND experiment,
NSI with massless neutrinos was also proposed as the mechanism behind this anomaly~\cite{Valle:1987gv,Roulet:1991sm, Guzzo:1991hi,Barger:1991ae, Guzzo:2001mi}.
After KamLAND confirmed the phenomenon of mass--induced electron neutrino (antineutrino) oscillation, NSI was excluded as the main mechanism behind the solar neutrino oscillations,
 although its presence  has been considered at subleading level in solar neutrino experiments, see for instance~\cite{Friedland:2004pp, Miranda:2004nb,Guzzo:2004ue,Escrihuela:2009up,Gonzalez-Garcia:2013usa}.
These analyses have found that a small amount of NSI,  $\epsilon_{ee}^{dV} \simeq 0.3$,  is in better agreement with data than the standard solution at the level of 2$\sigma$. Ref.~\cite{Palazzo:2011vg} finds the best fit value to lie at $\epsilon_{e\tau}^{dV}- \epsilon_{e\mu}^{dV}=0.23$.
On one hand, this result is due to the non--observation of the upturn of the solar neutrino spectrum predicted by the standard LMA--MSW solution at around 3 MeV~\cite{Friedland:2004pp,Miranda:2004nb,Palazzo:2011vg}. On the other hand,
 there exists a small tension between the preferred value of $\Delta m^2_{21}$ by KamLAND and by solar experiments that can be eased by introducing NSI.
 More surprisingly, these studies revealed an alternative solution to the standard LMA--MSW, known as LMA-Dark or LMA-D solution~\cite{Miranda:2004nb,Escrihuela:2009up,Gonzalez-Garcia:2013usa}, requiring NSI with strength $\epsilon_{\tau\tau}^{dV}-\epsilon_{ee}^{dV}\simeq 1$.
 \begin{figure}
 \centerline{
\includegraphics[width=0.7\textwidth]{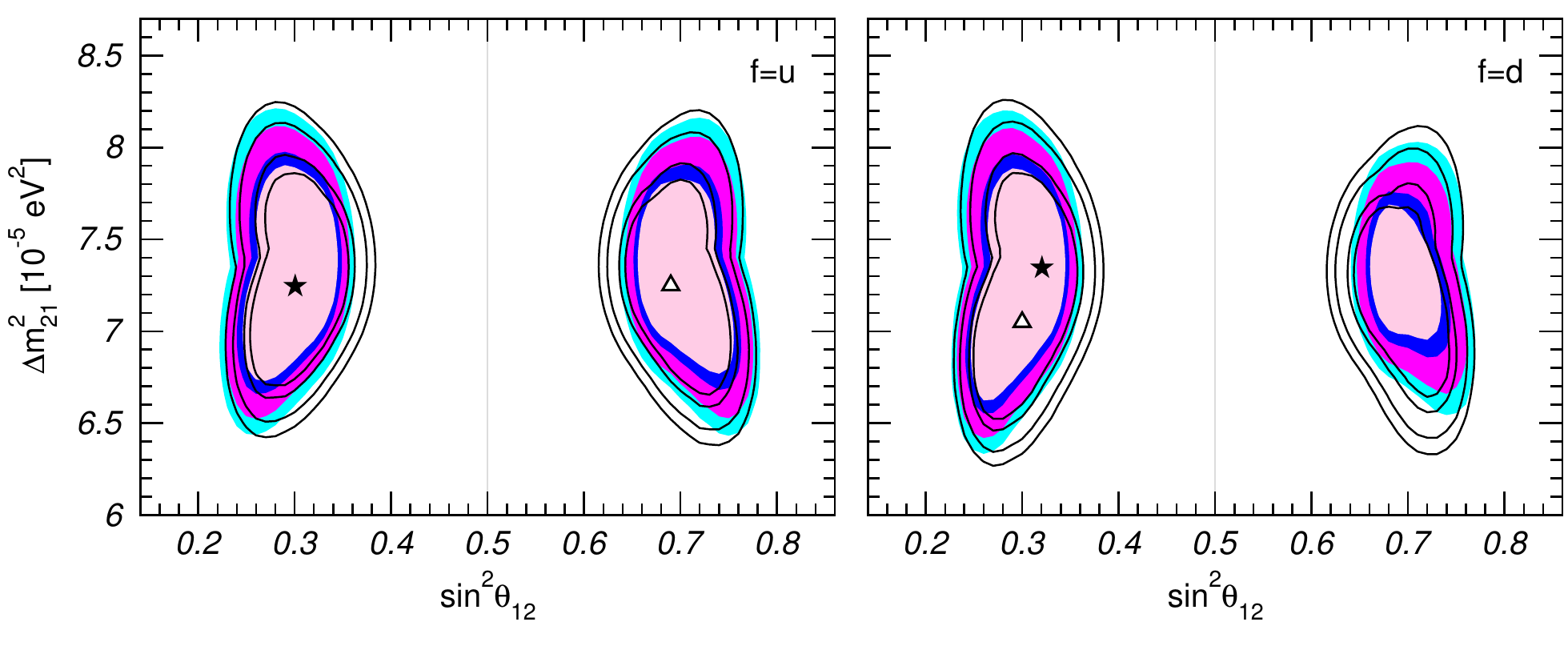}
}
\caption{
Allowed regions at 90\%, 95\%, 99\% and 3$\sigma$ C.L. (2 d.o.f.)
from the analysis of solar and KamLAND data in the presence of NSI with up (left) and down (right panel) quarks. The colored filled and contour regions in each panel
correspond to different analysis of solar SNO  data.  This figure has been taken from \cite{Gonzalez-Garcia:2013usa}, published under the terms of the Creative Commons Attribution Noncommercial License and therefore no copyright permissions were required for its inclusion in this manuscript. See this reference for further details.}
\label{fig:LMA-D}
 \end{figure}
 The presence of this new degenerate solution to the solar neutrino anomaly, shown in Fig.~\ref{fig:LMA-D}, can be understood in the framework of two-neutrino mixing as follows. Under this approximation (justified by the fact that $\sin^2 \theta_{13} \ll 1$), the two by two Hamiltonian matrix can be diagonalized with an effective mixing angle given by
\be
 \tan 2\theta_{12}^m = \frac{ \sin 2\theta_{12} (\Delta m_{12}^2/2E)}{\cos 2\theta_{12} (\Delta m_{12}^2/2E)-(H_m)_{ee}}\, .
\ee
The splitting between two eigenvalues is given by
\be
\Delta^m =\left( (H_m)_{ee}^2+ \left(\frac{\Delta m_{21}^2}{2E}\right)^2-\frac{\Delta m_{21}^2}{E} (H_m)_{ee} \cos 2 \theta_{12} \right)^{1/2}.
\ee

Under the simultaneous transformations $\cos 2  \theta_{12}  \to -\cos 2  \theta_{12} $  and $(H_m)_{ee} \to
-(H_m)_{ee}$, we find $\theta_{12}^m \to \pi /2 -\theta_{12}^m$ and $\Delta^m \to \Delta^m$ which means that  the off-diagonal elements of the $2 \times 2$ Hamiltonian remains the same but the diagonal elements (the $11$ and $22$ elements) flip. That is, $P(\nu_e \to \nu_e)$ changes to $P(\nu_c\to \nu_c)$ where $\nu_c\equiv c_{23}\nu_\mu-s_{23} \nu_\tau$ is the combination that $\nu_e$ converts to (that is $\langle \nu_c|\nu_3 \rangle=\langle \nu_c|\nu_e \rangle=0 $). Since in two neutrino approximation, we can write $P(\nu_e \to \nu_e)+P(\nu_e \to \nu_c)=1$, $P(\nu_c \to \nu_e)+P(\nu_c \to \nu_c)=1$ and $P(\nu_e \to \nu_c)= P(\nu_c \to \nu_e)$.  We therefore conclude $P(\nu_e \to \nu_e) =P(\nu_c \to \nu_c)$. As a result, under the transformation described above, $P(\nu_e \to \nu_e)$ remains invariant.
This transformation is not possible for the case of standard matter effects, where the value of $(H_m)_{ee}$ is fixed to $\sqrt{2}G_F N_e$. However, if one considers the presence of neutrino NSI with the matter field $f$, the effective Hamiltonian in the medium is modified to:
\be
 \label{eq:Hm-NSI}
H_m' = H_m + H_{NSI}= \sqrt{2} G_F N_e \left( \begin{matrix}
 1 &0 &0 \cr
 0& 0 &0 \cr
 0 & 0 & 0\end{matrix} \right)  + \sqrt{2} G_F\sum_f N_f
 \left( \begin{matrix}
 \epsilon_{ee}^{fV} & \epsilon_{e\mu}^{fV}  &  \epsilon_{e\tau}^{fV}  \cr
 \epsilon_{e\mu}^{fV \ast} & \epsilon_{\mu\mu}^{f V}  &  \epsilon_{\mu\tau}^{f V}  \cr
 \epsilon_{e\tau}^{fV \ast} & \epsilon_{\mu\tau}^{f V \ast}  &  \epsilon_{\tau\tau}^{f V}
 \end{matrix} \right) .
\ee
 Allowing for sufficiently large values of the $\epsilon_{ee}^{f V}$ coupling in the effective Hamiltonian in matter $H_m'$, it is now possible to apply the transformation described above, obtaining a  degenerate solution to the solar neutrino anomaly with $\cos 2  \theta_{12} < 0$.
 Notice that, letting $\cos 2 \theta_{12}$ to change sign, we are violating the historical choice of keeping  $\theta_{12}$ in the first octant and, therefore,  $\nu_1$ will not be anymore the state giving the largest contribution to $\nu_e$, but the lighter one between those two eigenstates that give the main contribution to $\nu_e$. This change of  definition is in fact  equivalent  to maintain  the same convention regarding the allowed range for the mixing angle, but allowing $\Delta m_{21}^2$ to be negative. Indeed, changing $\Delta m_{21}^2 \to -\Delta m_{21}^2 $ instead of $\cos 2 \theta_{12} \to -\cos 2 \theta_{12}$, we would find the same degeneracy. In other words, for a given $H_m$,  solar neutrino data only determine the sign of the product $\Delta m_{21}^2 \cos 2 \theta_{12}$, not the signs of $\Delta m_{21}^2$ and $ \cos 2 \theta_{12}$ separately, and therefore there is a freedom in definition. Since the LMA-D solution was introduced in the literature keeping $\Delta m_{21}^2$ positive while  allowing $\theta_{12}$ to vary in the range $(0,\pi/2)$ ~\cite{Miranda:2004nb} and this convention has become popular in the literature since then, we will use it along this review.
Note also that the degeneracy found at the neutrino oscillation probability is exact only for a given composition of matter ({\it i.e.,} for a given $N_n/N_p=N_n/N_e$). The composition slightly varies across the Sun radius and of course is quite different for the Sun and the Earth. Because of this, the allowed regions in the neutrino oscillation parameter space for the LMA and LMA-D solutions are not completely degenerate.
 A small $\chi^2$ difference between the best fit point of the LMA solution and the local minimum of LMA-Dark solution appears because the relevant data analyses take into account the varying composition of the Sun and the day-night asymmetry due to propagation in the Earth.

Unfortunately, the  degeneracy between the LMA and LMA-Dark solutions could not be lifted by the KamLAND reactor experiment because KamLAND was not sensitive to the octant of the solar mixing angle due to the lack of matter effects.
A possible way to solve this problem was proposed in Ref.~\cite{Escrihuela:2009up}. There, it was found that  the combination of solar experiments, KamLAND
and neutrino neutral--current scattering experiments, such as CHARM~\cite{Dorenbosch:1986tb}, may help to probe the LMA-D solution.
The relevance of the degeneracy in the solar neutrino parameter determination has been explored recently in Ref.~\cite{Coloma:2016gei}.
As discussed in this analysis,
the ambiguity of LMA-D does not affect only the octant of the solar mixing angle but it also
makes impossible the determination of the neutrino mass ordering at oscillation experiments.
More recently, a global analysis of neutrino scattering and solar neutrino experiments was performed to further investigate
the situation of the LMA-D solution~\cite{Coloma:2017egw}. Besides the accelerator experiment CHARM, the authors also
 considered the NuTeV experiment~\cite{Zeller:2001hh}. They found that the degenerate LMA-D solution may be lifted for NSI with down quarks, although
 it does not disappear for the case of neutrino NSI with  up quarks. As discussed in that work, constraints from CHARM and NuTeV
experiments can be however directly applied only for NSI with relatively heavy mediators.
For the case of NSI mediated by lighter particles (above 10 MeV), constraints coming from coherent
neutrino-nucleus scattering experiments may be used to resolve the degeneracy.
Indeed,  after the recent observation of such process at the COHERENT experiment~\cite{Akimov:2017ade}, a combined analysis
of neutrino oscillation data including the observed number of events in this experiment  has excluded the  LMA-D solution (for up and down quarks)  at the 3$\sigma$ level~\cite{Coloma:2017ncl}~\footnote{The analysis of atmospheric neutrino data performed in \cite{Coloma:2017ncl}  employs two
simplifying assumptions. First, the solar mass splitting is neglected.
Second,  rather than taking the most general matter potential, it is assumed that two of eigenvalues of this matrix are degenerate. As a result, the derived constraints on the
NSI couplings are  more stringent than what we expect in the most general case.}. One should, however, bear in mind that the analysis~\cite{Coloma:2017ncl} assumes the mediator of interaction in Eq. (\ref{eq:NC-NSI-Lagrangian}) is heavier than 50 MeV. As we shall see in sect V, for light mediator, their conclusion should be revised.
Besides that, COHERENT data along with neutrino oscillation  data has been used to improve the current bounds on the flavor--diagonal NSI parameters \cite{Coloma:2017ncl}\footnote{Notice that since the beam at COHERENT  does not contain $\nu_\tau$, this experiment cannot directly probe $\epsilon_{\tau \tau}$. The bounds on $\epsilon_{\tau\tau}$ come from combining the limits   on $\epsilon_{\mu\mu}$ and $\epsilon_{ee}$ by COHERENT with the bounds on $\epsilon_{\tau\tau}-\epsilon_{ee}$ and  $\epsilon_{\tau\tau}-\epsilon_{\mu\mu}$ from oscillation experiments. }:
\begin{equation}
-0.09 < \epsilon^{uV}_{\tau\tau} < 0.38 \, , \, -0.075 < \epsilon_{\tau\tau}^{dV} < 0.33 \, \quad \rm{(90\% \, C.L.)}\ .
\end{equation}
These limits on NC vector interactions of $\nu_\tau$  improve previous bounds by one order of magnitude~\cite{Davidson:2003ha,Escrihuela:2009up,Gonzalez-Garcia:2013usa}. For the flavor--changing NC NSI couplings, however, the improvement is much smaller~\footnote{Note that the existing bounds on  $\epsilon^{qV}_{e\mu}$  were revised in Ref.~\cite{Biggio:2009kv} showing that previously derived loop bounds do not hold in general.}:
\begin{equation}
-0.073 < \epsilon^{uV}_{e\mu} < 0.044 \, , \, -0.07 < \epsilon_{e\mu}^{dV} < 0.04 \, \quad \rm{(90\% \, C.L.)}\ ,
\end{equation}
\begin{equation}
-0.15 < \epsilon^{uV}_{e\tau} < 0.13 \, , \, -0.13 < \epsilon_{e\tau}^{dV} < 0.12 \, \quad \rm{(90\% \, C.L.)}\ .
\end{equation}
The spectrum of coherent elastic neutrino--nucleus scattering events at COHERENT has also been analyzed to constrain the amplitude of  NSI in Ref.~\cite{Liao:2017uzy}.

Besides their impact on solar neutrino propagation, NSI can also affect the detection processes at solar neutrino experiments.
In experiments like Super--Kamiokande and Borexino, for instance, the presence of NSI may modify the cross section of neutrino elastic
 scattering on electrons, used to observe solar neutrinos.
Analyzing data from solar neutrino experiments, and in particular the effect of NSI on neutrino detection in Super-Kamiokande,
 in combination with KamLAND, Ref.~\cite{Bolanos:2008km} reported limits on the NSI
parameters which are competitive and complementary to the ones obtained from laboratory experiments. For the case of $\nu_e$ NSI
interaction with electrons, the reported bounds (taking one parameter at a time) are:
\begin{equation}
-0.021 < \epsilon_{ee}^{eL} < 0.052 \, ,\quad -0.18 < \epsilon_{ee}^{eR} < 0.51 \quad \rm{(90\% \, C.L.)}\, ,
\end{equation}
 while for the case of $\nu_\tau$ NSI interaction with electrons, looser constraints are obtained:
 \begin{equation}
-0.12 < \epsilon_{\tau\tau}^{eL} < 0.060 \, ,\quad   -0.99< \epsilon_{\tau\tau}^{eR} < 0.23 \quad \rm{(90\% \, C.L.)}\,.
\end{equation}

The sensitivity of the Borexino solar experiment to NSI has also been investigated in Refs.~\cite{Berezhiani:2001rt,Agarwalla:2012wf}.
Using $^7$Be neutrino data from Borexino Phase I, the following 90\% C.L. bounds have been derived~\cite{Agarwalla:2012wf}
 \begin{eqnarray}
& -0.046 < \epsilon_{ee}^{eL} < 0.053 \, & , \quad  -0.21 < \epsilon_{ee}^{eR} < 0.16\, ,  \\
& -0.23 < \epsilon_{\tau\tau}^{eL} < 0.87 \, & ,\quad  -0.98 < \epsilon_{\tau\tau}^{eR} < 0.73  \, .
\end{eqnarray}
As can be seen, the NSI constraints obtained from Borexino and the combined analysis of solar (mainly Super-Kamiokande) and KamLAND data are comparable.
It is expected that future results from Borexino Phase II, as well as the combination of all solar data, including Borexino, plus KamLAND data would allow a significant
improvement on the current knowledge of neutrino NSI with matter \cite{Khan:2017oxw}.

%%%% atmospheric %%%%%

\subsection{NSI in atmospheric neutrino experiments}

The impact of non-standard neutrino interactions on atmospheric
neutrinos was originally considered in
Refs.~\cite{Fornengo:1999zp,Fornengo:2001pm,Friedland:2004ah,Friedland:2005vy}.
Assuming a two--flavor neutrino system, it was
shown~\cite{Fornengo:2001pm} that the presence of large NSI couplings
together with the standard mechanism of neutrino oscillation can
spoil the excellent description of the atmospheric neutrino anomaly
given by neutrino oscillations. Thus, quite strong bounds on the
magnitude of the non--standard interactions were derived.
Using atmospheric neutrino data from the first and second phase of the Super--Kamiokande experiment,
the following constraints were obtained, under the two--flavor neutrino approach~\cite{Mitsuka:2011ty}:
\begin{equation}
|\epsilon_{\mu\tau}^{dV}| < 0.011 \, ,\quad   |\epsilon_{\mu\mu}^{dV}-\epsilon_{\tau\tau}^{dV}| < 0.049 \, \quad \rm{(90\% \, C.L.)}\ .
\end{equation}
However, Ref.~\cite{Friedland:2004ah,Friedland:2005vy} showed that
 a three--family analysis  significantly relaxes  the previous bounds in such a way that the values
of the NSI couplings with quarks comparable to the standard neutral current
couplings can be still compatible with the Super--Kamiokande atmospheric data.
A more recent three--neutrino analysis of NSI in the atmospheric neutrino flux can be found in
Ref.~\cite{GonzalezGarcia:2011my}, where the following limits on the effective NSI couplings with electrons have been obtained:
\begin{equation}
- 0.035 \,(-0.035) < \epsilon_{\mu\tau}^{eV} < 0.018 \,(0.035) \, , |\epsilon_{\tau\tau}^{eV}-\epsilon_{\mu\mu}^{eV}| < 0.097 \,(0.11) \quad \rm{(90\% \, C.L.)}
\end{equation}
for the case of real (complex) $\epsilon_{\mu\tau}^{eV}$ coupling.\\

The IceCube extension to lower energies, DeepCore, has made possible the observation of atmospheric
neutrinos down to 5 GeV  with unprecedented statistics. Indeed, with only 3 years of data, DeepCore allows
the determination of neutrino oscillation parameters with similar precision as the one obtained from the long--lived
Super--Kamiokande or the long--baseline accelerator experiments~\cite{Aartsen:2017nmd}.
Focusing now on its sensitivity to NSI, the idea of using IceCube data to constrain the $\mu-\tau$ submatrix of $\epsilon$ was first proposed in \cite{Esmaili:2013fva}.
Using the most recent data release from DeepCore, the IceCube collaboration has reported the following constraints on the  flavor--changing NSI coupling~\cite{Aartsen:2017xtt}:
\begin{equation}
- 0.0067 < \epsilon_{\mu\tau}^{dV} < 0.0081 \, \, \quad \rm{(90\% \, C.L.)}\ .
\end{equation}
From a different data sample containing higher energy neutrino data from IceCube, the authors of Ref.~\cite{Salvado:2016uqu} have derived somewhat more restrictive
 bounds on the same NSI interactions:
\begin{equation}
- 0.006 < \epsilon_{\mu\tau}^{dV} < 0.0054 \, \, \quad \rm{(90\% \, C.L.)}\ .
\end{equation}
Both results are fully compatible and constitute the current best limits on NSI in the $\nu_\mu$--$\nu_\tau$ sector. Note, however,
that unlike in Ref.~\cite{GonzalezGarcia:2011my}, these results have been obtained under the assumption of no flavor diagonal NSI couplings, $\epsilon_{\alpha\alpha} = 0$.

Future prospects on NSI searches in atmospheric neutrino experiments have been considered in the context of PINGU,
the future project to further lower the energy threshold at the IceCube observatory.
Ref.~\cite{Choubey:2014iia} shows that, after three years of data taking in PINGU in the energy range between 2 and 100 GeV,
the Super--Kamiokande constraints on the NSI couplings may be improved by one order of magnitude:
\begin{equation}
- 0.0043 < \epsilon_{\mu\tau}^{eV} < 0.0047  \, ,\quad   -0.03 < \epsilon_{\tau\tau}^{eV} < 0.017 \, \quad \rm{(90\% \, C.L.)}\ .
\end{equation}

Likewise, the impact of NSI interactions on atmospheric neutrinos on the future India-based Neutrino Observatory (INO)
has been analyzed in Ref.~\cite{Choubey:2015xha}. Besides discussing its constraining potential towards NSI, this work studies
how the sensitivity to the neutrino mass hierarchy of INO, one of the  main goals of the experiment, may change  in the presence of NSI. \\

Notice that the above bounds have been derived from the study of the atmospheric neutrinos flux at neutrino telescopes.
Ref. \cite{Gonzalez-Garcia:2016gpq} discusses the effects of NSI on  high energy astrophysical neutrinos detected by IceCube when they propagate through the Earth.

%%%% reactor %%%%%

\subsection{NSI in reactor experiments}

Modern reactor neutrino experiments, like Daya Bay, RENO and Double Chooz, provide a very accurate determination of the
reactor mixing angle $\theta_{13}$~\cite{An:2015nua,RENO:2015ksa,Abe:2014bwa}.  Being at the precision era of the neutrino parameter determination,
it is imperative  to investigate the robustness of  this successful measurement in the presence of NSI.
Refs.~\cite{Ohlsson:2008gx,Leitner:2011aa,Agarwalla:2014bsa,Girardi:2014gna,Khan:2013hva} have addressed this point.
In principle, short--baseline reactor experiments may  be affected by the presence of new neutrino interactions in $\beta$ and inverse-$\beta$ decay processes,
relevant for the production and detection of reactor antineutrinos ~\cite{Kopp:2007ne}. The NSI  parameters relevant
for these experiments are the CC NSI couplings between up and down quarks, positrons and antineutrinos of flavor $\alpha$, $\epsilon^{ud}_{e\alpha}$.
Considering unitarity constraints on the CKM matrix as well as  the non-observation
of  neutrino  oscillations  in  the  NOMAD  experiment, one may find  the following constraints on these CC NSI couplings~\cite{Biggio:2009nt}:
\be
|\epsilon^{ud V}_{e\alpha}| < 0.041\, , \, |\epsilon^{ud L}_{e\mu}| < 0.026 \, , \, |\epsilon^{ud R}_{e\mu}| < 0.037Ê\, , \, \rm{(90\%\, C.L.)} \ .
\ee
Ref.~\cite{Agarwalla:2014bsa} explored the  correlations between the NSI parameters and the reactor mixing angle determination,
showing that  the presence of NSI may lead to relatively large deviations in the measured value of $\theta_{13}$ in Daya Bay, as it can be seen in Fig.~\ref{fig:NSIreac}.
%
%-------------------------------------------------------------------
\begin{figure}[!tb]
\centerline{
\includegraphics[width=0.33\textwidth]{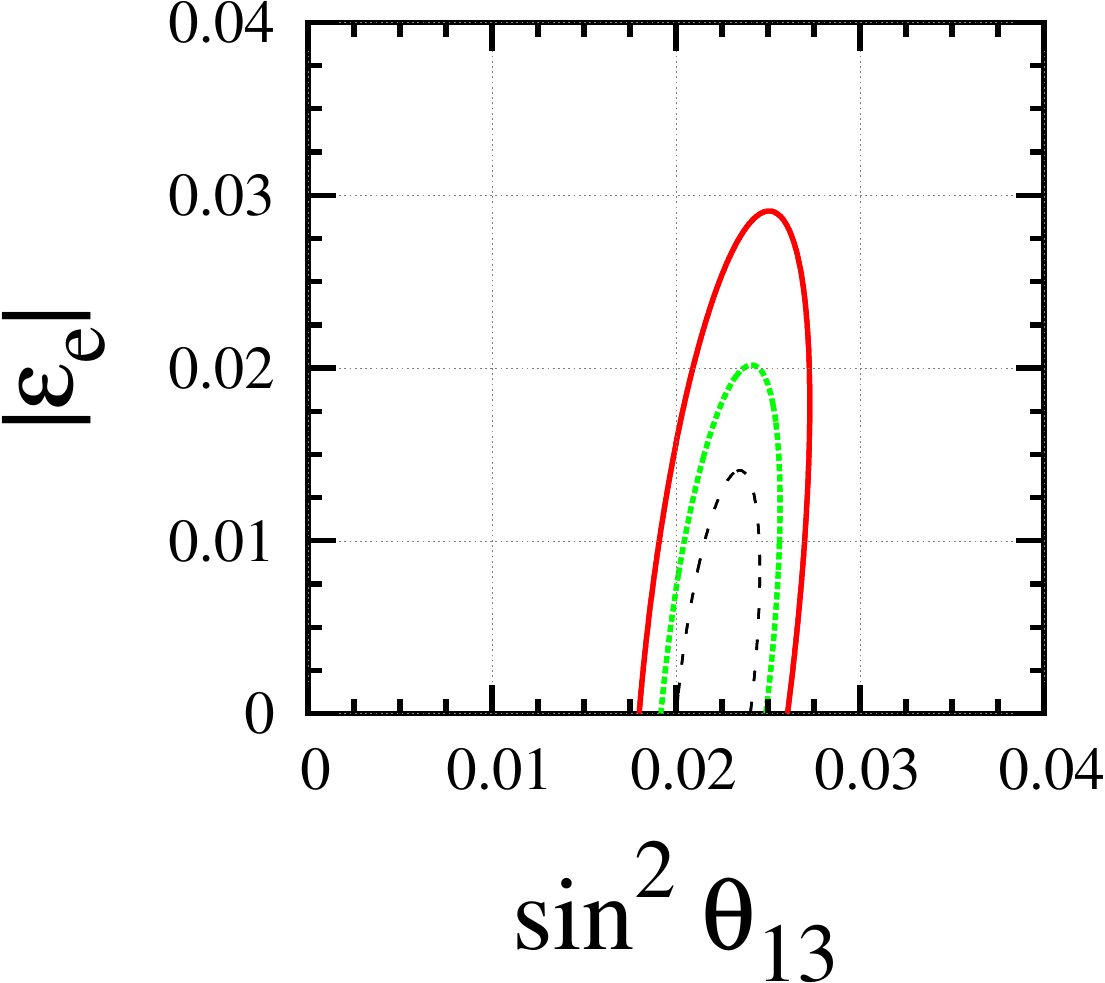}
\includegraphics[width=0.33\textwidth]{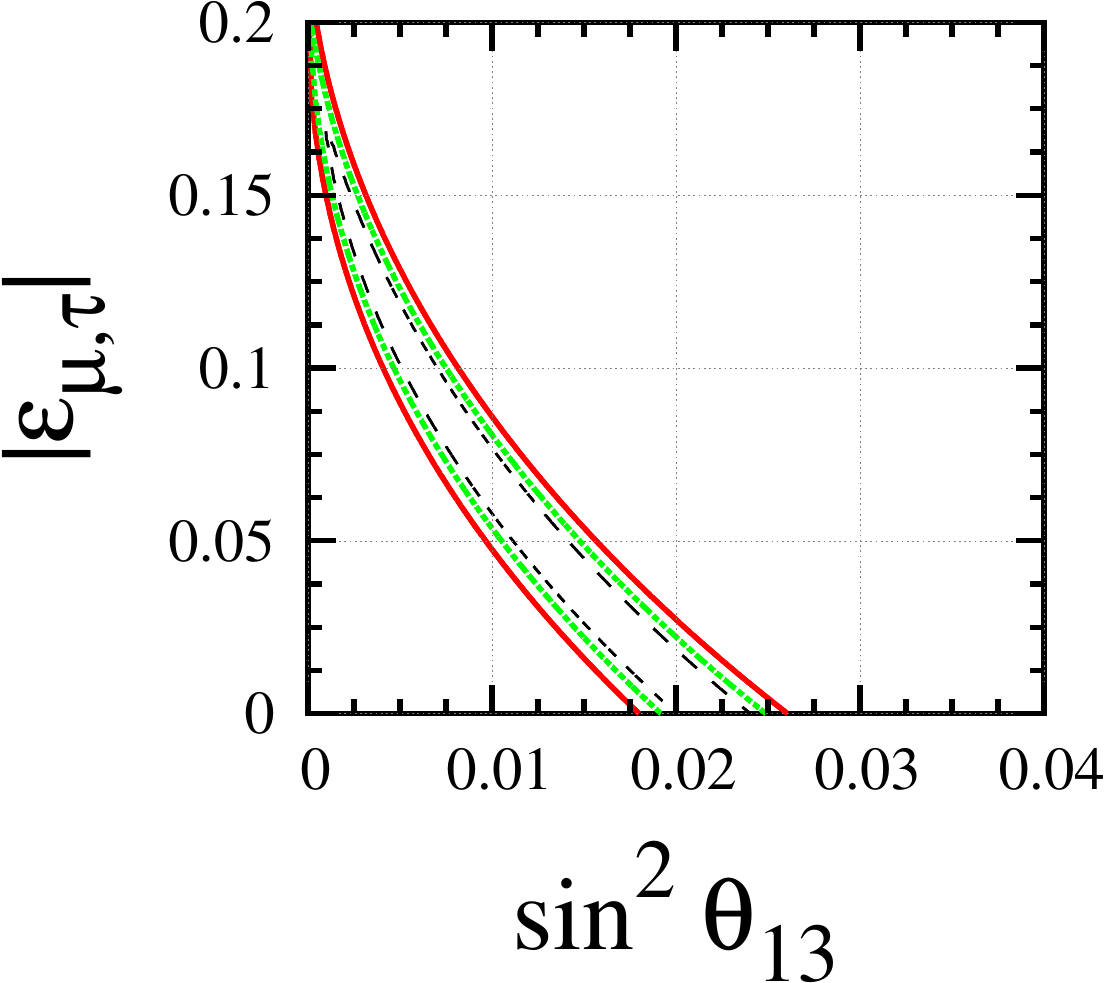}
\includegraphics[width=0.33\textwidth]{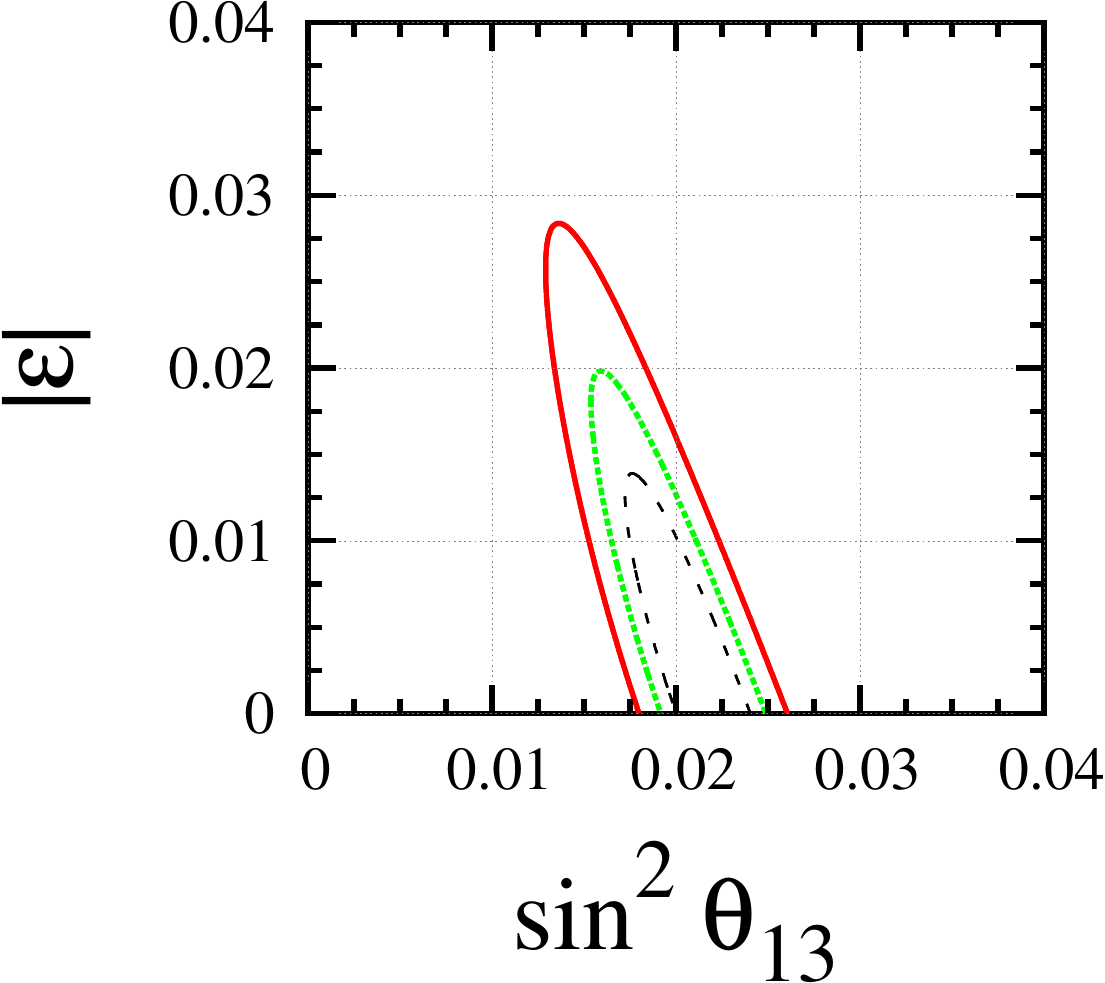}
}
\caption{
68\%, 90\% and 99\% C.L. allowed regions from Daya Bay for different scenarios involving CC NSI  (left) of only $\nu_e$; {\it i.e.,} $\epsilon^{ud P}_{ee}$, (middle) of  only  $\nu_\mu$ or $\nu_\tau$; {\it i.e.,}  $\epsilon^{ud P}_{\mu e}$ or $\epsilon^{ud P}_{\tau e}$,  and (right panel)  simultaneously of all neutrino flavors  with $\epsilon$ = $\epsilon^{ud P}_{ee}$ = $\epsilon^{ud P}_{\mu e}$ = $\epsilon^{ud P}_{\tau e}$. In drawing the figures, 5\% uncertainty on the total event rate normalization of Daya Bay events was assumed. Plots  are taken from Ref.~\cite{Agarwalla:2014bsa}, published under the terms of the Creative Commons Attribution Noncommercial License and therefore no copyright permissions were required for their inclusion in this manuscript.}
\label{fig:NSIreac}
\end{figure}
%-------------------------------------------------------------------
%
Conversely, the total number of events observed in Daya Bay  was used to constrain the corresponding NSI couplings under two assumptions: i) perfect theoretical knowledge of the reactor neutrino flux in absence of NSI and ii) assuming a conservative error on its total normalization. In the latter case, it was shown that assuming an uncertainty of 5\% on the reactor flux
 can relax the bounds by one order of magnitude, obtaining the following conservative limits on the NSI strengths~\footnote{
The NSI parameters probed in this kind of analysis obtain contributions from the (V$\pm$A) operators in Eq.~\ref{eq:CC-NSI-Lagrangian} so,
taking one parameter at a time, the derived bounds apply to all the chiralities.}
\be
|\epsilon^{ud P}_{ee}| < 0.015\, , \, |\epsilon^{ud P}_{e\mu}| < 0.18 \, , \, |\epsilon^{ud P}_{e\tau}| < 0.18Ê\, , \, \rm{(90\%\, C.L.)}\, ,
\ee
with P=L,R,V,A. Note that  these results improve the existing bounds on the $\epsilon^{ud}_{ee}$ coupling reported above. On the other hand,
one finds that an improved knowledge of the standard absolute neutrino flux from nuclear reactors together with a larger data sample from Daya Bay
 will result in a more stringent bound on the other two couplings in the near future. Notice also that previous results have been obtained assuming
  that the NSI couplings at neutrino production and detection satisfy $\epsilon_{\alpha\beta}^s$ =  $\epsilon_{\alpha\beta}^{d \ast}$. In this case, the presence of NSI
would only produce a shift in the oscillation amplitude without altering the L/E pattern of the oscillation probability, and therefore, the analysis of the total
 neutrino rate in Daya Bay provides enough information. The investigation of more exotic scenarios where  $\epsilon_{\alpha\beta}^s$ $\neq$  $\epsilon_{\alpha\beta}^{d \ast}$
 will require the spectral analysis of the Daya Bay data~\cite{Leitner:2011aa}.\\

NSI at future intermediate baseline reactor experiments like JUNO and RENO-50 (see, for instance, Refs.~\cite{Ohlsson:2013nna} and \cite{Khan:2013hva}) are discussed at Section V.

%%%%%   LBL %%%%%%%%

\subsection{NSI in long--baseline neutrino experiments}

Besides neutrino production and detection,  NSI can also modify the
neutrino propagation through the Earth in long--baseline accelerator experiments~\footnote{See for instance Ref.~\cite{Kopp:2007ne},
where the impact of NSI on long--baseline experiment is analyzed in detail.}.
This effect will be larger for experiments with  larger baselines such as MINOS or NO$\nu$A.
Using their neutrino and antineutrino data sample, the MINOS Collaboration reported the following bounds on the flavor-changing NC NSI with electrons~\cite{Adamson:2013ovz}:
\begin{equation}
- 0.20 < \epsilon_{\mu\tau}^{eV} < 0.07 \, \, \, \rm{(90\% \, C.L.)}\ .
\end{equation}
MINOS appearance data were also used to constrain NSI interactions between the first and third family~\cite{Adamson:2016yso}, although
the reported bound,  $|\epsilon_{e\tau}^{eV}| < 3.0$ (90\% C.L.) does not improve the previous limits on that parameter~\cite{Biggio:2009nt}.\\

Regarding the long--baseline experiment NO$\nu$A, the presence of NSI in the neutrino propagation has been proposed as a way to solve the mild tension
between the measured values of the atmospheric mixing angle in T2K and NO$\nu$A~\cite{Liao:2016bgf}.
Under this hypothesis, the deviation of the NO$\nu$A  preferred value for $\theta_{23}$ from maximal mixing  would be explained through  the NSI-modified
matter effects. The T2K experiment, with a  shorter baseline, has a limited sensitivity to  matter effects in the neutrino propagation so, its
$\theta_{23}$ measurement would be unaffected by NSI. Note, however, that the size of the
NSI required to reproduce the observed results is of the same order as the standard neutrino interaction (to be more precise $\epsilon_{e\tau}, (\epsilon_{\tau\tau}-\epsilon_{\mu\mu})\simeq (\epsilon_{\tau\tau}-\epsilon_{ee}) \sim O(1)$).\\

The presence of NSI has also been considered to reconcile the measured value of $\theta_{13}$ in reactor experiments and T2K~\cite{Girardi:2014kca}.
In that case, it is suggested that CC-NSI in the neutrino production and detection processes may be responsible for the different values of the reactor mixing
 angle measured in Daya Bay and T2K.\\

Finally, it has been shown that long--baseline neutrino facilities can also suffer from degeneracies in the reconstruction of some parameters
due to the existence of  new neutrino interactions with matter. For instance, Ref.~\cite{Forero:2016cmb} states that NC NSI may affect the sensitivity to the CP--violating
 phase $\delta$ in experiments like T2K and NO$\nu$A. According to this analysis,  it would be possible confusing signals of NSI with a discovery of CP violation, even
  if CP is conserved in nature. This result  is illustrated in Fig.~\ref{fig:del-deg}, where it is shown how the standard CP--violating scenario may be confused with an hybrid
   standard plus NSI CP--conserving scenario.\\
%-------------------------------------------------------------------
\begin{figure}[!tb]
\centerline{
\includegraphics[width=0.35\textwidth]{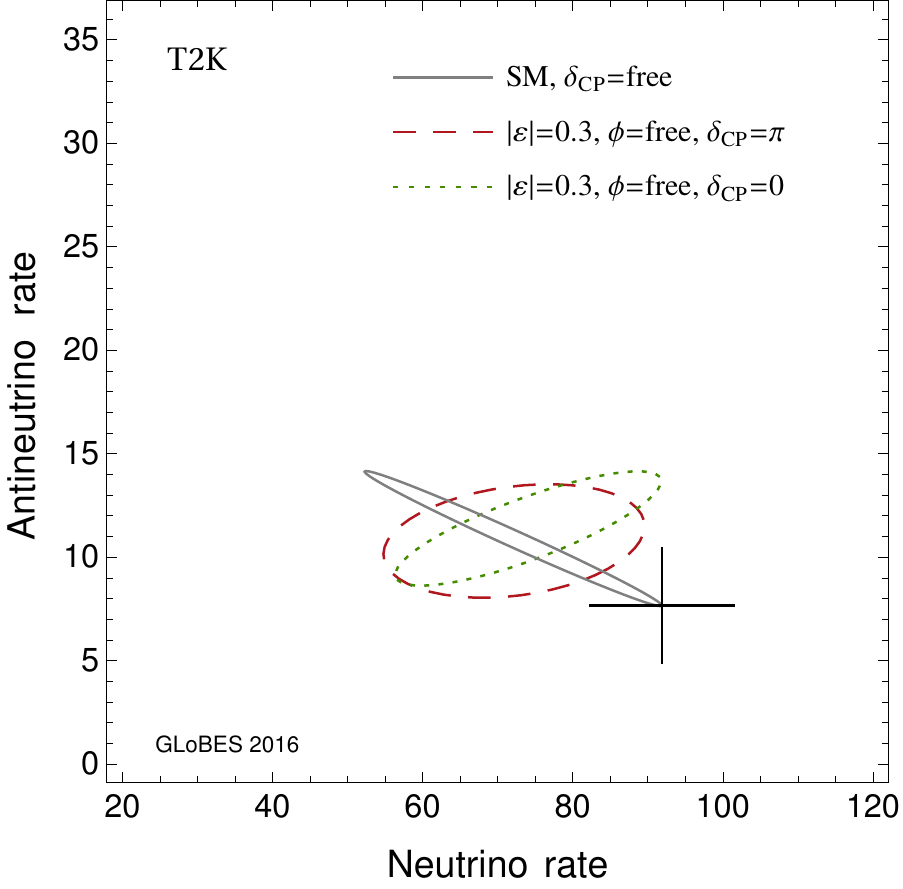}
\includegraphics[width=0.35\textwidth]{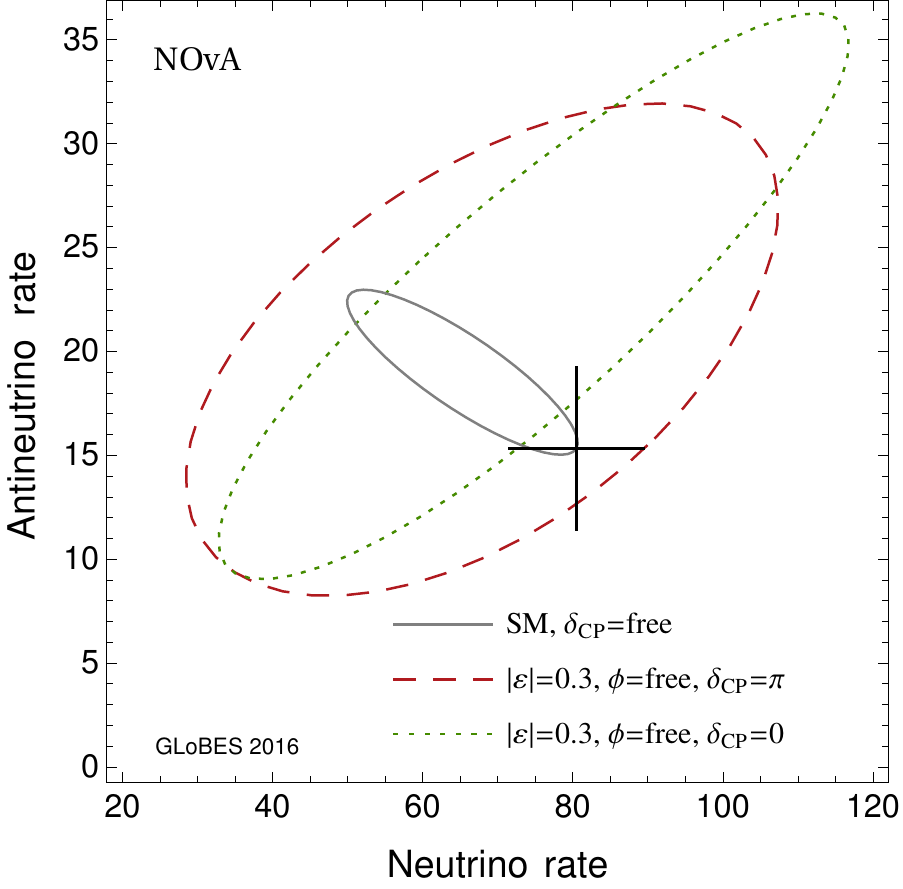}
}
\caption{
Bi-rate plots for T2K and NO$\nu$A for SM (solid line) and SM + NSI scenario (dashed and dotted line).  The cross indicates the SM point for $\delta = -\pi/2$ with the corresponding statistical uncertainty. Plots taken from Ref.~\cite{Forero:2016cmb} and reproduced with the permission of the American Physical Society.}
\label{fig:del-deg}
\end{figure}
%-------------------------------------------------------------------

 Future sensitivities to NSI as well as the presence of new degeneracies  due to NSI in future long--baseline experiments such as DUNE, T2HK and T2HKK
are analyzed in more detail in Sect. V.
It is worth mentioning that CC-NSI, affecting the production and detection of neutrinos can show up also in short baseline experiments \cite{Blennow:2015nxa,Ge:2016dlx,Khan:2016uon}.

%%%%%   non-oscillation experiments %%%%%%%%

\subsection{NSI in non-oscillation neutrino experiments}

Neutrino scattering experiments constitute a very precise tool towards the understanding of neutrino interactions with matter. Indeed, this kind of experiments has been often used to measure the electroweak mixing angle $\theta_W$~\cite{Patrignani:2016xqp}.
%
%%% neutrino--electron scattering %%%%
Non--standard neutrino interactions may contribute significantly to the neutrino--electron elastic scattering cross section and therefore they cannot be ignored when studying this process.
Ref.~\cite{Barranco:2007ej} compiled most of the neutrino scattering experiments potentially modified by the presence of NSI, from the neutrino accelerator--based experiments LSND
and CHARM to the short--baseline neutrino reactor experiments Irvine, Rovno and MUNU, including as well as the measurement of the process $e^+ e^- \to \nu \overline{\nu} \gamma$ at LEP.
From a combined analysis of all experimental data, allowed ranges on the $\epsilon_{\alpha\beta}^e$ were obtained. Some of these results are among the current strongest constraints on NSI couplings, and are reported in Table \ref{tab:boundsFD}.
The antineutrino--electron scattering data collected by the TEXONO Collaboration has been also used to constrain the presence of neutrino NC NSI with electrons~\cite{Deniz:2010mp} as well as CC NSI at neutrino production and detection~\cite{Khan:2014zwa}.\\

%%%% neutrino-nucleus scattering %%%%

In order to constrain the NSI between neutrinos and quarks, one may use data from the neutrino--nucleus experiments
NuTeV, CHARM and CDHS.
From the combination of atmospheric and accelerator data from NuTeV, CHARM and CDHS, the following limits
on the non--universal vectorial and axial NSI parameters were derived~\cite{Escrihuela:2011cf}:
\begin{equation}
|\epsilon^{dV}_{\mu\mu}| < 0.042 \, , \qquad  -0.072 < \epsilon^{dA}_{\mu\mu} < 0.057 \, \quad \rm{(90\% \, C.L.)}.
\end{equation}
For the case of the flavor changing NSI couplings (with $q=u,d$)
\begin{equation} |\epsilon^{qV}_{\mu \tau}| < 0.007 \, , \qquad |\epsilon^{qA}_{\mu \tau}| < 0.039  \, \quad \rm{(90\% \, C.L.)}.
\end{equation}
%

%%%% coherent neutrino-nucleus scattering  %%%%

Under this category we include also the first observation of coherent neutrino--nucleus scattering observed at the COHERENT
experiment recently~\cite{Akimov:2017ade}. As discussed above,  the COHERENT data have been used to constrain neutrino NSI
 with quarks in Refs.~\cite{Coloma:2017ncl,Liao:2017uzy}. The combination of solar neutrino oscillation data with COHERENT has been exploited
 to investigate the status of the solar degenerate solution LMA-D.

%%%% FD-NC NSI %%%%%

\begin{table}[th]
    \centering
    \begin{tabular}{lccc}
        \hline\hline\\[-2.5mm]
     & 90\% C.L. range & origin & Ref.\\[1mm]
      \hline \hline\\[-2.5mm]
      & & NSI with quarks & \\[1mm]
      \hline\hline\\[-2.5mm]
      $\epsilon^{dL}_{ee}$ & $[-0.3, 0.3]$ & CHARM & \cite{Davidson:2003ha}\\[1mm]
      $\epsilon^{dR}_{ee}$ & $[-0.6, 0.5]$ & CHARM & \cite{Davidson:2003ha}\\[1mm]
      $\epsilon^{dV}_{\mu\mu}$ & $[-0.042, 0.042]$ & atmospheric + accelerator & \cite{Escrihuela:2011cf}\\[1mm]
        $\epsilon^{uV}_{\mu\mu}$ & $[-0.044, 0.044]$ & atmospheric + accelerator & \cite{Escrihuela:2011cf}\\[1mm]
      $\epsilon^{dA}_{\mu\mu}$ & $[-0.072, 0.057]$ & atmospheric + accelerator & \cite{Escrihuela:2011cf}\\[1mm]
        $\epsilon^{uA}_{\mu\mu}$ & $[-0.094, 0.14]$ & atmospheric + accelerator & \cite{Escrihuela:2011cf}\\[1mm]
      $\epsilon^{dV}_{\tau\tau}$ & $[-0.075, 0.33]$ & oscillation data + COHERENT &\cite{Coloma:2017ncl}\\[1mm]
       $\epsilon^{uV}_{\tau\tau}$ & $[-0.09, 0.38]$ & oscillation data + COHERENT & \cite{Coloma:2017ncl}\\[1mm]
      $\epsilon^{qV}_{\tau\tau}$ & $[-0.037, 0.037]$ &  atmospheric & \cite{GonzalezGarcia:2011my}$^a$\\[1mm]
      \hline\hline\\[-2mm]
      & & NSI with electrons & \\[1mm]
      \hline\hline\\[-2mm]
       $\epsilon^{eL}_{ee}$ & $[-0.021, 0.052]$ & solar + KamLAND &\cite{Bolanos:2008km}\\[2mm]
       $\epsilon^{eR}_{ee}$ & $[-0.07, 0.08]$ & TEXONO & \cite{Deniz:2010mp}\\[2mm]
      $\epsilon^{eL}_{\mu\mu}$, $\epsilon^{eR}_{\mu\mu}$ & $[-0.03,  0.03]$ & reactor + accelerator &\cite{Davidson:2003ha,Barranco:2007ej}\\[2mm]
       $\epsilon^{eL}_{\tau\tau}$ & $[-0.12, 0.06]$ & solar + KamLAND &\cite{Bolanos:2008km}\\[2mm]
       $\epsilon^{eR}_{\tau\tau}$ & $[-0.98, 0.23]$ & solar + KamLAND and Borexino &\cite{Bolanos:2008km,Agarwalla:2012wf}\\
       &  [-0.25, 0.43] & reactor + accelerator  &\cite{Barranco:2007ej}\\[2mm]
       $\epsilon^{eV}_{\tau\tau}$ & $[-0.11, 0.11]$ & atmospheric &\cite{GonzalezGarcia:2011my}\\[2mm]
       \hline\hline \\[-1mm]
    \multicolumn{4}{l}{$^a$ Bound adapted from $\epsilon^{eV}_{\tau\tau}$.}
    \end{tabular}
      \caption{Bounds on Flavor Diagonal NC NSI couplings\label{tab:boundsFD}}
  \end{table}

\subsection{Summary of current bounds on NSI parameters}

Here we summarize the current constraints on the NSI couplings from different experiments discussed throughout this section.
 For more details about the assumptions considered in each
case, we refer the reader to the previous subsections as well as to the original references  where the constraints have been calculated.
The limits summarized in Tables \ref{tab:boundsFD}, \ref{tab:boundsFC} and \ref{tab:boundsCC} have been obtained assuming only
one nonzero NSI coupling at a time. \\

Table \ref{tab:boundsFD} contains the limits on the flavor diagonal NC NSI couplings between neutrinos and electrons $\epsilon_{\alpha\alpha}^{eP}$
and neutrinos and quarks $\epsilon_{\alpha\alpha}^{qP}$, with $P = L, R, V, A$ being the chirality index and $q = u, d$. The table indicates the origin
of the reported bound as well as the reference where it has been obtained as well.
Most of the limits have been derived from the combination of neutrino oscillation and detection or production experimental results. For instance, the joint analysis of atmospheric neutrino data
and accelerator measurements in NuTeV, CHARM and CDHS~\cite{Escrihuela:2011cf}, or solar and KamLAND data together with the recent bounds of
COHERENT~\cite{Coloma:2017ncl}.\footnote{The bounds in~\cite{Coloma:2017ncl} assume mediator mass to be heavier than $\sim 50$ MeV. As we shall discuss in the next section, these bounds do not apply for mediator mass lighter than $\sim 10$~MeV. } In other cases the constraints reported in the table come just from one type of experiment, as the limits derived  only from
CHARM~\cite{Davidson:2003ha}, TEXONO~\cite{Deniz:2010mp} or atmospheric data~\cite{GonzalezGarcia:2011my}. Note that, for the latter case, we have adapted
 the bound on  $\epsilon_{\tau\tau}^{eV}$
reported in Ref.~\cite{GonzalezGarcia:2011my} to the corresponding bound for quarks, $\epsilon_{\tau\tau}^{qV}$.\\

Table \ref{tab:boundsFC} collects the limits of the flavor changing NC NSI couplings between neutrinos and electrons $\epsilon_{\alpha\beta}^{eP}$
and neutrinos and quarks $\epsilon_{\alpha\beta}^{qP}$, with the same conventions indicated above for $P$ and $q$.
As discussed before, in this case most of the bounds also emerge from the complementarity of different types of experiments, as the combination of reactor and accelerator
non-oscillation experiments in Ref.~\cite{Barranco:2007ej}. On the other hand, the first analyses on NSI obtained from IceCube data~\cite{Salvado:2016uqu,Aartsen:2017xtt}
offer very strong bounds on  $\epsilon^{qV}_{\mu\tau}$. This last constraint has also been adapted to get the equivalent  bound for NSI with electrons,   $\epsilon^{eV}_{\mu\tau}$.\\

Finally, Table \ref{tab:boundsCC} contains the limits on the neutrino CC NSI with quarks and electrons ({\textit{semileptonic}} CC NSI) and the CC NSI with leptons only
 ({\textit{purely-leptonic} CC NSI) in terms of the couplings $\epsilon_{\alpha\beta}^{ud P}$  and  $\epsilon_{\alpha\beta}^{ll'P}$, respectively.
The former ones, have been discussed in the context of the neutrino production and detection in the Daya Bay reactor experiment, as analyzed in
Ref.~\cite{Agarwalla:2014bsa}. Previous bounds on this type of NSI have been derived using the negative searches for neutrino oscillations at short distances in
the NOMAD experiment~\cite{Astier:2001yj,Astier:2003gs}, as reported in the table~	\cite{Biggio:2009nt}.
Constraints on leptonic CC NSI using the results of the KARMEN experiment~\cite{Eitel:2000by} as well as the deviations of Fermi's constant $G_F$ in the presence of these interactions, have also been obtained in Ref.~\cite{Biggio:2009nt}. We refer the reader to that work for further details on the derivation of these constraints.

%%%% FC-NC NSI %%%%%

\begin{table}[th!]
    \centering
    \begin{tabular}{lccc}
      \hline\hline\\[-2.5mm]
    & 90\% C.L. range & origin & Ref.\\[1mm]
   \hline\hline\\[-2.5mm]
     &  \multicolumn{2}{c}{NSI with quarks} & \\[1mm]
   \hline\hline\\[-2.5mm]
     $\epsilon^{qL}_{e\mu}$ & $[-0.023, 0.023]$ & accelerator & \cite{Escrihuela:2011cf,Miranda:2015dra}\\[1mm]
      $\epsilon^{qR}_{e\mu}$ & $[-0.036, 0.036]$ & accelerator & \cite{Escrihuela:2011cf,Miranda:2015dra}\\[1mm]
      {\bf $\epsilon^{uV}_{e\mu}$} & $[-0.073, 0.044]$ & oscillation data + COHERENT & \cite{Coloma:2017ncl}\\[1mm]
       {\bf $\epsilon^{dV}_{e\mu}$} & $[-0.07, 0.04]$ &oscillation data + COHERENT & \cite{Coloma:2017ncl}\\[1mm]
      $\epsilon^{qL}_{e\tau}$,  $\epsilon^{qR}_{e\tau}$ & $[-0.5, 0.5]$ & CHARM & \cite{Davidson:2003ha}\\[1mm]
      $\epsilon^{uV}_{e\tau}$ & $[-0.15,0.13]$  &  oscillation data + COHERENT & \cite{Coloma:2017ncl} \\[1mm]
      $\epsilon_{e\tau}^{dV}$ & $[-0.13,0.12]$ & oscillation data + COHERENT &\cite{Coloma:2017ncl}\\[1mm]
     $\epsilon^{qL}_{\mu\tau}$ & $[-0.023, 0.023]$ & accelerator & \cite{Escrihuela:2011cf}\\[1mm]
      $\epsilon^{qR}_{\mu\tau}$ & $[-0.036, 0.036]$ & accelerator & \cite{Escrihuela:2011cf}\\[1mm]
      $\epsilon^{qV}_{\mu\tau}$ &$ [-0.006, 0.0054]$ & IceCube &\cite{Salvado:2016uqu}\\[1mm]
       $\epsilon^{qA}_{\mu\tau}$ & $[-0.039, 0.039]$ & atmospheric + accelerator & \cite{Escrihuela:2011cf}\\[1mm]
   \hline\hline\\[-2.5mm]
      &  \multicolumn{2}{c}{NSI with electrons} & \\[1mm]
       \hline\hline\\[-2.5mm]
      $\epsilon^{eL}_{e\mu}$, $\epsilon^{eR}_{e\mu}$ & $[-0.13, 0.13]$ & reactor + accelerator &\cite{Barranco:2007ej} \\[1mm]
      $\epsilon^{eL}_{e\tau}$ &  $[-0.33, 0.33]$ & reactor + accelerator &\cite{Barranco:2007ej} \\[1mm]
      $\epsilon^{eR}_{e\tau}$ & $[-0.28, -0.05]$ \& $[0.05, 0.28]$ & reactor + accelerator &\cite{Barranco:2007ej} \\
      &   [-0.19, 0.19] & TEXONO  & \cite{Deniz:2010mp}    \\[1mm]
      $\epsilon^{eL}_{\mu\tau}$,  $\epsilon^{eR}_{\mu\tau}$&  $ [-0.10, 0.10]$ & reactor + accelerator &\cite{Davidson:2003ha,Barranco:2007ej} \\[1mm]
       $\epsilon^{eV}_{\mu\tau}$ & $[-0.018, 0.016]$ & IceCube &\cite{Salvado:2016uqu}$^a$ \\[1mm]
      \hline\hline \\[-1mm]
    \multicolumn{4}{l}{$^a$ Bound adapted from $\epsilon^{qV}_{\mu\tau}$.}
    \end{tabular}
      \caption{Bounds on Flavor changing NC NSI couplings\label{tab:boundsFC}}
  \end{table}

%%%% CC NSI %%%%%

\begin{table}[th!]
    \centering
    \begin{tabular}{lccc}
     \hline\hline\\[-2.5mm]
     & 90\% C.L. range & origin & Ref.\\[1mm]
      \hline\hline\\[-2.5mm]
     & & semileptonic NSI & \\[1mm]
     \hline\hline\\[-2.5mm]
      $\epsilon^{ud P}_{ee}$ & $[-0.015, 0.015] $& Daya Bay & \cite{Agarwalla:2014bsa}\\[1mm]
      $\epsilon^{ud L}_{e\mu}$ & $[-0.026, 0.026] $& NOMAD &\cite{Biggio:2009nt}\\[1mm]
       $\epsilon^{ud R}_{e\mu}$ & $[-0.037, 0.037] $& NOMAD &\cite{Biggio:2009nt}\\[1mm]
       $\epsilon^{ud L}_{\tau e}$ & $ [-0.087, 0.087]$ & NOMAD &\cite{Biggio:2009nt}\\[1mm]
       $\epsilon^{ud R}_{\tau e}$ & $ [-0.12, 0.12]$ & NOMAD &\cite{Biggio:2009nt}\\[1mm]
       $\epsilon^{ud L}_{\tau\mu}$ & $ [-0.013, 0.013]$ & NOMAD &\cite{Biggio:2009nt}\\[1mm]
       $\epsilon^{ud R}_{\tau\mu}$ &  $[-0.018, 0.018]$ & NOMAD &\cite{Biggio:2009nt}\\[1mm]
     \hline\hline\\[-2.5mm]
          & & purely leptonic NSI  & \\[1mm]
   \hline\hline\\[-2.5mm]
      $\epsilon^{\mu e L}_{\alpha e}$, $\epsilon^{\mu e R}_{\alpha e}$  & $[-0.025, 0.025]$ & KARMEN &\cite{Biggio:2009nt}\\[1mm]
  $\epsilon^{\mu e L}_{\alpha \beta}$, $\epsilon^{\mu e R}_{\alpha \beta}$  & $[-0.030, 0.030]$ & kinematic $G_F$ &\cite{Biggio:2009nt}\\[1mm]
       \hline\hline
    \end{tabular}
      \caption{Bounds on CC NSI couplings\label{tab:boundsCC}}
  \end{table}

%%%%%%%%%%%%%%%%%%%%%%%%%%

\section{Viable models leading to sizeable NSI\label{viableMODELS}}
As we saw in the previous section, neutral current NSI of neutrinos with matter fields can lead to observable effect on neutrino oscillation provided that the  NSI parameters $\epsilon_{\alpha \beta}$ are large enough. As briefly discussed in the introduction, it is possible to build viable models for NSI by invoking an intermediate state of relatively light mass
($\sim 10$ MeV) which has escaped detection so far because of its very small coupling. In this chapter, we review the models that give rise to sizeable NSI through integrating out a new gauge boson $Z^\prime$ with a mass smaller than $\sim 100~{\rm MeV}$.  We however note that an alternative model has been suggested \cite{Forero:2016ghr} in which NSI  are obtained from $SU(2)_L$ scalar doublet-singlet mixing. We shall not cover this possibility in the present review. The models described in this chapter introduce a new $U(1)^\prime$ gauge interaction which is responsible for NSI between neutrinos and quarks.

In section \ref{direct}, we describe the general features of the model gauging a linear combination of lepton flavors and Baryon number with a light $O(10~{\rm MeV})$ gauge boson. We then outline general phenomenological consequences. We show how a simple economic model can be reconstructed to reproduce the NSI pattern that gives the best fit to neutrino data, solving the small tension between KamLAND and solar neutrino by explaining the suppression of the upturn in the low energy part of the solar neutrino spectrum.
In section \ref{LFv}, we describe another model which can provide arbitrary flavor structure $\epsilon_{\alpha \beta}^u=\epsilon_{\alpha \beta}^d$ (both lepton flavor violating and lepton flavor conserving) without introducing new interactions for charged leptons.
In sect. \ref{Coh}, the impact of the recent results from the COHERENT experiment is outlined.
\subsection{NSI from new $U(1)^\prime$ \label{direct}}
In this section, we show how we can build a model based on $U(1)^\prime \times SU(2)_L \times U(1)_Y$ gauge symmetry which gives rise to NSI for neutrinos.
 Notice that the NSI of interest for neutrino oscillation involves only neutrinos and quarks of first generation which make up the matter.
However, to embed the scenario within a gauge symmetric theory free from anomalies, the interaction should involve other fermions.

  Let us first concentrate on quark sector and discuss the various possibilities of $U(1)^\prime$ charge assignment. Remember that, in the flavor basis by definition, the interaction of $W_\mu$ boson with quarks is diagonal: $W_\mu \sum_{i=1}^3 \bar{u}_{iL} \gamma^\mu d_{i L}$, where $i$ is the flavor index. To remain invariant under $U(1)^\prime$, $u_{iL}$ and $d_{iL}$ should have the same values of $U(1)^\prime$ charge.  As discussed in sect II.A, the SNO experiment has measured the rate of neutral current interaction of solar neutrinos by Deuteron dissociation $\nu+D\to \nu+p+n$. In general, a large contribution to neutral current interaction from new physics should have affected the rate measured by SNO but this process, being a Gamow-Teller transition, is only sensitive to the axial interaction. In order to maintain the SM prediction for the total neutrino flux measured at the SNO experiment via NC  interactions, the coupling to (at least the first generation of) quarks should be non-chiral. Thus, the $U(1)^\prime$ charges of $u_{1L}$, $u_{1R}$, $d_{1L}$ and $d_{1 R}$ should be all equal. In principle, different generations of quarks can have different $U(1)^\prime$ charges. Such a freedom opens up abundant possibilities for anomaly cancelation.
However, if the coupling of the new gauge bosons to different quark generations is non-universal, in the quark mass basis, off-diagonal couplings of form $Z^\prime_\mu \bar{q}_i \gamma^\mu q_j |_{i \ne j}$ appear which can lead to $q_i \to Z^\prime q_j$ with a rate enhanced by $m_{q_i}^3/m_{Z^\prime}^2$ due to longitudinal component of $Z^\prime$. These bounds are discussed in great detail in Ref. \cite{Babu:2017olk}.  To avoid these decays, we assume the quarks couple to $Z^\prime$ universally. In other words, the $U(1)^\prime$ charges of quarks are taken to be proportional to baryon number, $B$. Yukawa couplings of quarks to the SM Higgs will  then be automatically invariant under $U(1)^\prime$.

 Let us now discuss the couplings of leptons to the new gauge boson. There are two possibilities:
1) $U(1)^\prime$ charges are assigned to a combination of lepton numbers of different flavors. In this case, the $U(1)^\prime$ charges of charged leptons and neutrinos will be equal. (2) Neutrinos couple to $Z^\prime$ through mixing with a new fermion with mass larger than $m_{Z^\prime}$. In this case charged leptons do not couple to $Z^\prime$ at tree level. We shall return to the second case in sect. \ref{LFv}.  In the present section, we focus on the first case.
As discussed in \cite{Farzan:2015hkd}, it is possible to assign $U(1)^\prime$ charge to linear combinations of leptons which  do not even correspond to charged lepton mass eigenstates. However, let us for the time being study the charge assignment as follows
\be
\label{charge-as} a_e L_e+a_\mu L_\mu+a_\tau L_\tau+B .
\ee

\begin{figure}[t]
{\includegraphics[scale=1]{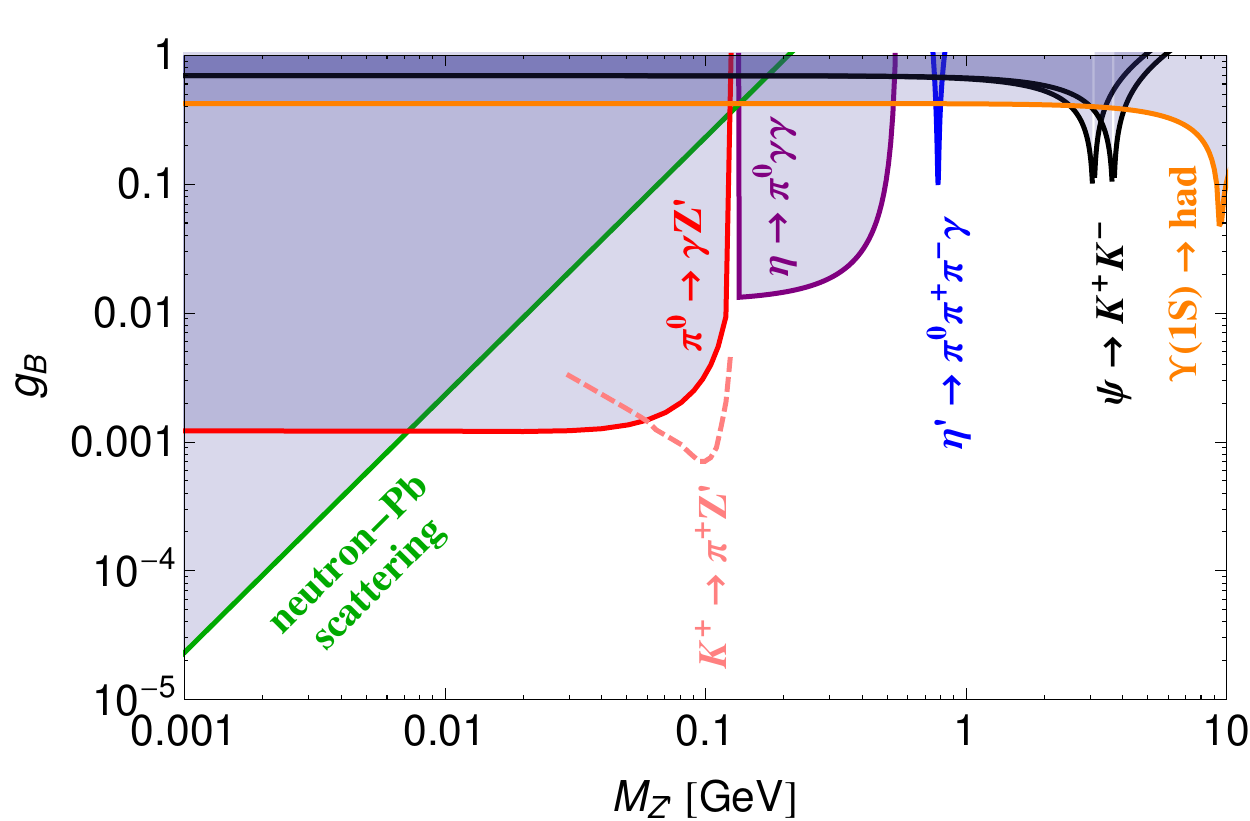}}
\caption{
Parameter space of a gauge boson $Z'$ coupled to quarks with coupling $g^\prime/3=g_B/3$. The shaded areas are excluded at $90\%$~C.L. Plot taken from Ref.~\cite{Farzan:2016wym} and reproduced with the permission of the American Physical Society.}
\label{fig:limits}
\end{figure}
Denoting the new gauge coupling  by $g^\prime$, the coupling of each generation of leptons  and quarks to $Z^\prime$  are, respectively, $g^\prime a_\alpha$ and $g^\prime/3$. There are strong bounds on the new couplings of the electrons. If $a_e \ne 0$, $Z^\prime$ with a mass of $\sim 10~$MeV will dominantly decay into $e^-e^+$ so strong bounds from beam dump experiments apply. These bounds combined with supernova cooling study yield $g^\prime a_e <3 \times 10^{-11}$ (see Fig. 4 of \cite{Harnik:2012ni}). On the other hand, for $m_{Z^\prime}<m_\pi$, the bound from $\pi^0 \to \gamma Z^\prime$
is  $g^\prime <3\times 10^{-3}$ \cite{Batley:2015lha}. (See Fig. \ref{fig:limits} which is taken from \cite{Farzan:2016wym}.)  These bounds are too stringent to lead to a discernible $\epsilon_{ee}$. We therefore set $a_e =0$ which means at tree level, neither electron nor $\nu_e$ couple to $Z^\prime$.
With such charge assignment, we obtain
\be \epsilon_{\alpha \alpha}^u= \epsilon_{\alpha \alpha}^d= \frac{g^{\prime 2} a_\alpha}{6\sqrt{2} G_F m_{Z^\prime}^2} \ \ \ {\rm and} \ \ \  \epsilon_{\alpha \beta}^u=0 |_{\alpha \ne \beta} .\label{eps}\ee
Notice that, with this technique, we only obtain lepton flavor conserving NSI. For neutrino oscillation not only the absolute value of $\epsilon_{\alpha \alpha}-\epsilon_{\beta \beta}$ but also its sign is important.  In fact, neutrino oscillation data favor positive value of $\epsilon_{ee}-\epsilon_{\mu \mu} \simeq \epsilon_{ee}-\epsilon_{\tau \tau}\sim 0.3$.  If $a_\mu+a_\tau=-3$, the anomalies cancel without any need for new generations of leptons and/or quarks. However, just like in $B-L$ and $L_\mu-L_\tau$ gauge theories, the presence of right-handed neutrinos is necessary to cancel the $U(1)^\prime- U(1)^\prime- U(1)^\prime$ anomaly.
Let us take $a_\mu=a_\tau =-3/2$ so that anomalies cancel; moreover, we obtain $\epsilon_{\mu \mu}=\epsilon_{\tau \tau}$. We can then accommodate the best fit with
\be g^\prime=4\times 10^{-5} \frac{m_{Z^\prime}}{10~{\rm MeV}} \left(\frac{ \epsilon_{ee}-\epsilon_{\mu \mu}}{0.3}\right)^{1/2}. \label{3}\ee
For the LMA-Dark solution $\epsilon_{ee}-\epsilon_{\mu \mu} <0$ is required, so the value of $a_\mu \simeq a_\tau $ should be positive. As a result, more chiral fermions are needed to be added to cancel anomalies. We will return to this point later.

Since the $U(1)^\prime$ charges of the left-handed and right-handed charged leptons are equal, their Yukawa coupling (and therefore their mass terms) preserve $U(1)^\prime$ automatically.
We should however consider the mass matrix of neutrinos with more care. While the flavor diagonal elements of neutrino mass matrix can be produced without any need for $U(1)^\prime$ breaking, if $a_\alpha$ is not universal, obtaining the neutrino mass mixing requires symmetry breaking. As mentioned above, right-handed neutrinos are also required to cancel anomalies.
If the masses of neutrinos are of Dirac type,  right-handed neutrinos will be as light as  left-handed neutrinos. They can be produced in the early universe via $U(1)^\prime$ coupling so, if they are light, they can contribute to the relativistic degrees of freedom.  To solve both problems at one shot, we can invoke the seesaw mechanism. For simplicity, we take $a_\mu =a_\tau$ so that the mixing between the second and third generation does not break $U(1)^\prime$.  Generalization to $a_\mu \ne a_\tau$ will be straightforward. Let us denote the right-handed neutrino of generation ``$i$" by $N_i$. Under $U(1)^\prime$,
\be N_1 \to N_1, \ N_2 \to e^{i a_\mu \alpha} N_2 \ {\rm and} \ N_3 \to e^{i a_\tau \alpha} N_3= e^{i a_\mu \alpha} N_3. \ee
Dirac mass terms come from
\be \lambda_1 N_1^T H^c cL_e+\lambda_2 N_2^T H^c cL_\mu+\lambda_3 N_3^T H^c cL_\tau+\lambda_4 N_2^T H^c cL_\tau+\lambda_5 N_3^T H^c cL_\mu+H.c. \ee
By changing the basis, either of $\lambda_4$ and $\lambda_5$ can be set to zero, but the nonzero one will mix the second and the third generations.
Moreover, we add electroweak singlet scalars $S_1$ and $S_2$ with $U(1)^\prime$ charges $-2 a_\mu$ and $-a_\mu$, respectively.
We can then write the following potential
\be M_1 N_1^TcN_1 +S_1 (A_2 N_2^TcN_2+A_3 N_3^TcN_3+A_{23} N_2^TcN_3)+S_2(B_2 N_1^TcN_2+B_3 N_1^TcN_3)+H.c \ee
Once $S_1$ and $S_2$ develop a vacuum expectation value (VEV), $U(1)^\prime$ will be broken leading to the desired neutrino mass and mixing scheme. The VEVs of $S_1$ and $S_2$ induce a mass of
\be g^\prime a_\mu ( 4 \langle S_1 \rangle^2 + \langle S_2 \rangle^2)^{1/2}  \ee
for the $Z^\prime$ boson.
  Taking $g^\prime a_\mu\sim 10^{-5}-10^{-4}$, we find that as long as  $\langle S_1 \rangle \sim \langle S_2 \rangle \sim 100 ~{\rm GeV} (10^{-4}/g^\prime a_\mu)$, the contribution to the $Z^\prime$ mass will be $\sim 10$ MeV as desired.
  In case that more scalars charged under $U(1)^\prime$ are added to the model (we shall see examples in sect \ref{LFv}), the $Z^\prime$ mass receives further contributions.

  For $m_{Z^\prime}<m_\pi$, the $Z^\prime$ can decay only to neutrino pair at tree level with a lifetime of
  $$ c \tau_{Z^\prime} \sim 10^{-9} ~{\rm km} \left( \frac{7 \times 10^{-5}}{g^\prime} \right)^2 \left(\frac{10~{\rm MeV}}{m_{Z^\prime}} \right) \frac{1}{a_\mu^2+a_\tau^2} . $$
  As a result, $Z^\prime$ evades the bounds from the beam dump experiments. In the following, we go through possible experiments that can search for the $Z^\prime$ boson.
  \begin{figure}[t]
\centering
\includegraphics[width=0.5\textwidth]{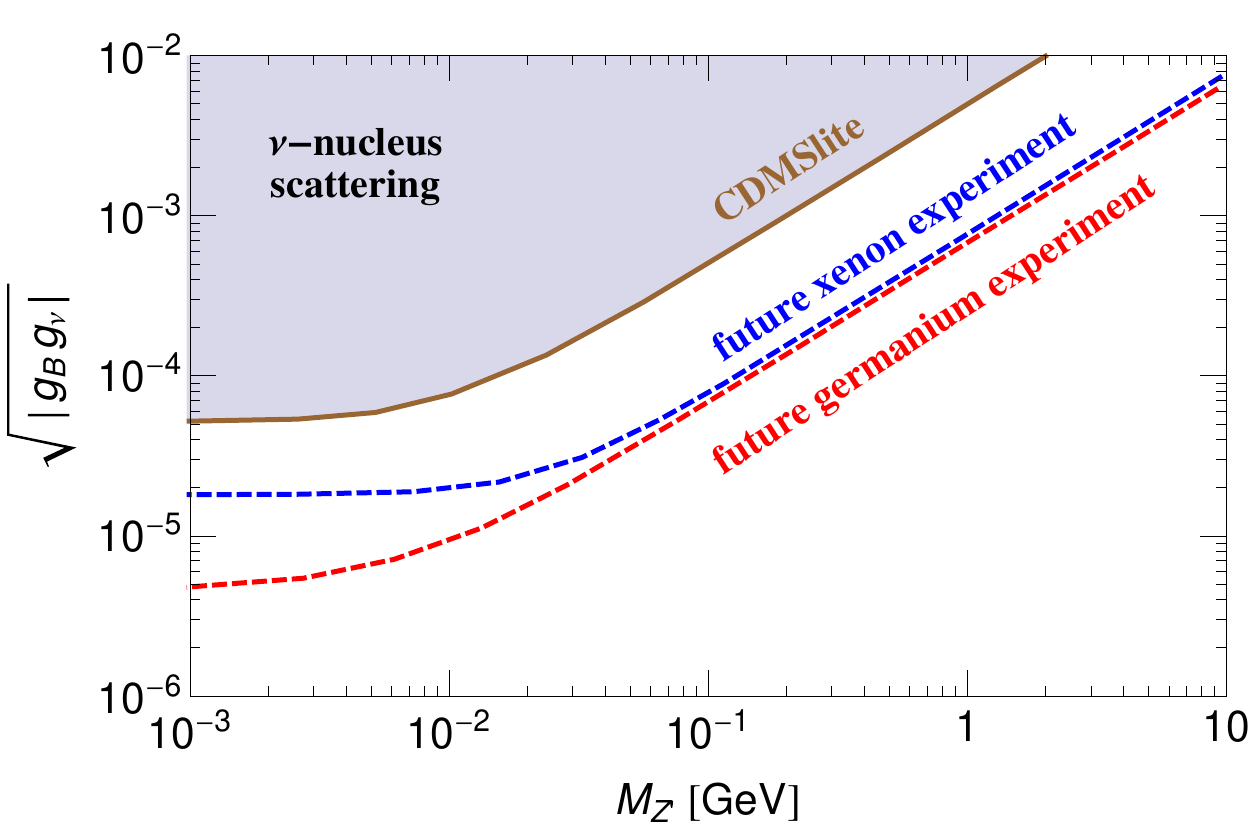}
\caption{\label{fig:neutrino_nucleon_scattering}
Approximate $90\%$~C.L.~bounds on the product of couplings of neutrinos and quarks to $Z^\prime$ from solar-neutrino nuclear recoils in CDMSlite and optimistic projections for second-generation Xenon (e.g.~LUX--ZEPLIN) and Germanium experiments (e.g.~SuperCDMS SNOLAB), adapted from Ref.~\cite{Cerdeno:2016sfi}. Plot taken from Ref.~\cite{Farzan:2016wym} and reproduced with the permission of the American Physical Society.
}
\end{figure}
\begin{figure}[t] \centering
	\includegraphics[width=0.6\textwidth]{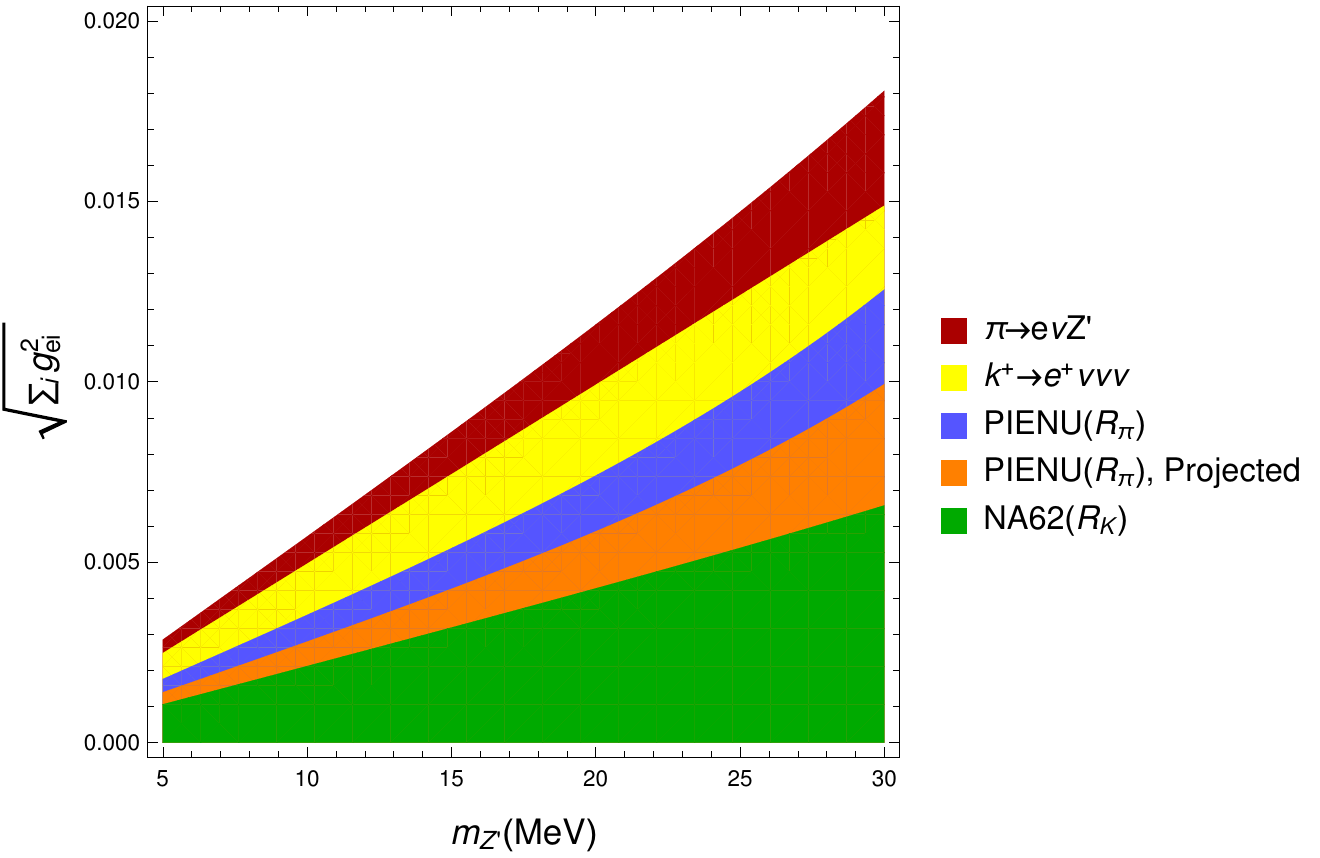}
       \caption{\label{Ei} 90$\%$ C.L. constraints on   $\sqrt{\sum_i g_{ei}^2}$ versus $m_{Z^\prime}$ from constraints on $\pi\longrightarrow e\nu Z'$ \cite{Picciotto:1987pp}  and  $K^+\longrightarrow e^+\nu\nu\nu$ \cite{Heintze:1977kk}  branching ratios, from current and projected $R_\pi$  measurement by PIENU \cite{Aguilar-Arevalo:2015cdf} and from the $R_K$ measurement by  NA62 \cite{Lazzeroni:2012cx}. $g_{e i}$ is the coupling of $Z^\prime_\mu \bar{\nu}_e \gamma^\mu \nu_i$ where $\nu_i$ can be any neutrino state much lighter than $\sim 100$ MeV. Figure taken  from Ref.~\cite{Bakhti:2017jhm} and reproduced with the permission of the American Physical Society.}
\end{figure}

\begin{figure}[t] \centering
	\includegraphics[width=0.5\textwidth]{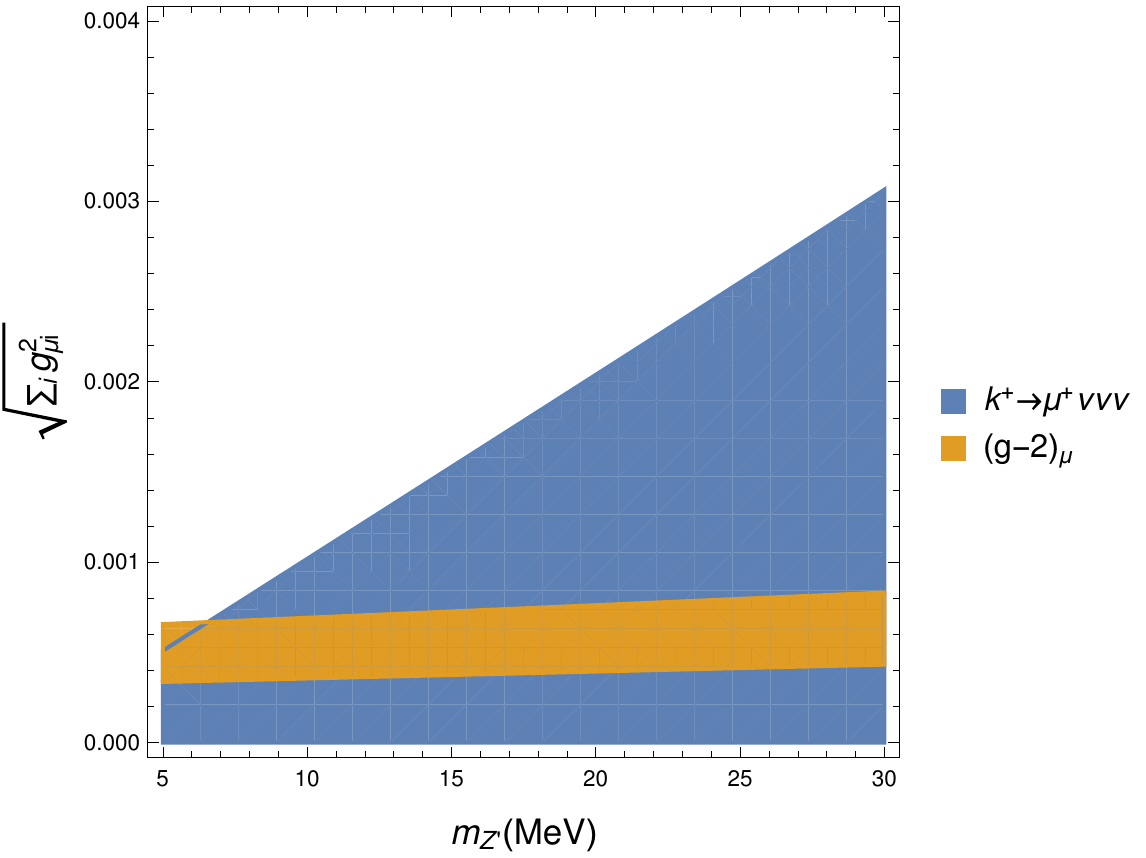}
        \caption{\label{Ei1}  90$\%$ C.L. constraints on   $\sqrt{\sum_i g_{\mu i}^2}$  versus $m_{Z^\prime}$ from $K^+\longrightarrow \mu^+\nu\nu\nu$  branching ratio \cite{Pang:1989ut}. The band shows the parameter space within  $L_\mu$ gauge models (giving rise to equal couplings to $\mu$ and $\nu_\mu$) that can explain the $(g-2)_\mu$ anomaly \cite{Pospelov:2008zw}.  $g_{\mu i}$ is the coupling of $Z^\prime_\mu \bar{\nu}_\mu \gamma^\mu \nu_i$ where $\nu_i$ can be any neutrino state much lighter than $100$~MeV. Figure taken  from Ref.~\cite{Bakhti:2017jhm} and reproduced with the permission of the American Physical Society.}
\end{figure}
  %\begin{itemize}
  %\item
  %{\it{Impact on meson decay:}}
  In the presence of new interactions, new decay modes for charged mesons open up: $K^+ \to l^++\nu+Z^\prime$ and $\pi^+ \to l^++\nu+Z^\prime$.  The typical upper bounds from meson decay are of order of $O(0.001)$
  \cite{Bakhti:2017jhm,Belotsky:2001fb,Laha:2013xua,Barger:2011mt} which are too weak to be relevant for our models; see Figs.~\ref{Ei} and \ref{Ei1}, which are taken from \cite{Bakhti:2017jhm}. As shown in \cite{Bakhti:2017jhm}, the bound  on the $\nu_e$ coupling to the $Z^\prime$ boson can be dramatically improved by customized searches for three body decays ($K^+ \to e^++{\rm missing~ energy}$) and ($\pi^+ \to e^++{\rm missing~ energy}$).
  %Decay of $Z^\prime$ to $\nu_\mu \bar{\nu}_\mu$ and $\nu_\tau \bar{\nu}_\tau$ can warm up the neutrino background in early universe leading to a positive contribution $\Delta N_{eff}$. The bound %from CMB and BBN on $\Delta N_{eff}$ implies $m_{Z^\prime}>5$ MeV \cite{Kamada:2015era}.
 % \item
  %{\it{The $Z^\prime-\gamma$ mixing:}}
  
  In principle, $Z^\prime$ can  kinetically  mix with the hypercharge gauge boson which gives rise to  $Z^\prime$ mixings both with the photon and the $Z$ bosons. Even if we set the kinetic mixing to zero at tree level, it can be produced at loop level as long as there are particles charged under both $U(1)$ gauge symmetries. Going to a basis where the kinetic terms of gauge bosons is canonical, the $Z^\prime$ boson obtains a coupling to the electron given by $e\epsilon$ where $\epsilon$ is the kinetic mixing between $Z^\prime$ and the photon. This coupling can affect neutrino  interaction with the electron on which there are strong bounds from solar experiments (mostly Super-Kamiokande and Borexino)~\cite{Bolanos:2008km,Agarwalla:2012wf}. Ref \cite{Kamada:2015era}, setting the tree level kinetic mixing equal to zero, has  calculated the kinetic mixing for the $L_\mu -L_\tau$ models and has found it to be finite and of order of $e g^\prime/8 \pi^2$.  The Borexino bound~\cite{Agarwalla:2012wf} can then be translated into $g^\prime e\stackrel{<}{\sim} 10^{-4}$ which can be readily satisfied for $g^\prime <10^{-4}$. The loop contribution to the photon $Z^\prime$ mixing from a charged particle is very similar to its contribution  to the vacuum polarization (photon field renormalization)  replacing $(q e)^2$ with $(q e)(a_\alpha g^\prime)$. In case of the $L_\mu -L_\tau$ gauge symmetry,  $a_e=0$, $ a_\mu=-a_\tau$ and since the electric charges of $\mu$ and $\tau$ are the same, the infinite parts of their contribution to the mixing cancels out. In general, we do  not however expect such a cancelation and  counter terms are therefore required. Once we open up the possibility of tree level kinetic mixing, the sum of tree level and loop level mixing can be set to arbitrarily small value satisfying any bound.

 %  Even if we set the kinetic mixing equal to zero, a kinetic mixing of order of $\epsilon \sim e g^\prime/(8\pi^2)$ appears in one loop %level. The bound from Borexino on such a mixing (coming from neutrino electron interaction) is $(g^\prime e \epsilon)^{1/2}<10^{-5}$ which %can be easily satisfied for $g^\prime <10^{-4}$.
%\item
%{\it{$Z^\prime-Z$ mixing:}}
The above discussion on the $Z^\prime-\gamma$ kinetic mixing also applies to the $Z^\prime-Z$ kinetic mixing. Here, we should also check the $Z-Z^\prime$ mass mixing~\cite{Davoudiasl:2014kua}. It is straightforward to show that, since the $Z^\prime$ couplings are taken to be non-chiral, there is no contribution to the $Z-Z^\prime$ mass mixing at one loop level.
  If the model contains scalars that are charged both under electroweak and $U(1)^\prime$ and develop VEV, mass mixing between $Z$ and $Z^\prime$ appears even at tree level. In the minimal version of the model that is described above there is no such scalar but we shall come back to this point in sect. \ref{LFv}.

 % \item
  %{\it{ Big Bang Nucleosynthesis (BBN):}}
  Decay of $Z^\prime$ to neutrino pairs can warm up the neutrino background during and right after the Big Bang Nucleosynthesis (BBN) era. The effect can be described by the contribution to the effective extra relativistic degrees of freedom $\Delta N_{eff}$. As shown in Ref.~\cite{Kamada:2015era}, BBN bounds rule out $m_{Z^\prime}< 5$~ MeV. Of course, this lower bound on $m_{Z^\prime}$ applies only if
  the coupling is large enough to bring $Z^\prime$ to thermal equilibrium with neutrinos before they decouple from the plasma at $T\sim 1$ MeV. That is, for $ 1 ~{\rm MeV}<m_{Z^\prime}<5 ~{\rm MeV}$ the coupling should be smaller than $\sim 3 \times 10^{-10} $ \cite{Huang:2017egl}.

 % \item {\it{Supernova cooling:}}
 NSI can leave its imprint on the flavor composition of supernova neutrino flux~\cite{Das:2011gb,EstebanPretel:2007yu,Das:2017iuj}. Moreover, $Z^\prime$ particle can be produced and decay back to neutrinos within the supernova core. This leads to a shortening  of the mean free path of neutrinos inside the supernova core~\cite{Kamada:2015era}. This, in turn, results in prolonging the duration of neutrino emission from supernova. To draw a quantitative conclusion and bound, a full simulation is required.

 % \item {\it{Dips in the spectrum of high energy neutrinos:}}
 Once we introduce the new interaction for neutrinos, high energy neutrinos (or antineutrinos) traveling across the universe resonantly  interact with cosmic background antineutrinos (or neutrinos) producing $Z^\prime$ which decays back to $\nu \bar{\nu}$ pair with  energies lower than that of initial neutrino (or antineutrino). This will result in a dip in the spectrum of high energy neutrinos.
      Taking the cosmic background neutrinos as non-relativistic, we expect the position of the dip to be given by $E_\nu\sim {\rm PeV} (m_{Z^\prime}/10~{\rm MeV})^2 (0.05~{\rm eV}/m_\nu)$. The value is tantalizingly close to the observed (but by no means established) gap in the high energy IceCube data.
      Moreover, as shown in \cite{Kamada:2015era} with $g^\prime \sim  10^{-5}-10^{-4}$ (the range of interest to us), the optical depth is larger than one. Thus, this rather robust prediction can be eventually tested by looking for the dip in the high energy neutrino data.

 %     \item {\it{ Magnetic dipole moment of the muon:}}
  The contribution from the $Z^\prime$ loop to $(g-2)_\mu$ can be estimated as $g^{\prime 2}/(8\pi)$ up to  corrections of order $O(m_{Z^\prime}^2/m_\mu^2)\sim 0.01$.  For $g^\prime <10^{-4}$, the contribution is too small to explain the claimed discrepancy~\cite{Jegerlehner:2009ry}.

  %        \item  {\it{Neutrino scattering experiments:}}
  Let us now discuss the neutrino scattering experiments.
  The amplitude of the contribution from $t$-channel $Z^\prime$ exchange to neutrino quark scattering is suppressed relative to that from SM by a factor of $m_{Z^\prime}^2/(t- m_{Z^\prime}^2)$, where $t$ is the Mandelstam variable. At CHARM and NuTeV experiments, the energy momentum exchange was about 10 GeV ($t \gg m_{Z^\prime}^2$), so the new effects were suppressed. As a result, the bound found in \cite{Davidson:2003ha,Escrihuela:2009up,Coloma:2017egw} does not apply to the model with a light gauge boson. However, as discussed in \cite{Coloma:2017egw,Farzan:2016wym,Dutta:2017nht}, low energy scattering experiments can be sensitive to low mass gauge interactions. Three categories of scattering experiments have been studied in this regard: (1) Scattering of solar neutrino at direct dark matter search experiments \cite{Farzan:2016wym,Dutta:2017nht,Cerdeno:2016sfi,AristizabalSierra:2017joc}. As shown in \cite{Farzan:2016wym}, the upcoming Xenon based experiments such as LUX-Zeplin and the future Germanium based experiments such as superCDMS at SNOLAB can test most of the parameter space of our interest (see Fig.~\ref{fig:neutrino_nucleon_scattering}, adapted from \cite{Farzan:2016wym}). (2) As shown in detail in \cite{Coloma:2017egw,Shoemaker:2017lzs}, the running COHERENT experiment \cite{Akimov:2015nza} is an ideal setup to probe NSI with a light mediator.
              At this experiment, low energy $\nu_\mu$ and $\nu_e$ fluxes are produced via pion and muon decay at rest. The LMA-Dark solution can be entirely probed by this experiment \cite{Coloma:2017egw}. The COHERENT experiment has recently released its preliminary results, ruling out a significant part of the parameter space. We shall discuss the new results in sect. \ref{Coh}.  (3) Scattering of reactor $\bar{\nu}_e$ flux off nuclei can also probe NSI of the type we are interested in \cite{Wong:2008vk,Aguilar-Arevalo:2016khx,Aguilar-Arevalo:2016qen,Agnolet:2016zir,Billard:2016giu,Lindner:2016wff,Barranco:2005yy,Kerman:2016jqp}.

  %\item {\it{Neutrino trident production:}}
  The $Z^\prime$ gauge boson coupled to $\nu$ and $\mu$ can contribute to the so-called neutrino trident production
  $\nu+A \to \nu+A+\mu^++\mu^-$, where $A$ is a nucleus. The rate of such interaction was measured by the CCFR~\cite{Mishra:1991bv} and CHARM II~\cite{Geiregat:1990gz} collaborations, and is found to be consistent with the SM prediction. This observation sets the bound $g^\prime a<9 \times 10^{-4}$ for $m_{Z^\prime} \sim 10$~MeV \cite{Altmannshofer:2014pba,Ge:2017poy}.
  %\end{itemize}

As we saw earlier, taking $a_\mu=a_\tau=-3/2$, the contributions from  the field content of the SM to anomalies cancel out. We can then obtain any negative values of
$\epsilon_{\mu \mu}-\epsilon_{ee}=\epsilon_{\tau \tau}-\epsilon_{ee}\sim -1$ by choosing $g^\prime\sim 10^{-4} (|  \epsilon_{\mu \mu}-\epsilon_{ee}|)^{1/2}$ (see Eq.(\ref{3})).
Let us now discuss if with this mechanism we can reconstruct a model that embeds the LMA-Dark solution with  positive $\epsilon_{\mu \mu}-\epsilon_{ee} =\epsilon_{\tau\tau}-\epsilon_{ee} \sim 1$.
 The condition  $\epsilon_{\mu \mu}-\epsilon_{ee}=\epsilon_{\tau \tau}-\epsilon_{ee}\sim 1$ can be satisfied if $a_e=0$, $a_\mu=a_\tau>0 $ and
\be
\label{failed}g^\prime\sim 10^{-4} \left( \frac{m_{Z^\prime}}{10~{\rm MeV}}\right) \left( \frac{1}{a_\mu}\right)^{1/2}.
\ee
We should however notice that with $a_\mu=a_\tau>0$, the cancelation of $U(1)^\prime-SU(2)-SU(2)$ and $U(1)^\prime-U(1)-U(1)$ anomalies require
new chiral fermions. A new generation of leptons with $U(1)^\prime$ charge equal to $-(3+a_\mu+a_\tau)$ can cancel the anomalies but in order for these new fermions to acquire masses large enough to escape bounds from direct production at  colliders, their Yukawa
couplings enter the non-perturbative regime.  Similar argument holds if we add a new generation of quarks instead of leptons. Another option is to add a pair of new generations of leptons (or quarks) with opposite $U(1)_Y$ charges but equal $U(1)^\prime$ charge of $-(3+a_\mu+a_\tau)/2$. Let us denote the field content of the fourth generation with $\nu_{R 4}$, $e_{R 4}$ and $L_4$, and similarly that of the fifth generation with $\nu_{R 5}$, $e_{R 5}$ and $L_5$. As pointed out, the hypercharges of fourth and fifth generation are opposite so we can write Yukawa terms of type
$$Y_1 S e_{R4}^Tc e_{5 R}+Y_2 S L_{4}^TcL_5+{\rm H.c.}\, ,$$
where $S$ is singlet of the electroweak  symmetry group, $SU(2)\times U(1)$ with a $U(1)^\prime$ charge of $3+a_\mu+a_\tau$. Even for $Y_1\sim Y_2 \sim 1$, in order to obtain heavy enough mass, $\langle S\rangle$ should be of order of TeV. On the other hand, $\langle S\rangle$ contributes to $Z^\prime$ mass so
$$ {\rm Masses~ of ~4th ~and~ 5th~ generation}\lesssim \langle S\rangle <5~{\rm TeV}\frac{m_{Z^\prime}}{10~{\rm MeV} }\frac{2\times 10^{-6}}{g^\prime (3+a_\mu+a_\tau)}.$$
In other words, $g^\prime \lesssim \frac{5~{\rm TeV}}{M_{4,5}} \frac{2 \times 10^{-6}}{3+a_\mu +a_\tau}\frac{m_{Z^\prime}}{10~{\rm MeV}}$, where $M_{4,5}$ are the typical masses of the fourth and fifth generation leptons. Inserting this in Eq.~(\ref{eps}), we find $\epsilon_{\mu \mu}=\epsilon_{\tau \tau}\lesssim 0.01$.\\

In general, the cancelation of $U(1)^\prime-SU(2)-SU(2)$ and $U(1)^\prime-U(1)-U(1)$ anomalies  requires new chiral fermions charged under $U(1)^\prime$ and $SU(2)\times U(1)^\prime$ (or both). In the former case, we need new $U(1)^\prime$ charged scalars whose VEV contribute to the $Z^\prime$ mass.  The lower bounds on the masses of new particles set a lower bound on the VEV of new scalars which, in turn, can be translated into an upper bound on $g^\prime/m_{Z^\prime}$ which leads to $\epsilon_{\mu \mu}\stackrel{<}{\sim} 0.01$.
In the second case, large masses of the 4th and 5th generations requires non-perturbative Yukawa coupling to the Higgs. If masses of the new fermions could be about few hundred GeV, none of these obstacles would exist. Fortunately, there is   a trick to relax the strong lower bounds from colliders on the masses of new particles. Let us suppose the charged particles are just slightly heavier than their neutral counterparts. Their decay modes can be then of type $e^-_{4(5)} \to \nu_{4(5)} l \nu, \nu_{4(5)} q^\prime \bar{q}$ with a final charged lepton or jet too soft to be detected at colliders. In this case, the new generation can be as light as few 100 GeV so their mass can come from a perturbative Yukawa coupling to the SM Higgs or new scalars charged under $U(1)^\prime$ and VEV of $\sim 100$ GeV opening up a hope for $g^\prime/m_{Z^\prime} \sim 10^{-5} ~{\rm MeV}^{-1}$ and therefore for $\epsilon_{\mu \mu}=\epsilon_{\tau\tau}\sim 1$.

\subsection{A model both for LF conserving and LFV NSI\label{LFv}}
As mentioned in section \ref{direct}, the coupling of $Z^\prime$ to neutrinos can be achieved with two mechanisms: (i) The $(\nu_\alpha \ \ell_{L \alpha})$ doublet is assigned  a charge under $U(1)^\prime$, so neutrinos directly obtain a gauge coupling to the $Z^\prime$ boson. This route was discussed in section \ref{direct}. (ii) Active neutrinos mix with a new fermion singlet under electroweak {group symmetry} but charged under new $U(1)^\prime$. In this section, we focus on the second route. Using the notation in Eq.(\ref{charge-as}), in the present scenario {one has} $a_e=a_\mu=a_\tau=0$ so to cancel the $U(1)^\prime-SU(2)-SU(2) $ and $U(1)^\prime-U(1)-U(1)$ anomalies, we should add new fermions. As discussed in the previous section, in order to make these new fermions heavier than $\sim 1 $~TeV, we need new scalars charged under $U(1)^\prime$ with a VEV of 1~TeV. To keep the contribution from the new VEV to $Z^\prime$ mass under control, \be g^\prime<10^{-5} \frac{m_{Z^\prime}}{10~{\rm MeV}}.\ee Notice that this tentative bound is stronger than the bound from $\pi \to Z^\prime \gamma$ (see Fig. \ref{fig:limits}).  Let us introduce a new Dirac fermion $\Psi$ which is neutral under electroweak symmetry but charged under $U(1)^\prime$. Its $U(1)^\prime$ charge denoted by $a_\Psi$ can be much larger than one. Since we take equal $U(1)^\prime$ charges for $\Psi_L$ and $\Psi_R$, no anomaly is induced by this Dirac fermion. Let us denote the mixing of $\Psi$ with neutrino of flavor $\alpha$ with $\kappa_\alpha$. Such a mixing of course breaks $U(1)^\prime$. Mixing can be obtained in
two ways:
\begin{itemize}

\item We add a sterile Dirac $N$ (neutral both under electroweak and under $U(1)^\prime$) and a scalar ($S$) to break $U(1)^\prime$. The $U(1)^\prime$ charge of the $S$ is taken to be equal to that of $\Psi$. We can then add terms like the following to the Lagrangian:
\be m_\Psi \bar{\Psi}\Psi+m_N \bar{N}N+ Y_\alpha \bar{N}_R H^T cL_\alpha +\lambda_L S \bar\Psi_R N_L.\ee
Notice that we were allowed to add a term of $\lambda_R S \bar{\Psi}_L N_R$ too, but this term is not relevant for our discussion. Taking $Y_\alpha \langle H\rangle, \lambda_L \langle S \rangle, m_\Psi \ll m_N$, we can integrate out $N$ and obtain
\be \label{kappa} \kappa_{\alpha}= \frac{Y_\alpha \langle H\rangle \lambda_L \langle S\rangle}{m_Nm_\Psi}.\ee
Since we take $\lambda_L \langle S\rangle<m_N$, in order to have sizeable $\kappa_\alpha$, the mass of $\Psi$ cannot be much larger than $Y_\alpha \langle H\rangle$. On the other hand, $Y_\alpha$ determines the new decay mode of $H \to \nu N$ which is observationally constrained \cite{Trocino:2016zde}. We therefore find an upper bound on $m_\Psi$ of few GeV. For example, taking $m_\Psi=2$~GeV, $m_N=20$~GeV, $Y_\alpha \langle H\rangle=0.1$~GeV, $\lambda_L=1$ and $\langle S\rangle \sim 4$ GeV, we obtain $\kappa_\alpha \simeq 0.01$. With such small $Y_\alpha$, the rate of the  Higgs decay  into $N$ and $ \nu$ will be as small as $\Gamma(H\to \mu \mu)$ and  therefore negligible. With $g^\prime<10^{-4}$, the contribution from $\langle S \rangle$ to $m_{Z^\prime}$ will also be negligible.

\item Another scenario which has been  proposed in  \cite{Farzan:2016wym} invokes a new Higgs doublet $H^\prime$ which has $U(1)^\prime$ charge equal to that of $\Psi$. The Yukawa coupling will be then equal to $\mathcal{L}=-\sum_\alpha y_\alpha \bar{L}_\alpha H^{\prime T}c\Psi$ which leads to $$\kappa_\alpha =\frac{y_\alpha \langle H^\prime\rangle}{\sqrt{2}m_\Psi}$$ where $\tan \beta= \langle H\rangle /\langle H^\prime \rangle$.
The VEV of $H^\prime$ can contribute to the $Z^\prime$ mass so we obtain
\be \cos \beta \leq 4 \times 10^{-5} (\frac{m_{Z^\prime}}{10~{\rm MeV}})\frac{1}{g_\Psi}. \label{cos}\ee
Thus, to obtain sizeable $\kappa_\alpha$  ({\it e.g.,} $\kappa_\alpha>0.03$), we find
\be M_\Psi<{\rm few}~{\rm GeV}\frac{m_{Z^\prime}}{10~{\rm MeV}} \frac{0.2}{g^\prime a_\Psi} \frac{0.03}{\kappa_\alpha}. \ee

Moreover, the VEV of $H^\prime$ can induce $Z-Z^\prime$ mixing  on which there are strong bounds
\cite{Davoudiasl:2012ag}. These bounds can be translated into $\cos \beta < 10^{-4} ({m_{Z^\prime}}/{10~{\rm MeV}})({1}/{g_\Psi})$
which is slightly weaker than the bound in (\ref{cos}). The smallness of $\langle H^\prime\rangle$, despite its relatively large mass, can be explained by adding a singlet scalar(s) of charge $a_\Psi$ with $\mathcal{L}=-\mu S_1^\dagger H^\dagger H^\prime$ which induces
$\langle H^\prime \rangle =-\mu \langle S_1 \rangle /(2 M_{H^\prime}^2)$. Taking $\langle S_1 \rangle \mu \ll M_{H^\prime}^2$, we find $\cos \beta \ll 1$. The components of $H^\prime$ can be pair produced at colliders via electroweak interactions. They will then decay to $\Psi$ and leptons. In particular, the charged component $H^-$ can decay into charged lepton plus $\Psi$ which appears as missing energy. Its signature will be similar to that of a charged slepton \cite{Farzan:2016wym}. According to the present bounds \cite{Farzan:2016wym,Khachatryan:2014qwa}, $m_{H^\prime}\gtrsim 300$~GeV.
\end{itemize}
Regardless of the mechanism behind the mixing between $\Psi$ and $\nu$, it will lead to the coupling of $Z^\prime$ to active neutrinos as follows
\be g^\prime a_\Psi Z^\prime_\mu \left( \sum_{\alpha,\beta} \kappa_\alpha^* \kappa_\beta \bar{\nu}_\alpha \gamma^\mu P_L \nu_\beta- \kappa_\alpha^* \bar{\nu}_\alpha \gamma^\mu P_L \Psi-\kappa_\alpha \bar{\Psi} \gamma^\mu P_L \nu_\alpha\right) \label{12}\, ,\ee
which leads to
$ \epsilon_{\alpha \beta}^u=\epsilon_{\alpha \beta}^d=\frac{g^{\prime 2} a_\Psi \kappa_\alpha^* \kappa_\beta}{6 \sqrt{2} G_F m_{Z^\prime}^2}.$
Notice that if the mixing of $\Psi$ with more than one flavor is nonzero, we can have lepton flavor violating NSI with $\epsilon_{\alpha \alpha}^{u (d)}
\epsilon_{\beta \beta}^{u (d)}=|\epsilon_{\alpha \beta}^{u (d)}|^2$. If more than one $\Psi$ is added, we may label the mixing of $i$th $\Psi$ to $\nu_\alpha$ with $\kappa_{i \alpha}$. The Schwartz inequality $(\sum_i \kappa_{i\alpha}^* \kappa_{i\beta})^2<(\sum_i |\kappa_{i \alpha}|^2)(\sum_i |\kappa_{i \beta}|^2)$ then still applies
$$ \epsilon_{\alpha \alpha}^{u(d)} \epsilon_{\beta \beta}^{u(d)}> ( \epsilon_{\alpha \beta}^{u(d)})^2.$$
Taking $\kappa_{i \alpha}=\delta_{i \alpha}$, meaning that each $\Psi_i$ mixes with only one $\nu_\alpha$, only diagonal elements of $\epsilon_{\alpha \beta}$ will be nonzero, preserving lepton flavor.\\

Notice that $\Psi$ in our model decays into $Z^\prime$ and $\nu$ and appears as missing energy. $\Psi$ should be heavier than MeV; otherwise, it can contribute to extra relativistic degrees of freedom in the early universe.
Remember that we have found that $m_\Psi<$ few GeV.
The mixing of active neutrinos with $\Psi$ results in the violation of the unitarity of $3\times 3$ PMNS matrix on which there are strong bounds \cite{Fernandez-Martinez:2016lgt,Escrihuela:2015wra,Escrihuela:2016ube}
 \be
 |\kappa_e|^2<2.5 \times 10^{-3}~ , ~~ |\kappa_\mu|^2<4.4 \times 10^{-4}~ ~~ {\rm and} ~~~ |\kappa_\tau|^2<5.6 \times 10^{-3}~{\rm at}~2\sigma
\ee
 which immediately give
 \be
 |\kappa_\mu \kappa_e|< 10^{-3}, \ \ \ |\kappa_\mu \kappa_\tau|<1.6 \times 10^{-3} \ \ \ {\rm and} \ \ \ |\kappa_e\kappa_\tau|<3.7 \times 10^{-3} \ \ \ {\rm at} \ 2\sigma.
 \ee
Under certain assumptions, Ref. \cite{Fernandez-Martinez:2016lgt} also derives independent bound on $\kappa_\alpha \kappa_\beta^*|_{\alpha \ne \beta}$ from lepton flavor violation (LFV) processes $l_\alpha^- \to l_\beta^- \gamma$, but these bounds are valid only for $m_\Psi \gg m_W$.
For our case with $m_\Psi\ll m_W$, a GIM mechanism is at work and suppresses the contribution to $l_\alpha^- \to l_\beta^- \gamma$ from $\nu-\Psi$ mixing.
In the case that the mixing comes from Yukawa coupling to $H^\prime$, because of the LFV induced by  $H^\prime$ and $\Psi$ coupling to more than one flavor, a new contribution to $l_\alpha^- \to l_\beta^- \gamma$ appears. As shown in \cite{Farzan:2016wym}, from $Br(\tau \to e \gamma)<3.3 \times 10^{-8}$, $Br(\tau \to \mu \gamma)<4.4 \times 10^{-8}$~\cite{Olive:2016xmw} and $Br(\mu \to e \gamma)<4.2 \times 10^{-13}$\cite{TheMEG:2016wtm}, one can respectively derive $|y_ey_\tau|<0.46 (m_{H^-}/(400~ {\rm GeV}))^2$,  $|y_\mu y_\tau|<0.53 (m_{H^-}/(400~ {\rm GeV}))^2$ and  $|y_ey_\mu|<7 \times 10^{-4} (m_{H^-}/(400~ {\rm GeV}))^2$. As demonstrated in \cite{Farzan:2016wym}, except for $\epsilon_{e \mu}$ which is strongly constrained by the bound from $\mu \to e \gamma$ within the model described in
  \cite{Farzan:2016wym}, all components of $\epsilon_{\alpha \beta}$ can be within the reach of current and upcoming long baseline neutrino experiments.
  If mixing is achieved via the mechanism described in  Eq.~(\ref{kappa}), no new bound from LFV rare decay applies and we can obtain all  $\epsilon_{\alpha \beta}$ (including $\epsilon_{e\mu}$) of the order of
  \be \epsilon_{\alpha \beta}^u=\epsilon_{\alpha \beta}^d=g^\prime a_\Psi \left(\frac{g^\prime}{10^{-5}}\right) \frac{\kappa^*_\alpha \kappa_\beta}{10^{-3}} \left( \frac{10~{\rm MeV}}{m_{Z^\prime}}\right)^2.\ee
 Notice that in this model, the coupling of $Z^\prime$ to neutrino pairs can be much larger than the coupling to quarks: $|g^\prime a_\Psi \kappa_\alpha \kappa_\beta|\gg g^\prime$, see Eq.(\ref{12}). The bounds from meson decays on $Z^\prime$ coupling to neutrino pairs have been studied in \cite{Bakhti:2017jhm}. The results are shown in Figs.~\ref{Ei} and \ref{Ei1}.
  The strongest bound  for $m_{Z^\prime}\sim 10$~MeV is of order of 0.001,  which can be readily satisfied if $\kappa_\alpha \kappa_\beta<10^{-3}$. However, further data on $[\pi^+\to e^+(\mu^+)+{\rm missing~energy}]$ or on $[K^+\to e^+(\mu^+)+{\rm missing~energy}]$ can probe  parts of the parameter space of interest to us.

  \subsection{Impact of recent results from COHERENT experiment \label{Coh}}

  Recently, the COHERENT experiment has released its first results confirming the SM prediction of elastic scattering of neutrinos off nuclei at 6.7$\sigma$, studying the interaction of $\nu_\mu$, $\bar{\nu}_\mu$ and $\nu_e$ flux from Spallation Neutron Source (SNS) at the Oak Ridge National Laboratory on a 14.6 kg CsI[Na] scintillator detector~\cite{Akimov:2017ade}. The preliminary results already set strong bounds on NSI.

  Assuming the validity of the contact interaction approximation ({\it i.e.,} assuming the mass of the mediator is heavier than $\sim 10$ MeV), Ref. \cite{Coloma:2017ncl} shows that the recent COHERENT data rules out LMA-Dark solution. Ref. \cite{Liao:2017uzy}, taking a universal coupling of $Z^\prime$ to SM fermions finds that
  \be \label{Danny} \left( g_\nu g_q\right)^{1/2} <6 \times 10^{-5} \ \ {\rm for} \ \ m_{Z^\prime}<30 ~{\rm MeV} \ \ {\rm at} \ 2 \sigma . \ee

  Let us discuss how this bound can constrain our model(s) for NSI.
Regardless of the details of the underlying theory, we can write
$$    \left( g_\nu g_q\right)^{1/2}=5.47 \times 10^{-5} \sqrt{\epsilon_{\alpha \beta}} \left( \frac{m_{Z^\prime}}{10~{\rm MeV}}\right)$$
where $g_q =g_B/3$ is the coupling of $Z^\prime$ boson to quarks and $(g_\nu)_{\alpha \beta}$ is its coupling to $\nu_\alpha$ and $\nu_\beta$. In the model described in sect \ref{direct}, $(g_\nu)_{\alpha \beta} =\delta_{\alpha \beta}a_{\alpha}g^\prime$ and in the model of sect \ref{LFv}, $(g_\nu)_{\alpha \beta}= g^\prime a_\Psi \kappa_\alpha \kappa_\beta$. Remember that, in order for $\epsilon_{\alpha \beta}$ to show up in neutrino oscillation experiments, it should be non-universal. For example,  the LMA-Dark solution
requires $\epsilon_{\mu\mu}-\epsilon_{ee}=\epsilon_{\tau\tau}-\epsilon_{ee}\sim 1$.
Setting $\epsilon_{ee}=0$ and $\epsilon_{\mu\mu} \ne 0$, we expect the bound on $|\epsilon_{\mu\mu}|$ from the COHERENT experiment to be slightly weaker than that found in Ref. \cite{Liao:2017uzy} taking $\epsilon_{\mu\mu}=\epsilon_{ee}\ne 0$. Thus, the LMA-Dark solution  still survives with the  present COHERENT data but, further data from COHERENT as well as the data from  upcoming reactor neutrino-nucleus coherent scattering experiments such as the setup described in \cite{Lindner:2016wff} can probe the most interesting part of the parameter space.

%%%%%%%%%%%%%%%%%%%%%%%%%%%%
%%%%%%%%%%%%%%%%%%%%%%%%%%%%
%%%%%%%%%%%%%%%%%%%%%%%%%

\section{NSI at upcoming long baseline experiments: T2HK, T2HKK, DUNE, JUNO and MOMENT\label{upcoming}}
In recent years, rich literature has been developed on the possibility of detecting the NSI effects in upcoming long baseline neutrino experiments. In particular, degeneracies induced by the presence of NSI in the DUNE experiment have been scrutinized \cite{Blennow:2016etl,Masud:2016bvp,Masud:2016gcl,Coloma:2015kiu,Agarwalla:2016fkh,Oki:2010uc,Fukasawa:2016lew,Liao:2016hsa,Abe:2016ero,Masud:2015xva,deGouvea:2015ndi,Huitu:2016bmb,Liao:2016orc,Masud:2017bcf,Rout:2017udo}.
In sect. \ref{NSI-up}, we  review the effects of NSI at DUNE, T2HK and T2HKK experiments. In sect \ref{NSI-Dark}, we show  how intermediate baseline reactor experiments such as JUNO and RENO-50 can help to determine ${\rm sign}(\cos 2 \theta_{12})$ and therefore test the LMA-Dark solution. In sect \ref{Moment}, we show how the MOMENT experiment can help to determine the octant of $\theta_{23}$ and the true value of $\delta$ despite the presence of  NSI.
Throughout this section, we set $\epsilon_{\mu \mu}=0$ for definiteness and consistency with  the majority of our references.

\subsection{NSI at upcoming long baseline neutrino experiments\label{NSI-up}}
Let us first briefly review the setups of the three upcoming state-of-the-art long baseline neutrino experiments which are designed to measure the yet unknown neutrino parameters with special focus on the Dirac CP-violating phase of the PMNS matrix.
\begin{itemize}
\item \textbf{DUNE:}  The source of the DUNE experiment will be at the Fermilab and the detector will be located at Sanford Underground Research Facility (SURF) at Homestake mine in South Dakota \cite{Acciarri:2015uup}. The baseline will be 1300 km. The far detector will be a 40 kton liquid Argon detector sitting on axis with the beam so the spectrum will be broad band. The energy of the neutrino beam will be around 3 GeV which comes from an 80 GeV proton beam with $1.47\times 10^{21}$ POT per year. A reasonable assumption for data taking is 3.5 years in each neutrino and antineutrinos modes.
    \item \textbf{T2HK:} The source of T2HK \cite{Abe:2015zbg} will be upgraded 30 GeV JPARC beam with $2.7 \times 10^{21}$ POT per year. The Hyper-Kamiokande detector with fiducial volume of 0.56 Mton  \cite{Kearns:2013lea} (25 times that of Super-Kamiokande) will be located in Kamiokande, $2.5^\circ$ off axis so the spectrum  will be narrow band. The energy of neutrinos will be around 0.6 GeV. The baseline of this experiment is 295 km.  A reasonable assumption for data taking is the 2TankHK-staged configuration \footnote{KEK Preprint 2016-21 and ICRR-Report-701-2016-1, https://lib-extopc.kek.jp/preprints/PDF/2016/1627/1627021.pdf} for which the data taking time is 6 years for one tank plus another 4 years with second tank \cite{Liao:2016orc}. The ratio of running time in neutrino mode to that in the antineutrino mode is 1:3.
        \item \textbf{T2HKK:} This project is an extension of T2HK \cite{Abe:2016ero} with an extra detector in Korea with a baseline of 1100 km.  Two options with $2.5^\circ$ off-axis-angle and $1.5^\circ$ off-axis-angle have been discussed which respectively correspond to neutrino energies of 0.6 GeV and 0.8 GeV.
\end{itemize}
Notice that at T2HK, both energy of the neutrino beam and the baseline are lower than those at DUNE. We therefore expect the DUNE experiment to be more sensitive to both standard and non-standard matter effects than T2HK and T2HKK. Although the baseline for the Korean detector of T2HKK is comparable to the DUNE baseline, the DUNE experiment will be more sensitive to matter effects than T2HKK, because  the energy of the beam at T2HKK is lower. Detailed simulation confirms this expectation \cite{Liao:2016orc}.
In the presence of NSI, new degeneracies will appear in long baseline neutrino experiments for determination of the value of $\delta$, mass ordering and the octant of $\theta_{23}$. One of the famous degeneracies is the so called generalized mass ordering degeneracy
\cite{Agarwalla:2014bsa,Gonzalez-Garcia:2013usa,Coloma:2016gei,Bakhti:2014pva,Liao:2016orc}.  The oscillation probability remains invariant under the following simultaneous transformations
\be \label{trans} \theta_{12} \to \frac{\pi}{2}-\theta_{12}, \ \delta \to \pi -\delta,  \ \Delta m_{31}^2 \to -\Delta m_{31}^2+\Delta m_{21}^2, \ {\rm and} \ V_{eff} \to -S \cdot V_{eff}^* \cdot S \ee
where $S={\rm Diag}(1,-1,-1)$ and $(V_{eff})_{\alpha\beta}=\sqrt{2} G_F N_e [(\delta_{\alpha 1} \delta_{\beta 1})+\epsilon_{\alpha\beta}]$ in which  $\epsilon_{\alpha \beta}=\sum_{f \in \{ e,u,d \}} (N_f/N_e)\epsilon^f_{\alpha \beta}$ depends on the composition of medium. Notice that the LMA-Dark solution with $\theta_{12}>\pi/4$ and $\epsilon^f\sim -1$ \cite{Agarwalla:2014bsa} is related to the generalized mass ordering transformation from the standard LMA solution with $\epsilon=0$.
For  the Earth (with $N_p\simeq N_n$), we can write $N_u/N_e \simeq N_d/N_e=3$. Notice however that the  transformation  in Eq. (\ref{trans}) does not depend on the beam energy or baseline.
As  a result, by carrying out long baseline neutrino experiments on Earth with different baseline and beam energy configurations, this degeneracy cannot be resolved. Resolving this degeneracy requires media with different $N_n/N_e$ composition. Notice that although the nuclear compositions of the Earth core and mantle are quite different, $N_n/N_e$ is uniformly close to 1 across the Earth radius \cite{Lisi:1997yc}.
  However, $N_n/N_e$ in the Sun considerably differs from that in the Earth. Moreover, it varies from the Sun center (with $N_n/N_e \simeq 1/2$) to its outer region (with $N_n/N_e \simeq 1/6$) \cite{Vinyoles:2016djt,Serenelli:2009yc}.
As a result, the solar neutrino data can in principle help to solve this degeneracy. In fact, \cite{Gonzalez-Garcia:2013usa} by analyzing solar  data shows that the LMA solution with $\epsilon_{ee}^u \simeq 0.3$ is slightly favored over the LMA-Dark solution.
The global analysis of solar, atmospheric and (very) long baseline data can in principle help to solve degeneracies.
For the time being, however, since the terrestrial experiments are not precise enough to resolve the effects of ${\rm sign}(\Delta m_{31}^2)$ and/or ${\rm sign}(\cos 2\theta_{12})$, the generalized mass ordering degeneracy cannot be resolved.

At relatively low energy long baseline experiments such as T2HK and T2HKK for which the contribution to the oscillation probability from higher orders of $O(V_{eff} \epsilon/|\Delta m_{31}^2|)$ can be neglected,  the appearance oscillation probability along the direction $\epsilon_{e \mu}/\epsilon_{e\tau}=\tan \theta_{23}$ will be equal to that for standard $\epsilon=0$ \cite{Liao:2016orc}. The DUNE experiment being sensitive to higher orders of $ (V_{eff} \epsilon/|\Delta m_{31}^2|)$ can solve this degeneracy  \cite{Liao:2016orc}.
At the DUNE experiment, another degeneracy appears when $\epsilon_{ee}$ and $\epsilon_{\tau e}$  are simultaneously turned on and the phase of $\epsilon_{e \tau}$ is allowed to be nonzero. As shown in Fig. \ref{4Coloma}, (which corresponds to Fig 4 of \cite{Coloma:2015kiu})  and confirmed in Fig 10 of \cite{Liao:2016orc}, in the  presence of cancelation due to the phase of $\epsilon_{e\tau}$ for $|\epsilon_{e\tau}|\sim 0.2- 0.3$, $|\epsilon_{ee}|$ as large as 2 cannot be disentangled from standard case with $\epsilon_{ee}=\epsilon_{e\tau}=0$ at the DUNE experiment. However, this figure also demonstrates that when information on $\epsilon$ from already existent data is used as prior, the degeneracy can be considerably solved, ruling out the $\epsilon_{ee}<0$ wing of solutions. That is because solar data rules out $\epsilon_{ee}<0$ for $\theta_{12}< \pi/4$ \cite{Gonzalez-Garcia:2013usa}. NSI can induce degeneracies in deriving ${\rm sign}(\cos 2 \theta_{23})$.
In principle, even with $\epsilon_{e\mu}$ ($\epsilon_{e\tau}$) as small as $O(0.01)$, the degeneracy due to  the phase of $\epsilon_{e\mu}$ ($\epsilon_{e\tau}$) makes the determination of the octant of $\theta_{23}$ problematic \cite{Agarwalla:2016fkh}. Because of the generalized mass ordering degeneracy, the presence of NSI can also jeopardize determination of  ${\rm sign} (\Delta m_{31}^2 )$ \cite{Deepthi:2016erc}.

 In summary, NSI induces degeneracies that makes determination of the true value of $\delta$ at DUNE impossible at 3 $\sigma$ C.L. The T2HKK experiment can considerably solve this degeneracy as demonstrated in Fig \ref{unknown} and in Fig \ref{known} (corresponding to Fig 11 and 12 of \cite{Liao:2016orc}).
  \begin{figure}
\begin{center}
\subfigure[]{\includegraphics[width=0.49\textwidth]{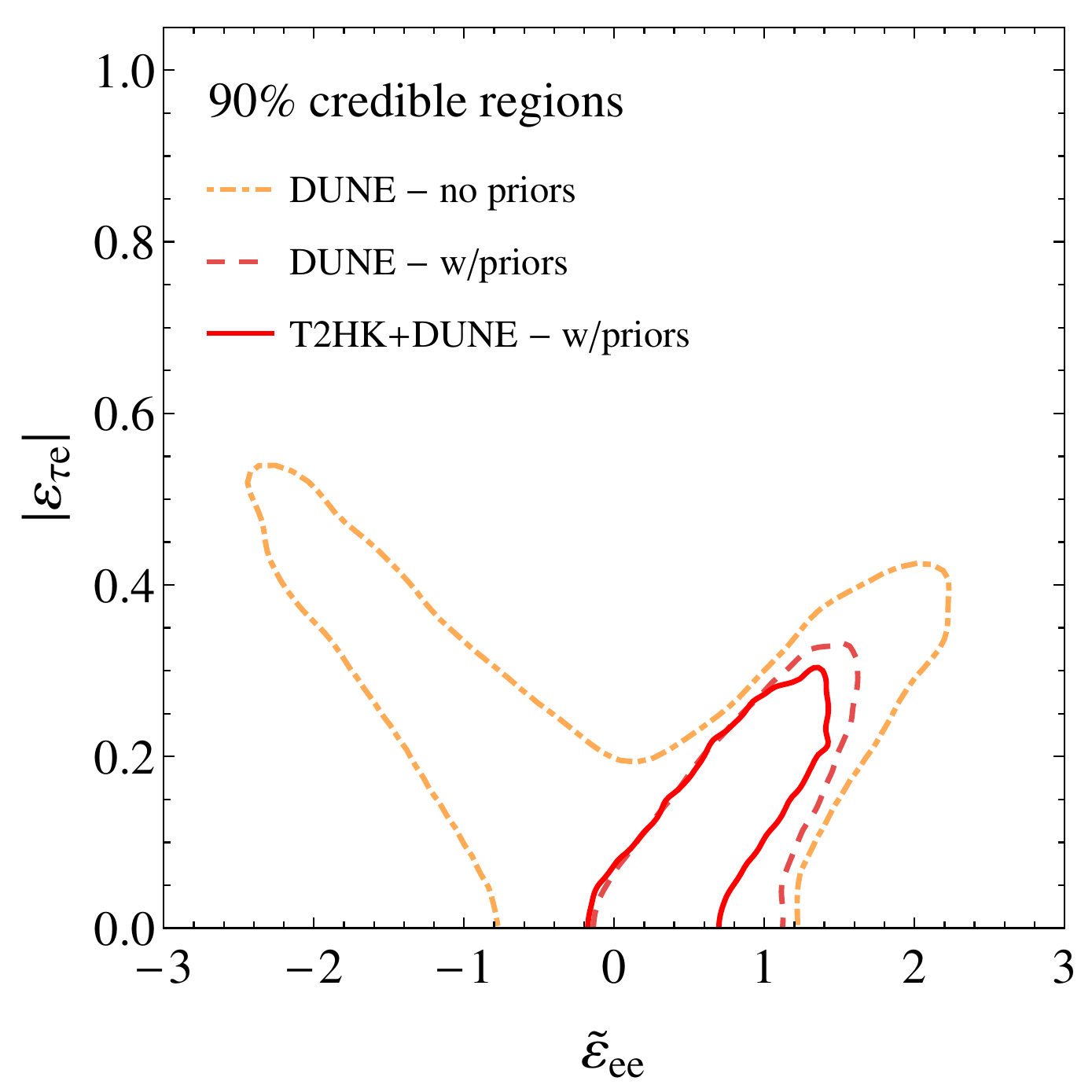}}
\end{center}
\vspace{0.5 cm} \caption[]{90 \% CL contours in the $|\epsilon_{\tau e}|$ and $\tilde{\epsilon}_{ee}=\epsilon_{ee}-\epsilon_{\mu\mu}$ plane for DUNE and DUNE+T2HK. (See Eq. (\ref{combination}) for the definition of $\epsilon$.) The phase of $\epsilon_{\tau e}$ is allowed to vary. Contours with priors take into account the present bounds from various neutrino oscillation experiments. This plot is taken from \cite{Coloma:2015kiu}, published under the terms of the Creative Commons Attribution Noncommercial License and therefore no copyright permissions were required for its inclusion in this manuscript.. ({We would like to thank P. Coloma for sending the original figure.})}
\label{4Coloma}
\end{figure}
%%%%%%%%%%%%%%%%%%%%%%%%%%%%%%%%%%%%%%%%%%%%%%%%%%
%%%%%%%%%%%%%%%%%%%%%%%%%%%%%%%%%%%%%%%%%%%%%%%%%%%%
  %\begin{figure}
%\begin{center}
%\subfigure[]{\includegraphics[width=0.49\textwidth]{epsee-epset-3eps}}
%\end{center}
%\vspace{2cm} \caption[]{Dependence of the projected MOMENT sensitivity to  $\delta-\theta_{23}$ on background  Suppression Factor (SF). Thick and thin lines respectively show SF=0.1 \% and 10\%. In %Fig a, standard oscillation with no NSI is assumed. In Fig b, the true values of $\epsilon$ are set to zero and uncertainties of $\epsilon$ shown in Eq. (\ref{range1},\ref{range2})   are treated by %the pull method. Plots are taken from \cite{Bakhti:2016prn}.}
%\label{SF}
%\end{figure}
%%%%%%%%%%%%%%%%%%%%%%%%%%%%%%%%%%%%%%%%%%%%%%%%%
%%%%%%%%%%%%%%%%%%%%%%%%%%%%
 \begin{figure}
\begin{center}
\subfigure[]{\includegraphics[width=0.49\textwidth]{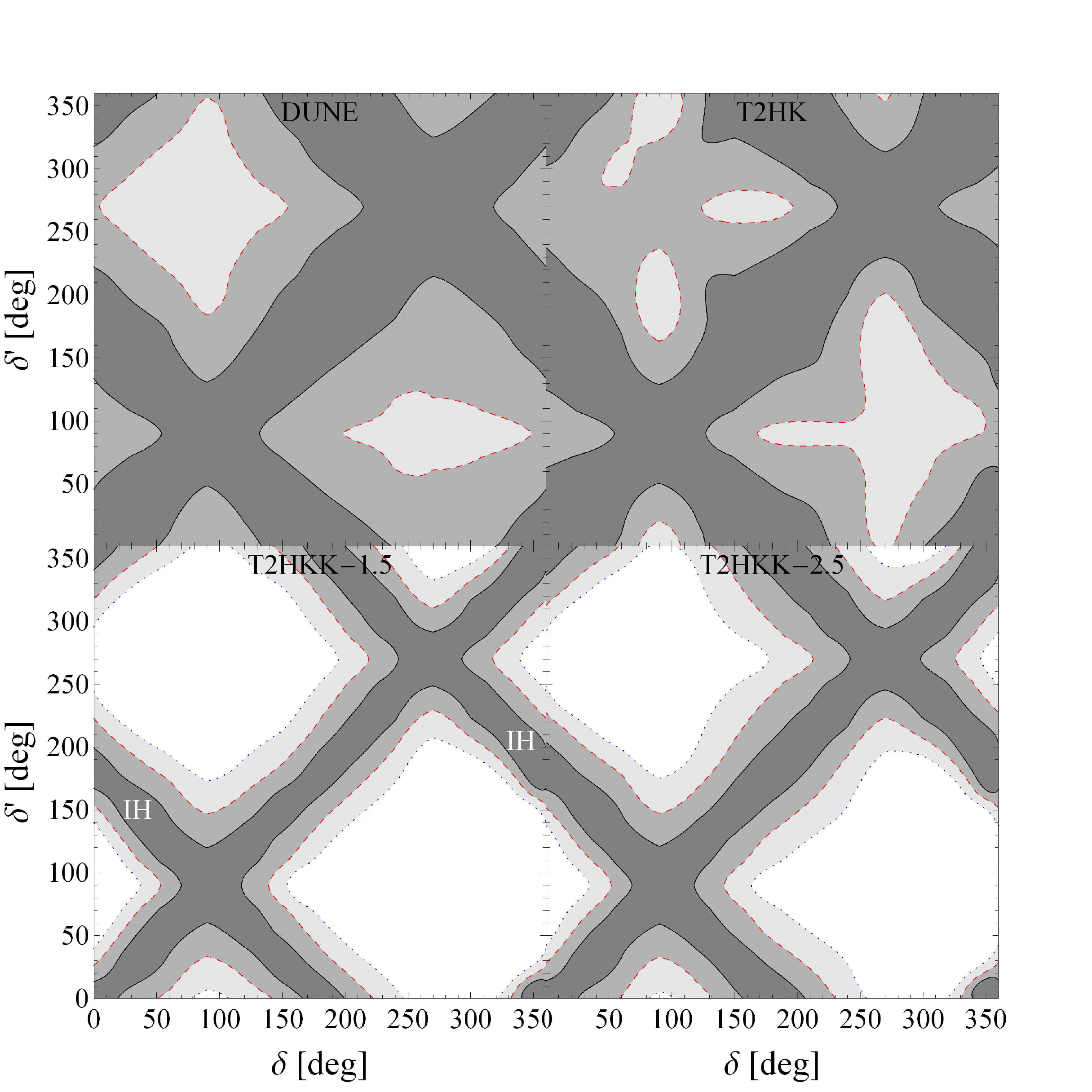}}
\end{center}
\vspace{0.5 cm} \caption[]{ $1 \sigma$, $2 \sigma$ and $3\sigma$ contours for extracted value of CP-violating phase $\delta^\prime$ at DUNE, T2HK, T2HKK with $1.5^\circ$ off-axis angle and T2HKK with $2.5^\circ$ off-axis angle versus true value of $\delta$. Normal mass ordering is assumed and taken to be unknown. The values of $\epsilon_{ee}$, $\epsilon_{e\mu}$ and $\epsilon_{e\tau}$ are allowed to vary. The parameters that are not shown are marginalized. This plot is taken from \cite{Liao:2016orc}, published under the terms of the Creative Commons Attribution Noncommercial License and therefore no copyright permissions were required for its inclusion in this manuscript. ({We would like to thank D. Marfatia for sending the original figure.})}
\label{unknown}
\end{figure}
%%%%%%%%%%%%%%%%%%%%%%%%%%%%%%%%%%%%%%
%%%%%%%%%%%%%%%%%%%%%%%%%%%%%%%%%%%%%%%
 \begin{figure}
\begin{center}
\subfigure[]{\includegraphics[width=0.49\textwidth]{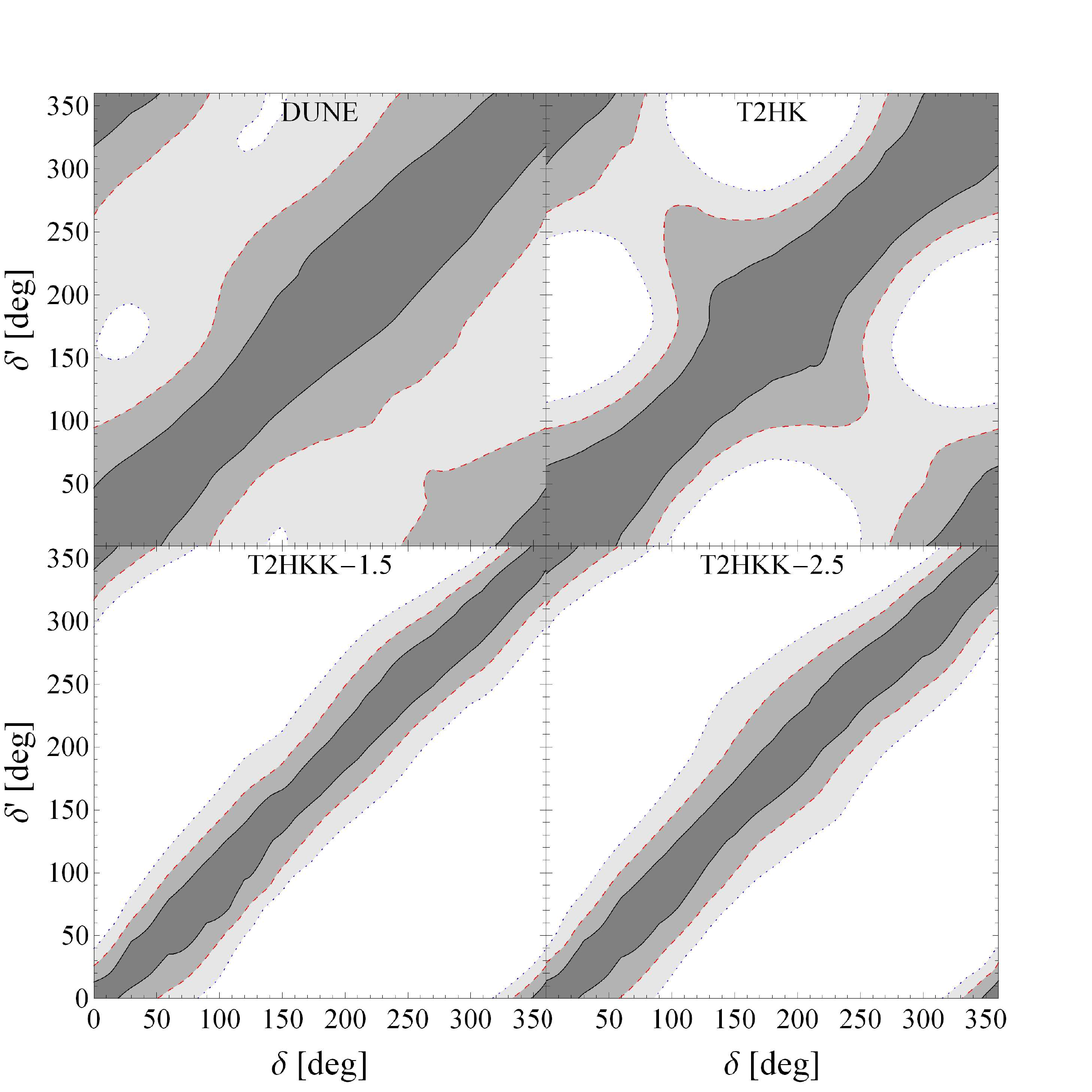}}
\end{center}
\vspace{0.5 cm} \caption[]{The same as Fig \ref{unknown} except that the mass ordering is assumed to be known. This plot is taken from \cite{Liao:2016orc}, published under the terms of the Creative Commons Attribution Noncommercial License and therefore no copyright permissions were required for its inclusion in this manuscript. (We would like to thank D. Marfatia for sending the original figure.)}
\label{known}
\end{figure}

\subsection{ JUNO and RENO-50 shedding light on LMA-Dark\label{NSI-Dark}}
To determine the sign of $\Delta m_{31}^2$, two reactor neutrino experiments with  baseline of $\sim 50$ km are proposed: The JUNO experiment in China which is planned to start data taking in 2020 and RENO-50 which is going to be an upgrade of the RENO experiment in South Korea\footnote{Ref. \cite{Joo:2017rmi} reports the current status of RENO-50.}. In this section, we show that for known mass ordering, these experiments can determine the octant of $\theta_{12}$.  At reactor experiments, since the energy is low, $|\Delta m_{31}^2|/E \gg \sqrt{2} G_F N_e$. Thus, the matter effects can be neglected and the survival probability can be written as
\begin{eqnarray}
\label{sur-prob}P(\bar{\nu}_e\to \bar{\nu}_e) & = &\left| |U_{e1}|^2+
|U_{e2}|^2e^{i \Delta_{21}}+|U_{e3}|^2e^{i
\Delta_{31}}\right|^2=\left|c_{12}^2 c_{13}^2+s_{12}^2 c_{13}^2e^{i
\Delta_{21}} +s_{13}^2 e^{i \Delta_{31}}\right|^2=\\
\nonumber
  & = & c_{13}^4\left(1-\sin^2 2 \theta_{12}\sin^2
\frac{\Delta_{21}}{2}\right)+s_{13}^4+2s_{13}^2 c_{13}^2\left[\cos
\Delta_{31}(c_{12}^2+s_{12}^2 \cos \Delta_{21})+s_{12}^2 \sin
\Delta_{31}\sin\Delta_{21}\right]\, ,
  \end{eqnarray}
where $\Delta_{ij}=\Delta m_{ij}^2 L/(2E_\nu)$ in which $L$ is the baseline. Notice that the first parenthesis  (which could be resolved at KamLAND) is only sensitive to $\sin^2 2\theta_{12}$ and cannot therefore
resolve the octant of $\theta_{12}$.  The  terms in the last parenthesis, however, are sensitive to the octant of $\theta_{12}$. To solve these terms two main challenges have to be overcome: (i)
These terms are suppressed by $s_{13}^2\sim 0.02$ so high statistics is required in order to resolve them. (ii) In the limit, $\Delta_{12} \to 0$, we can write $P(\bar{\nu}_e \to \bar{\nu}_e)= c_{13}^4+ s_{13}^4 +2 s_{13}^2 c_{13}^2 \cos \Delta_{31}$ so the sensitivity to $\theta_{12}$ is lost. To determine $\theta_{12}$ baseline should be large enough ({\it i.e.,} $L \gtrsim 10~{\rm  km}$). (iii) Condition $\Delta_{12} \gtrsim 1$ naturally implies $\Delta_{13} \gg 1$ so the terms sensitive to the octant of $\theta_{12}$ (and sign of $\Delta m_{31}^2$) oscillate rapidly.  To resolve these terms, the energy resolution and accuracy of reconstruction of the total energy scale must be high.  Notice that reactor experiments such as Daya Bay satisfy the first condition and resolve  the terms proportional to $s_{13}^2$, but cannot overcome the second challenge because at these experiments, $\Delta_{12}\ll 1$. At KamLAND, $\Delta_{12}>1$ but the statistics was too low to resolve the $s_{13}^2$ terms. JUNO and RENO-50, being designed to be sensitive to these terms to determine ${\rm sign}(\Delta m_{31}^2)$, can overcome all these three challenges.
The detectors at JUNO and RENO-50 experiments will employ liquid scintillator technique with an impressive energy resolution of $$\frac{\delta E_\nu}{E_\nu}\simeq 3 \%\times \left(\frac{E_\nu}{\rm MeV}\right)^{1/2}. $$ Moreover, the energy calibration error can be as low as  3 \%.
Using the GLoBES software \cite{Huber:2004ka,Huber:2007ji}, Ref \cite{Bakhti:2014pva} shows how JUNO and RENO-50 experiments can test LMA-Dark solution with $\theta_{12}>\frac{\pi}{4}$. Results are shown in Fig. \ref{Contour-bright} and Fig. \ref{Contour-dark}. The star denotes the true value of $\Delta m_{31}^2$ and $\theta_{12}$.  In Fig. \ref{Contour-bright} and Fig. \ref{Contour-dark}, normal and inverted mass orderings are respectively assumed. Ellipses show $3 \sigma$ C.L. contours, after 5 years of data taking. As seen from these figures, these upcoming experiments will be able to determine $|\Delta m_{31}^2|$ with much better accuracy than the present global data analysis so no prior on $|\Delta m_{31}^2|$ is assumed. The uncertainties of other relevant neutrino parameters are taken from \cite{GonzalezGarcia:2012sz} and are treated by pull-method.

\begin{figure}
\begin{center}
\subfigure[]{\includegraphics[width=0.49\textwidth]{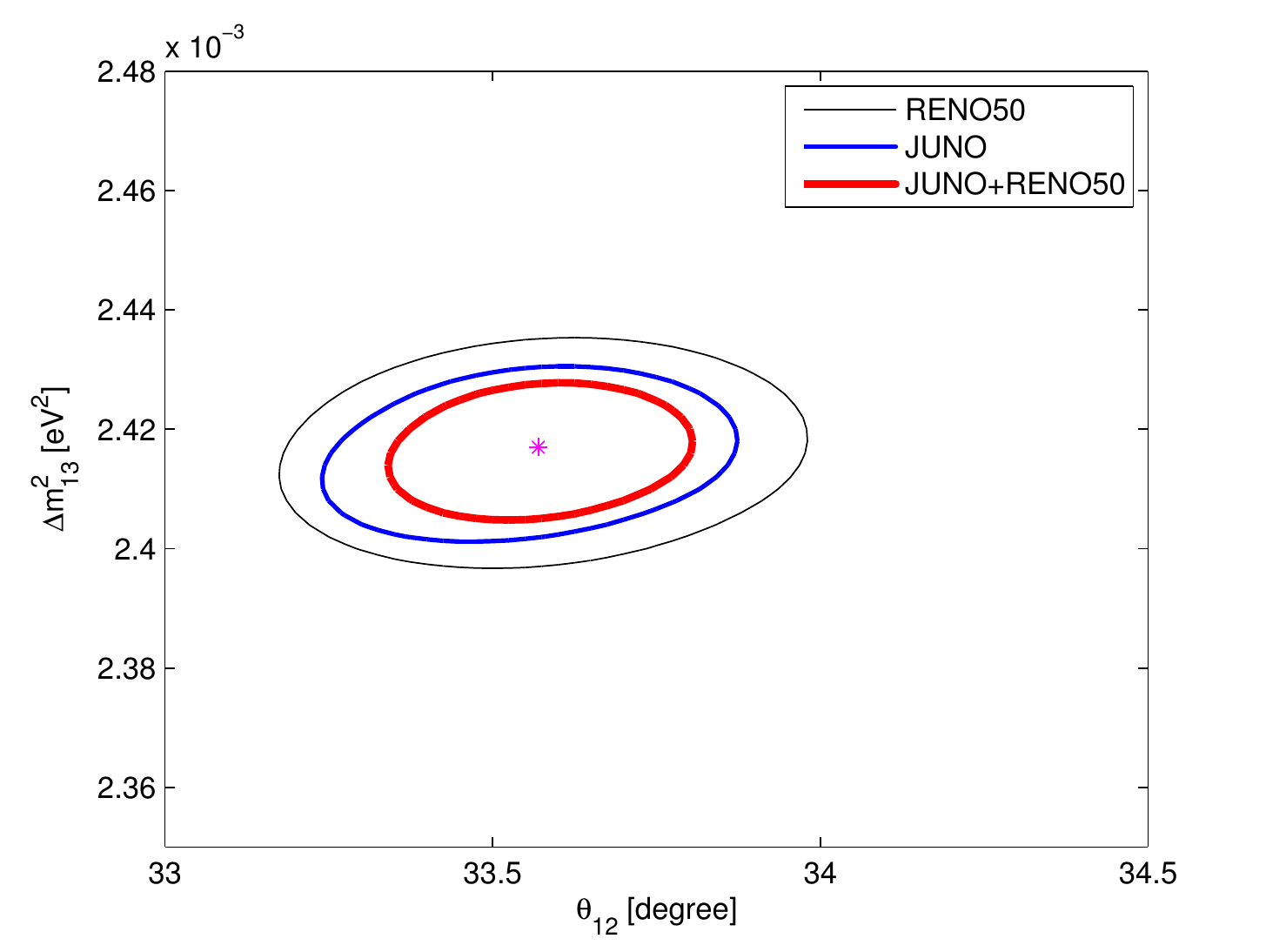}}
\subfigure[]{\includegraphics[width=0.49\textwidth]{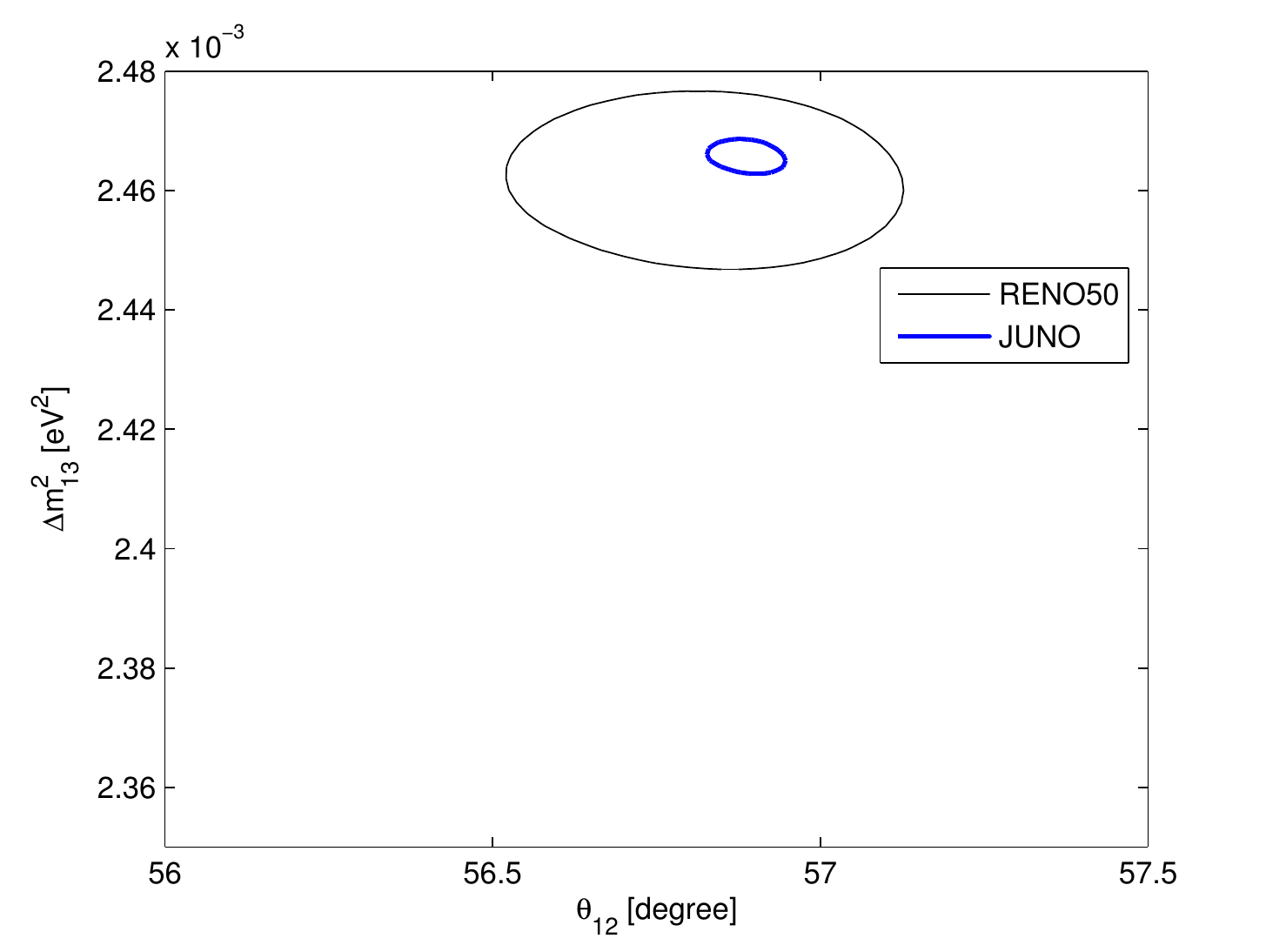}}
\subfigure[]{\includegraphics[width=0.49\textwidth]{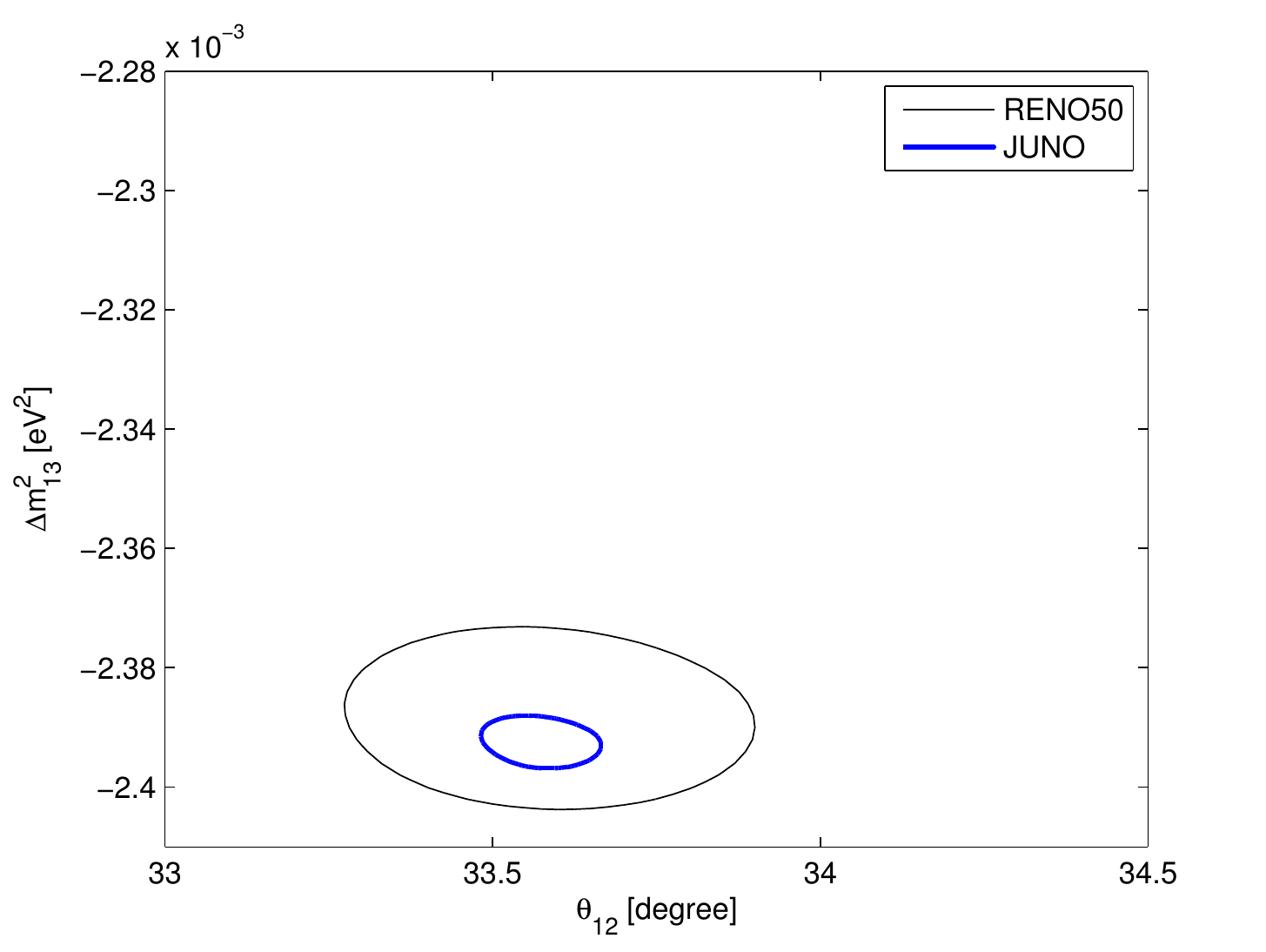}}
\subfigure[]{\includegraphics[width=0.49\textwidth]{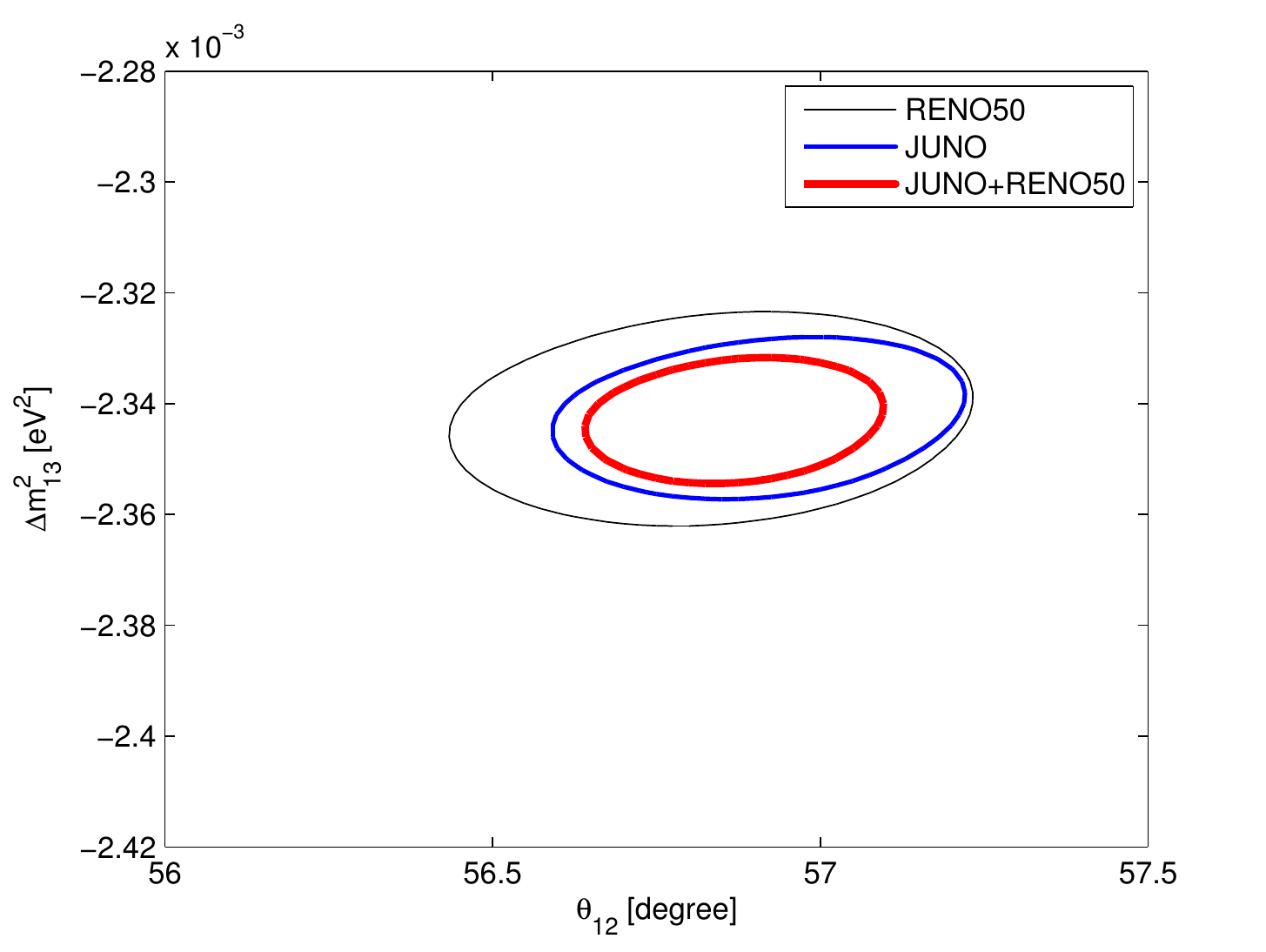}}
\end{center}
\vspace{0.5 cm} \caption[]{The 3 $\sigma$ C.L. contours for RENO-50 and JUNO after 5
years of data taking. The true values of the
neutrino parameters, marked with a star in Fig. (a), are taken to be
$\Delta m_{31}^2=2.417\times10^{-3}~{\rm eV}^2$,
$\theta_{12}=33.57^\circ$, $\Delta m_{21}^2=(7.45\pm
0.45)\times10^{-5}~{\rm eV}^2$ and $\theta_{13}=(8.75\pm
0.5)^\circ$.  Plots are taken from \cite{Bakhti:2014pva}, published under the terms of the Creative Commons Attribution Noncommercial License and therefore no copyright permissions were required for their inclusion in this manuscript.}
\label{Contour-bright}
\end{figure}
%%%%%%%%%%%%%%%%%%%%%%%%%%%%%%
%%%%%%%%%%%%%%%%%%%%%%%%%%%%%%%%%%%%
%%%%%%%%%%%%%%%%%%%%%%%%%%%%%%%%%%%%%%
\begin{figure}
\begin{center}
\subfigure[]{\includegraphics[width=0.49\textwidth]{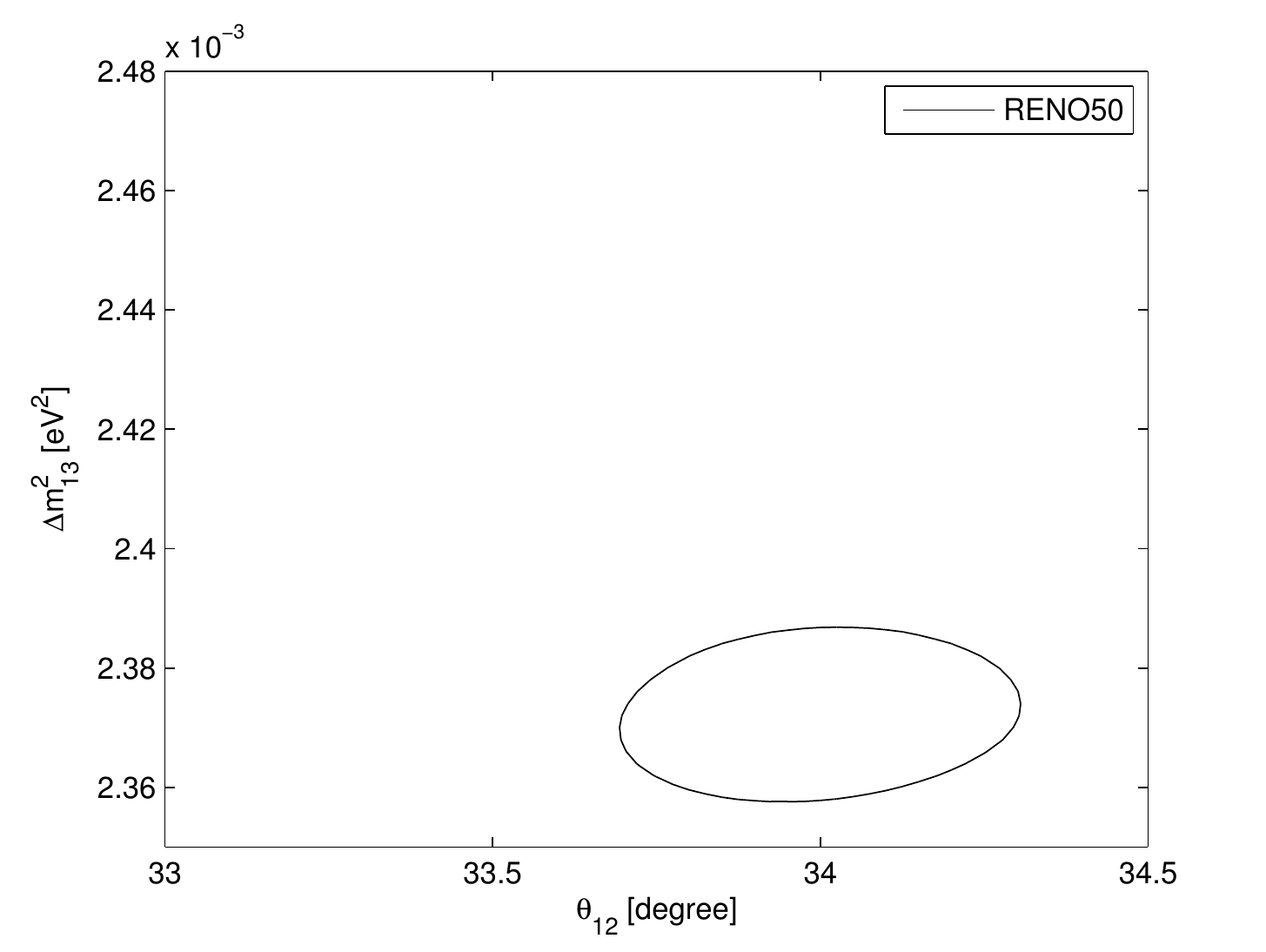}}
\subfigure[]{\includegraphics[width=0.49\textwidth]{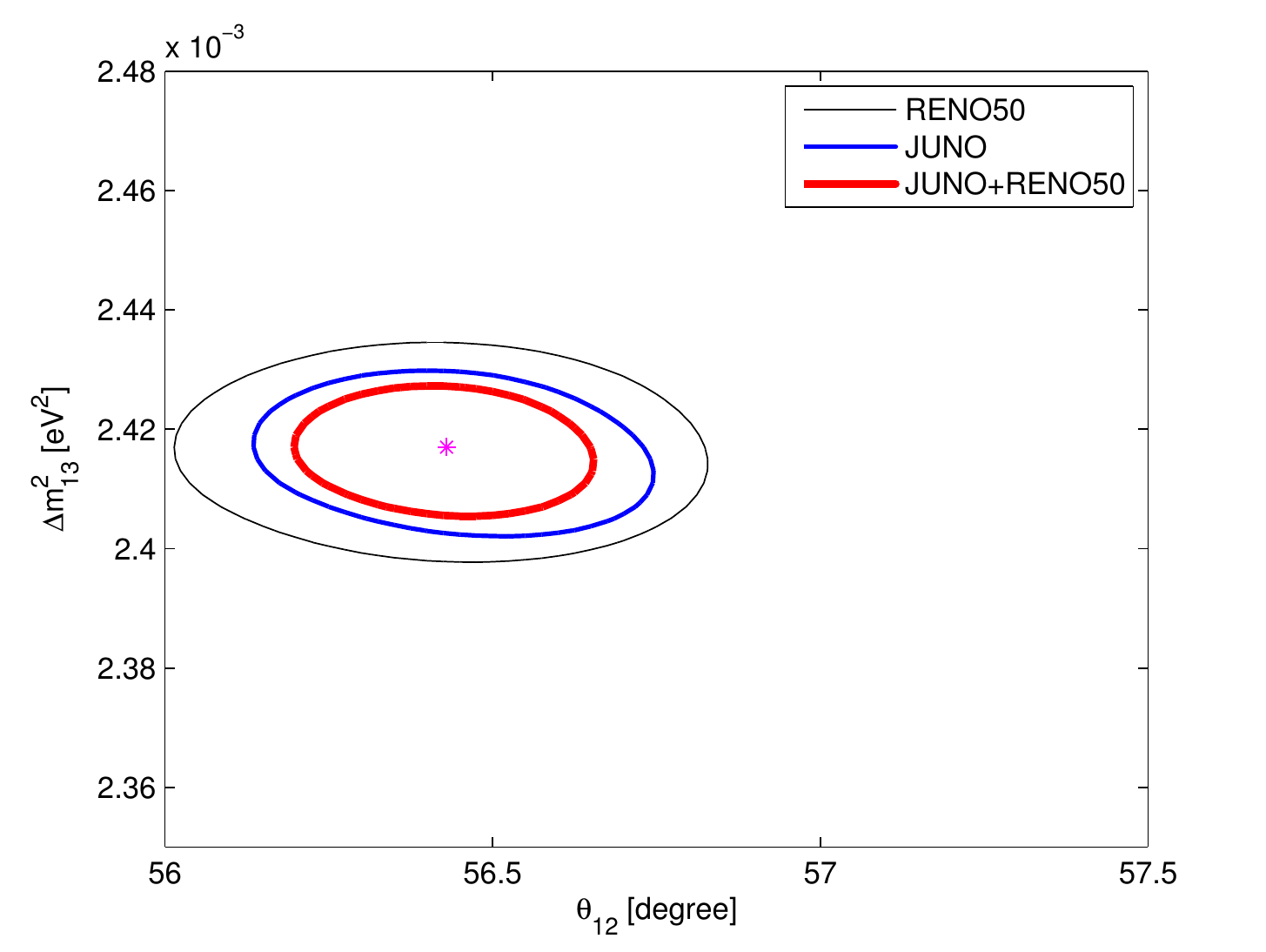}}
\subfigure[]{\includegraphics[width=0.49\textwidth]{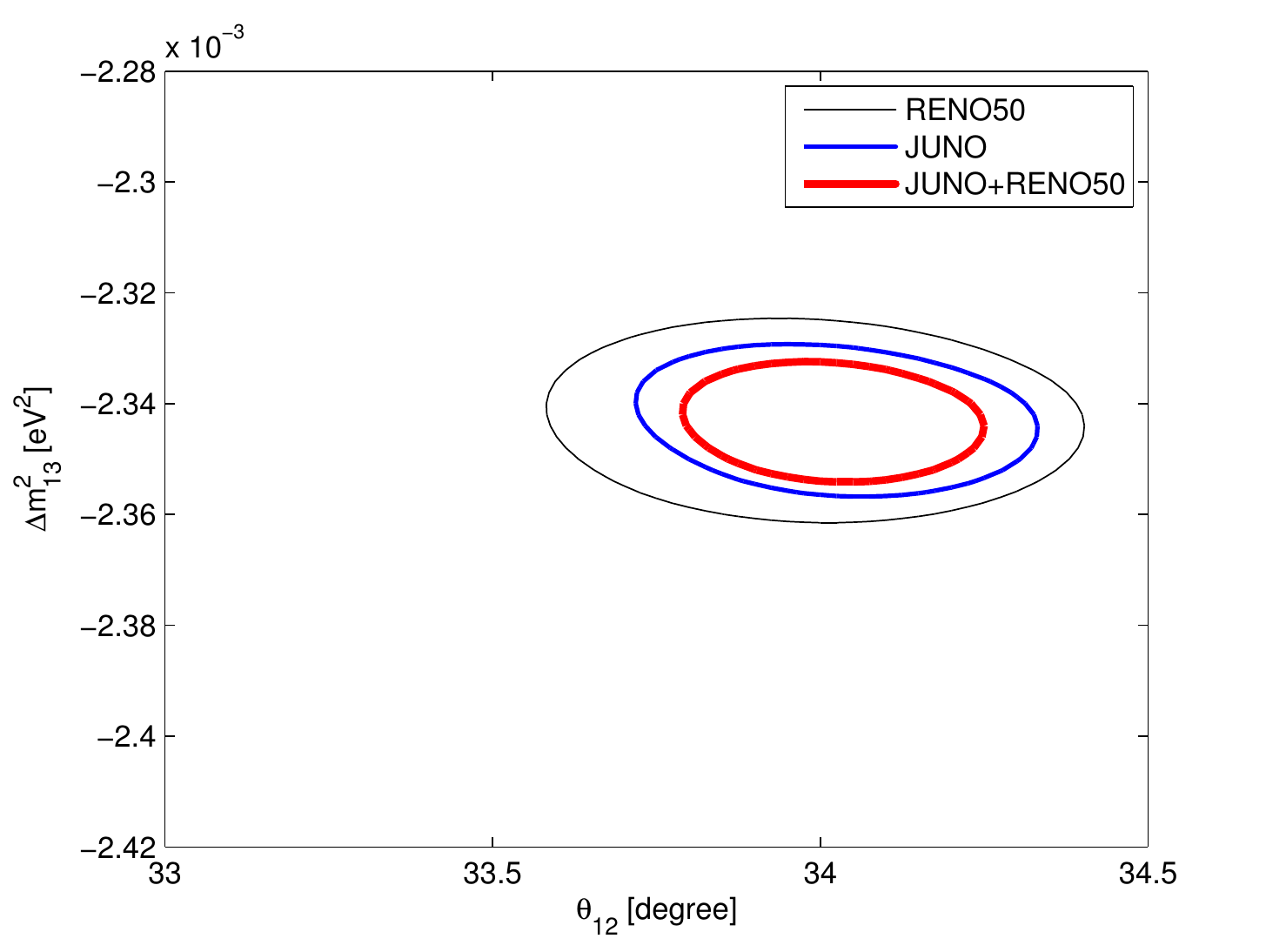}}
\subfigure[]{\includegraphics[width=0.49\textwidth]{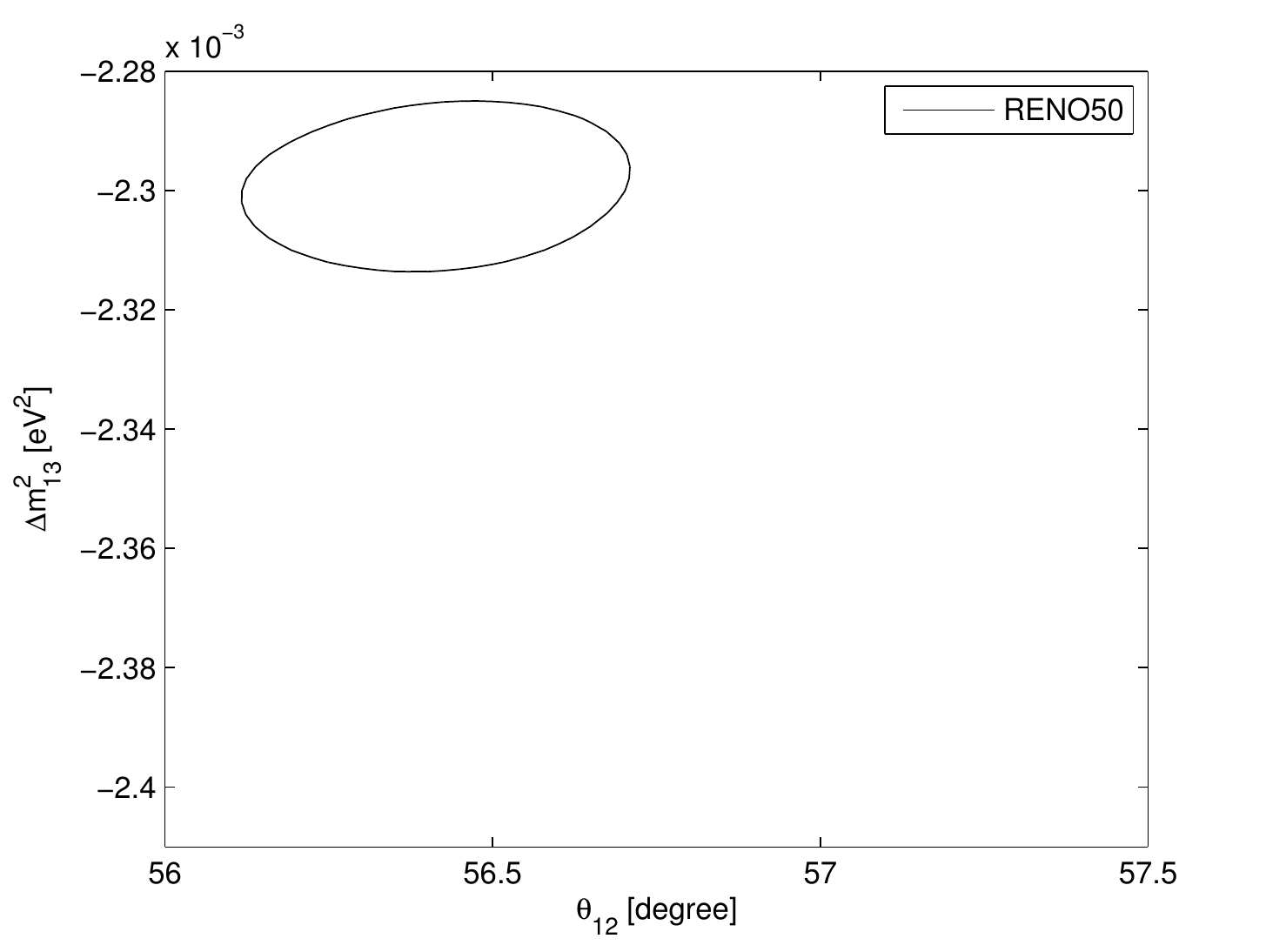}}
\end{center}
\vspace{0.5 cm} \caption[]{The same as Fig. \ref{Contour-bright} except
that the true values are taken to be $\Delta
m_{31}^2=2.417\times10^{-3}~{\rm eV}^2$ and
$\theta_{12}=56.43^\circ$. In other words,  the LMA-dark
solution is assumed to be true. Plots are taken from \cite{Bakhti:2014pva}, published under the terms of the Creative Commons Attribution Noncommercial License and therefore no copyright permissions were required for their inclusion in this manuscript.}

\label{Contour-dark}

\end{figure}

 For JUNO experiment, the uncertainties in the flux normalization and the initial energy spectrum at the source are taken respectively equal to $5 $ \% and 3 \%.
RENO-50 enjoys having a near detector (the detectors of present RENO) which can measure the flux with down to O$(0.3 \%)$ uncertainty. To perform the analysis, the energy range of 1.8 MeV to 8 MeV is divided to 350 bins of 17.7 keV size. The pull-method is applied by defining
\be \chi^2={\rm Min}|_{\theta_{pull}, \alpha_i}\left[ \sum_i \frac{[N_i(\theta_0,\bar{\theta}_{pull})-N_i(\theta,\theta_{pull})(1+\alpha_i)]^2}{N_i (\theta_0,\bar{\theta}_{pull})}+\sum_i \frac{\alpha_i^2}{(\Delta \alpha_i)^2} +\frac{(\theta_{pull}-\bar{\theta}_{pull})^2}{(\Delta \theta_{pull})^2}\right],\ee
where $N_i$ is the number of events at bin $i$. $\alpha_i$ is the pull parameter that accounts for the uncertainty in the initial spectrum at bin $i$. Pull parameters taking care of the other uncertainties are collectively denoted by $\theta_{pull}$.

As seen from these figures, JUNO and RENO-50 can determine the octant of $\theta_{12}$ for a given mass ordering. This result is relatively robust against varying the calibration error but as expected, is extremely sensitive to the energy resolution.  Increasing the uncertainty in energy resolution from $3 \%$ to $3.5 \%$, Ref. \cite{Bakhti:2014pva} finds that JUNO and RENO-50 cannot determine the octant at 3 $\sigma$ C.L. after five years. As seen from the figures, JUNO and RENO-50 experiments cannot distinguish two solutions which are related to each other with $\theta_{12} \leftrightarrow \pi/2 - \theta_{12}$ and $\Delta m_{31}^2 \to -\Delta m_{31}^2+\Delta m_{21}^2$ which stems from the generalized mass ordering degeneracy that we discussed in  subsect. \ref{NSI-up}.
\subsection{NSI at the MOMENT \label{Moment}}
The MOMENT experiment is a setup which has been proposed to measure the value of CP-violating phase, $\delta$ \cite{Cao:2014bea,Blennow:2015cmn}. MOMENT stands for MuOn-decay MEdium baseline NeuTrino beam. This experiment will be located in China. The neutrino beam in this experiment is provided by the  muon decay.  Beam can switch between muon decay
($\mu \to e \bar{\nu}_e \nu_\mu$) and antimuon decay ($\bar{\mu} \to e^+ \nu_e \bar{\nu}_\mu$).
The energies of neutrinos will be relatively low with a maximum energy at 700 MeV and peak energy at 150 MeV. The detector is going to be Gd doped water Cherenkov with fiducial mass of 500 kton, located at a distance of 150 km from the source.
The detection modes are
$$\nu_e +n \to p+e^- \ \ \ \ \ \bar{\nu}_\mu +n \to p+\mu^+$$
and
$$\bar{\nu}_e +n \to p+e^+ \ \ \ \ \ {\nu}_\mu +n \to p+\mu^-.$$
Gd at the detector can capture the final neutron so although the detector lacks magnetic field, it can distinguish between neutrino and antineutrino  with Charge Identification (CI) of 80 \% \cite{Huber:2008yx}.
The MOMENT experiment, with a baseline of 150 km and relatively low energy is not very sensitive to matter effects so it enjoys an ideal setup to determine $\delta$ and octant of $\theta_{23}$
without ambiguity induced by degeneracies with NSI. The potential of this experiment for determining $\delta$ and the octant of $\theta_{23}$ is studied in \cite{Bakhti:2016prn}  using GLoBES \cite{Huber:2004ka,Huber:2007ji}. The unoscillated flux of each neutrino mode is taken to be $4.7 \times 10^{11} {\rm m}^{-2} {\rm year}^{-1}$ and 5 years  of data taking in each muon and antimuon modes is assumed. Uncertainties in flux normalization of $\bar{\nu}_e$
and $\nu_\mu$ modes are taken to be correlated and equal to 5 \% but the uncertainties of fluxes
from muon  and antimuon modes are uncorrelated.

One of the main sources of background is atmospheric neutrinos. Since  the neutrino beam at the MOMENT experiment will
be sent in bunches, this source of background can be dramatically reduced. Reduction of background is parameterized by Suppression
Factor (SF). Results of Ref \cite{Bakhti:2016prn} are shown in Fig. \ref{MOMENT}. The assumed true value of $\delta$ and $\theta_{23}$ are shown with a star.
The mass ordering is taken to be normal and assumed to be known. All the appearance and disappearance modes are taken into account. In all these figures, the true
values of $\epsilon$ are taken to be zero. In Figs \ref{MOMENT}-b to \ref{MOMENT}-d, pull method  is applied on $\epsilon= \sum_f (N^f/N_e)\epsilon^f$, taking $1\sigma$ uncertainties on $\epsilon$ as follows \cite{Gonzalez-Garcia:2013usa}
\be |\epsilon_{e\mu}|<0.16, \ |\epsilon_{e \tau}|<0.26 \ {\rm and} \ |\epsilon_{\mu \tau}|<0.02 \label{range1}\ee
and
\be -0.018<\epsilon_{\tau \tau}-\epsilon_{\mu \mu} <0.054 \ \ {\rm and} \ \ 0.35 <\epsilon_{ee}-\epsilon_{\mu \mu} <0.93 . \label{range2}\ee
Results shown in Fig \ref{MOMENT}-c and \ref{MOMENT}-d  assume that T2K (NO$\nu$A) takes data in neutrino mode for 2 (3) years
and in antineutrino mode for 6 (3) years. For more details on the assumptions, see Ref \cite{Bakhti:2016prn}. As seen from Fig \ref{MOMENT}-b,
turning on NSI, NO$\nu$A and T2K (even combined) cannot establish CP-violation even at 1 $\sigma$ C.L.: while the true value of
$\delta$ is taken to be $270^\circ$ (maximal CP-violation), $\delta=0,360^\circ$ (CP-conserving) is within the 1 $\sigma$ C.L.
contour. At $3 \sigma$ C.L., these experiments cannot determine the octant of $\theta_{23}$. Moreover, in the presence of NSI, these experiments
cannot rule out the wrong octant even at 1 $\sigma$ C.L. But, comparing Fig a and b, we observe that turning on NSI within range
(\ref{range1},\ref{range2}) does not considerably  reduce the power of MOMENT to  measure the CP-violating phase and rule out the wrong
octant solution. In this figure, SF is taken to be 0.1 \% which is rather an optimistic assumption.  Fig \ref{SF} shows that increasing SF up to
10 \%, the power of octant determination is significantly reduced but the determination of $\delta$ is not dramatically affected.

Similar result holds valid when instead of normal mass ordering, inverted mass ordering is taken (and again assumed that the ordering is known) \cite{Bakhti:2016prn}.
Ref \cite{Blennow:2015cmn} shows that the MOMENT experiment itself can determine the mass ordering. According to \cite{Bakhti:2016prn}, as long as $\epsilon$ can vary in the range shown in Eqs.
(\ref{range1},\ref{range2}), MOMENT maintains its power to determine the mass ordering.
Of course, once we allow $\epsilon$ to vary in a wide range such that transformation in Eq. (\ref{trans}) can be made, the power of mass ordering determination is  lost due to the generalized mass ordering degeneracy.
 \begin{figure}
\begin{center}
\subfigure[]{\includegraphics[width=0.49\textwidth]{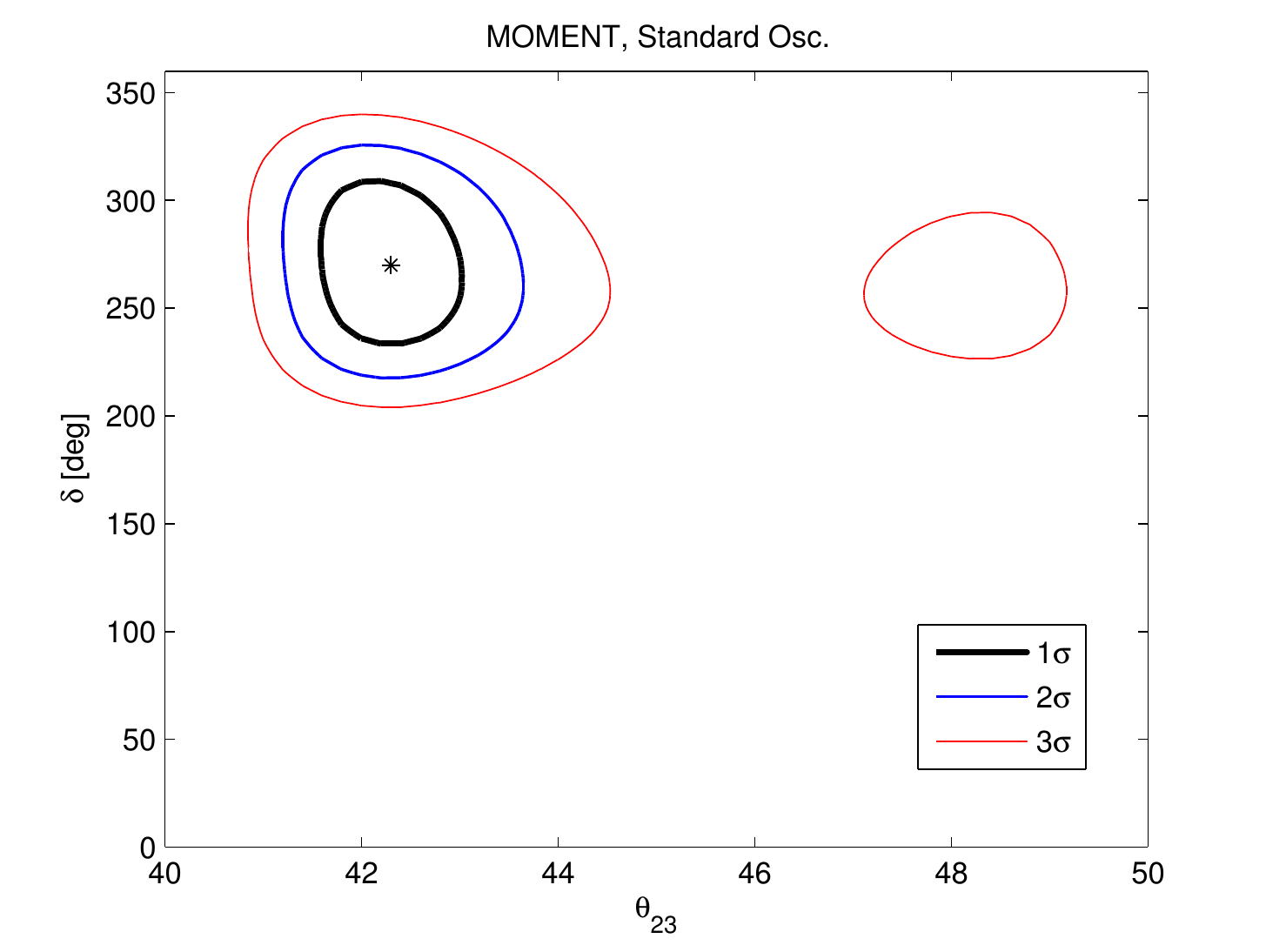}}
\subfigure[]{\includegraphics[width=0.49\textwidth]{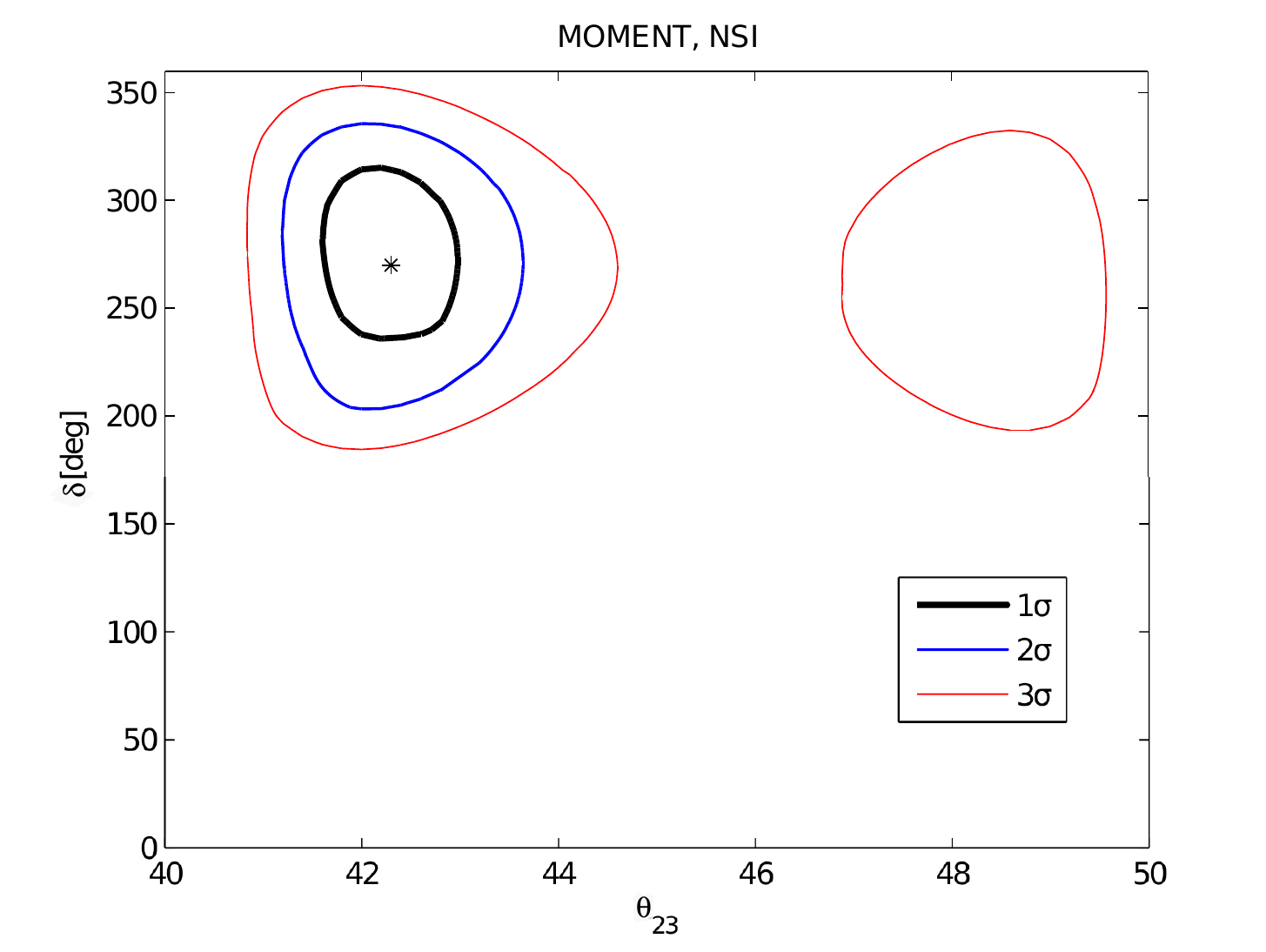}}
\subfigure[]{\includegraphics[width=0.49\textwidth]{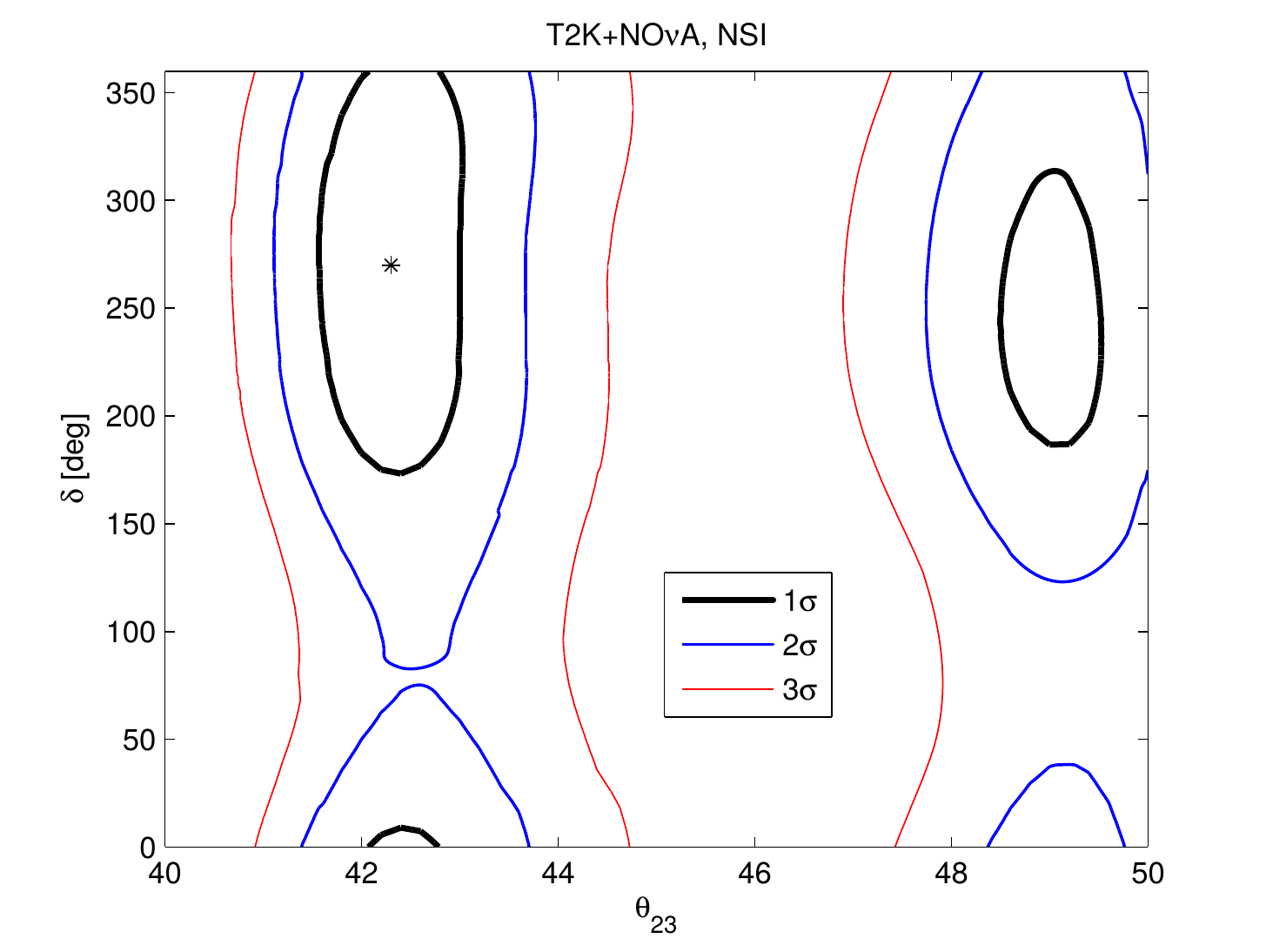}}
\subfigure[]{\includegraphics[width=0.49\textwidth]{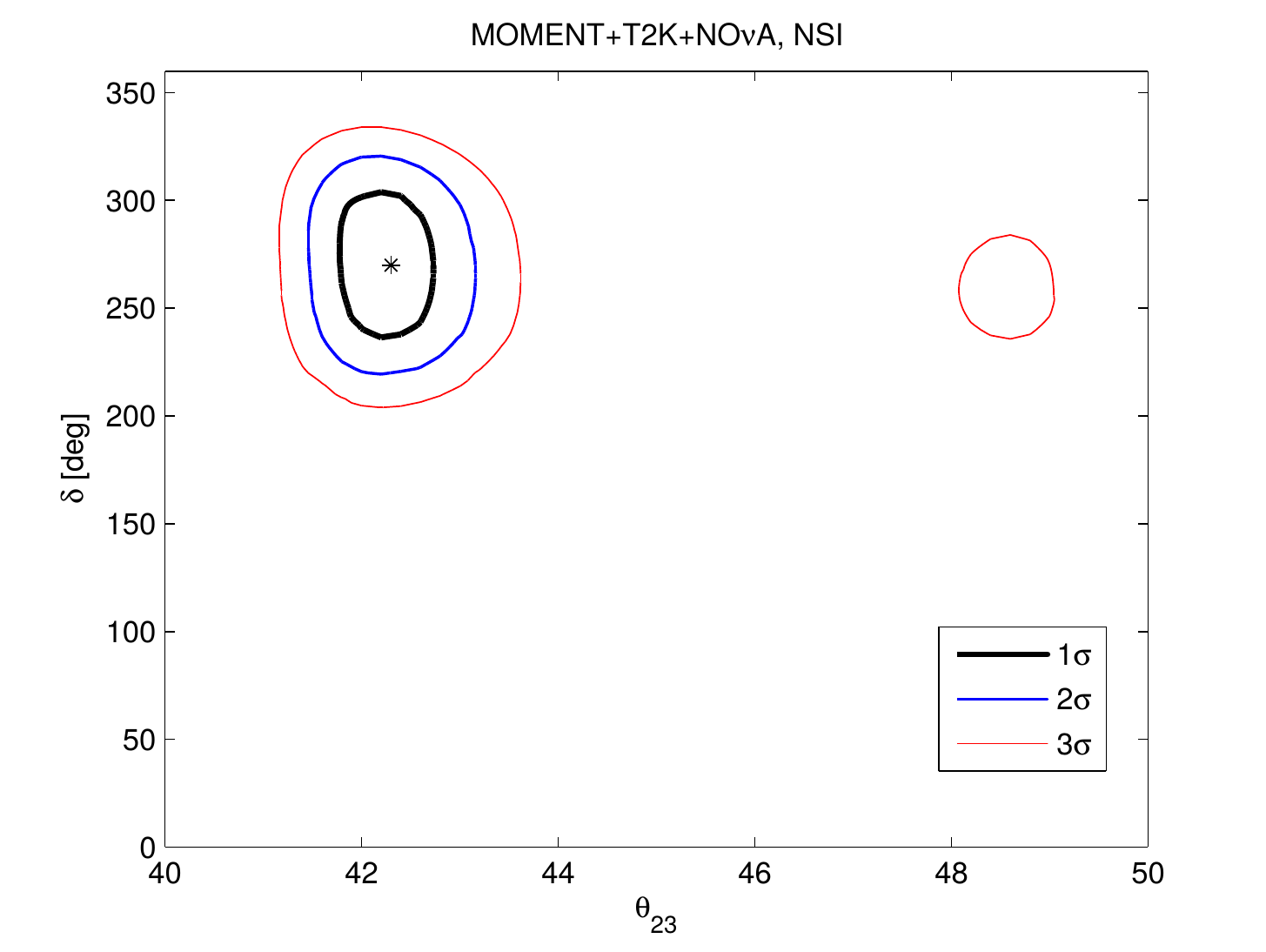}}
\end{center}
\vspace{0.5 cm} \caption[]{Sensitivity to $\delta-\theta_{23}$ projected for MOMENT, NO$\nu$A and T2K. The stars mark the assumed true values of $\delta$ and $\theta_{23}$ which are taken to be  their present best fit values \cite{Gonzalez-Garcia:2014bfa}.  Both appearance and disappearance modes are taken into account.  For MOMENT, SF= 0.1\%. Fig (a) shows the sensitivity of MOMENT for standard scenario without NSI. In Figs  (b), (c) and (d), pull method is used to treat the uncertainties of $\epsilon$ shown in Eqs. (\ref{range1},\ref{range2}). Fig (b) displays the sensitivity of the MOMENT experiment alone. Fig (c) shows the  sensitivity of the NO$\nu$A and T2K experiments combined and Fig (d) demonstrates the combined sensitivity of all three experiments.  Plots are taken from \cite{Bakhti:2016prn}.}
\label{MOMENT}
\end{figure}

 \begin{figure}
\begin{center}
\subfigure[]{\includegraphics[width=0.49\textwidth]{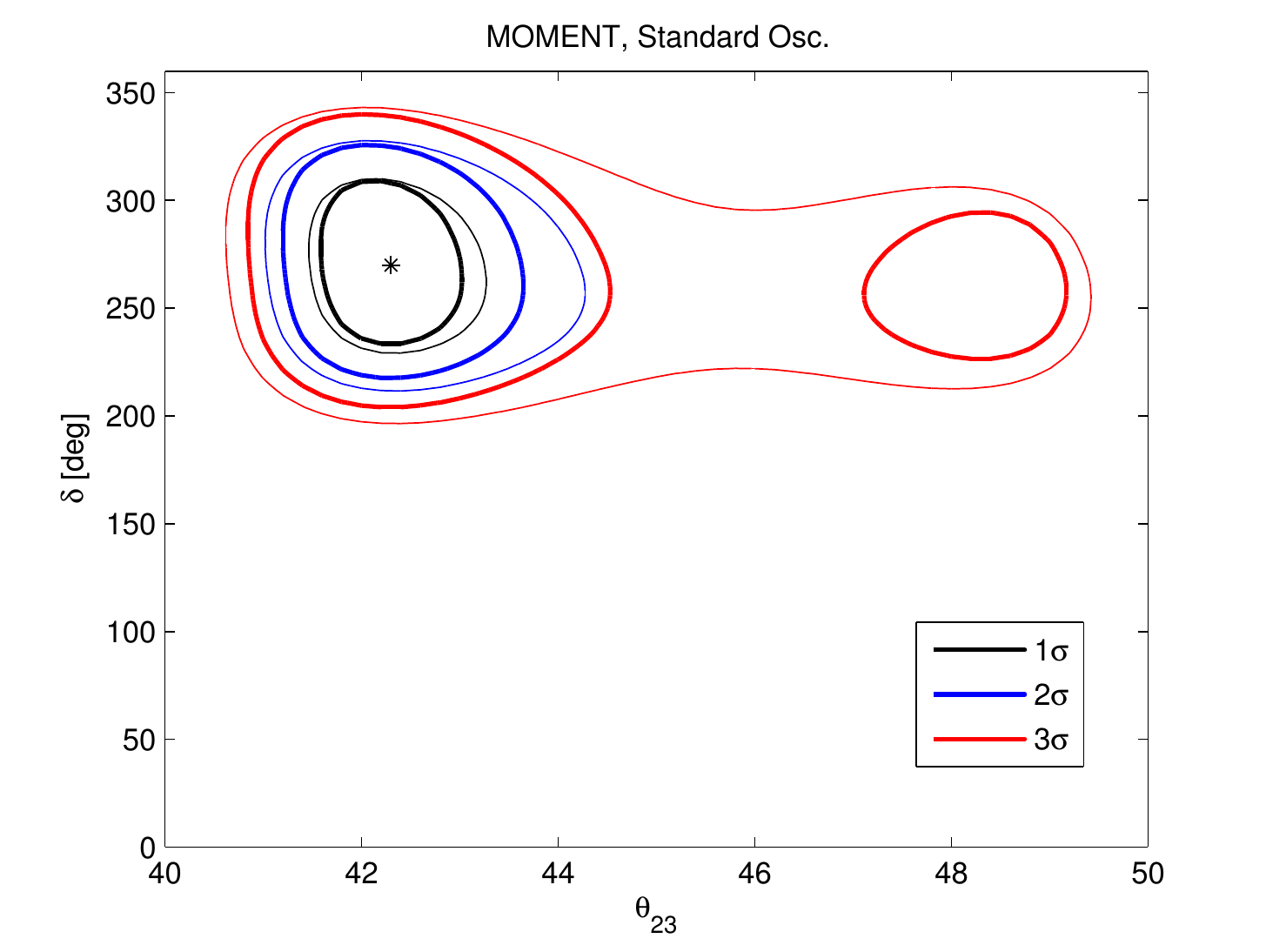}}
\subfigure[]{\includegraphics[width=0.49\textwidth]{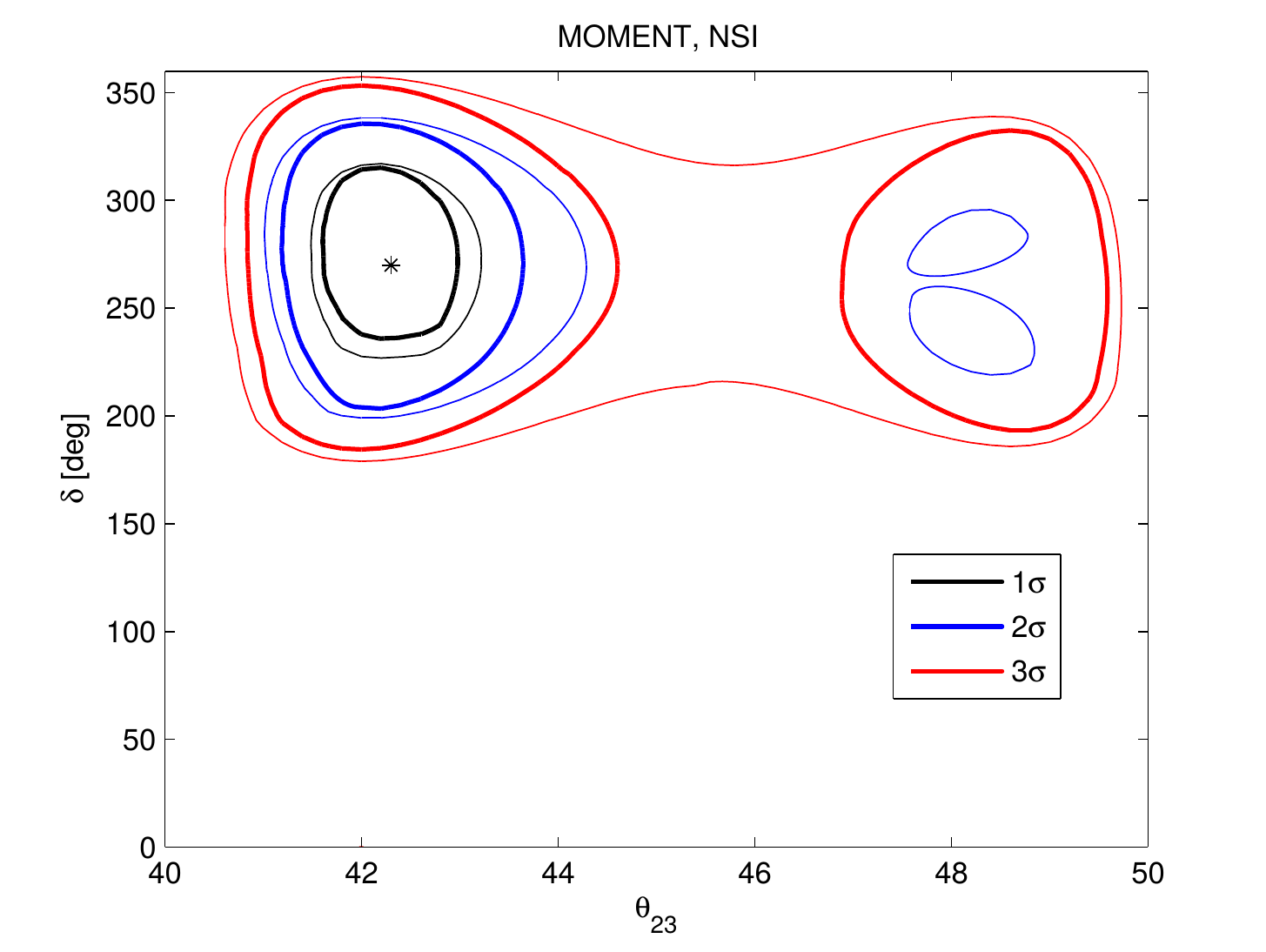}}
\end{center}
\vspace{0.5 cm} \caption[]{Dependence of the projected MOMENT sensitivity to  $\delta-\theta_{23}$ on background  Suppression Factor (SF). Thick and thin lines respectively show SF=0.1\% and 10\%. In Fig a, standard oscillation with no NSI is assumed. In Fig b, the true values of $\epsilon$ are set to zero and uncertainties of $\epsilon$ shown in Eqs. (\ref{range1},\ref{range2})   are treated by the pull method. Plots are taken from \cite{Bakhti:2016prn}.}
\label{SF}
\end{figure}

%%%%%%%%%%%%%%%%%

\section{Summary\label{summ}}

After multiple decades of experimental progress in the area of neutrino physics, the phenomenon of neutrino oscillations has been observed in a wide variety of experiments. The discovery of neutrino oscillations implies the existence of neutrino masses and therefore a need for an extension of the SM to include them.
Many possibilities have been proposed so far, see for instance Refs.~\cite{King:2003jb,Hirsch:2004he,Boucenna:2014zba,Cai:2017jrq}. Motivated by the two original {\it{anomalies}} in the solar and atmospheric neutrino sector, various other experiments have been proposed to search for neutrino oscillations in the solar and atmospheric neutrino flux as well as in man-made neutrino beams such as reactors or accelerators.
The large amount of experimental data collected over more than 20 years has allowed a very precise determination of some of the parameters responsible for the oscillations. These include the solar mass splitting,  $\Delta m^2_{21}$, the absolute value of the atmospheric mass splitting, $|\Delta m^2_{31}|$, as well as the solar ($\theta_{12}$) and reactor ($\theta_{13}$) mixing angles,  measured with relative accuracies below 5\%. Nevertheless,  the current precision of  atmospheric angle $\theta_{23}$ and the CP phase $\delta$ is not at that level. The sign of $\cos(2\theta_{23})$ or in other words the octant of $\theta_{23}$ is  yet unknown.
Moreover, although there are some hints for CP phase ($\delta$) to be close to $3\pi/2$, its value is not yet established.
 The sign of $\Delta m_{31}^2$ or equivalently the scheme of mass ordering (normal versus inverted) is also still unknown. In Section \ref{oscillation}, we have discussed the most relevant experimental information used in the global fits of neutrino oscillations~\cite{deSalas:2017kay,Capozzi:2017ipn,Esteban:2016qun} to obtain precise measurements of the oscillation parameters, exploiting the complementarity of the different data sets. The main results of these analysis have also been commented, with an emphasis on the still unknown parameters.
\\

Since their discovery, neutrinos have always surprised us by showing unexpected characteristics. In the dawn of the neutrino precision era, it is intriguing to ask whether neutrinos have new interactions beyond those expected within the standard model of particles.
Such new interactions can give a signal in different neutrino oscillation as well as non-oscillation experiments. No evidence for the presence of NSI has been reported so far. As a consequence of these negative searches, upper bounds on the magnitude of the new interactions can be set. In Section \ref{non-stan}, we have discussed the constraints on the NSI interactions, parameterized in terms of the $\epsilon_{\alpha\beta}$ couplings introduced in Eqs.~(\ref{eq:CC-NSI-Lagrangian}) and (\ref{eq:NC-NSI-Lagrangian}). The presence of NSI has been extensively analyzed in the literature, at the level of the production, detection and propagation of neutrinos in matter. The most restrictive limits on NSI are summarized in Tables  \ref{tab:boundsFD}, \ref{tab:boundsFC} and \ref{tab:boundsCC}.
\\

In principle, adding any new particle which couples both to neutrinos and to quarks will induce Non-Standard Interaction (NSI) for neutrinos. However, it is very challenging to build an electroweak symmetric model that leads to large enough NSI to be discernible at neutrino oscillation experiments without violating various bounds. We have discussed a class of models in which the new particle responsible
for NSI is a light $U(1)^\prime$ gauge boson $Z^\prime$ with mass $5~{\rm MeV}-{\rm few}~10 $~MeV with a coupling of order of $10^{-5}-10^{-4}$ to quarks and neutrinos. Within this range of parameter space, the NSI effective coupling can be as large as  the standard  effective Fermi coupling, $G_F$.

The total flux of solar neutrino has been measured by SNO experiment via dissociation of Deuteron through axial part of neutral current interaction and has been found to be consistent with the standard model prediction. To avoid a deviation from this prediction, the coupling of quarks to the new gauge boson is taken to be non-chiral with equal $U(1)^\prime$ charges for left-handed and right-handed quarks.
Moreover, the $U(1)^\prime$ charges of up and down quarks are taken to be equal to make the charged current weak interaction term invariant under  $U(1)^\prime$. The $U(1)^\prime$ charges of quarks is taken to be universal; otherwise, in the mass basis of quarks, we would have off-diagonal interactions leading to huge $q_i \to q_j Z^\prime$ rate enhanced by $(m_{q_i}/m_{Z^\prime})^2$. In summary, $U(1)^\prime$ charges of quarks is taken to be proportional to their baryon number.
We have discussed two different scenarios for
$U(1)^\prime$ charge assignment to leptons: (i) assigning $U(1)^\prime$ charge to the SM fermion as $a_e L_e +a_\mu L_\mu+a_\tau L_\tau+B$
where $L_\alpha$ denotes lepton flavor $\alpha$ and $B$ denotes Baryon number. With this assignment, lepton flavor will be preserved and both charged leptons and neutrinos
obtain lepton flavor conserving NSI.  A particularly interesting scenario  is $a_e=0$, $a_\mu=a_\tau=-3/2$ for which gauge symmetry anomalies automatically cancel without a need to add new specious. Choosing appropriate value of coupling [{\it i.e.,} $4 \times 10^{-5} (m_{Z^\prime}/10~{\rm MeV})$], the best fit to solar neutrino data with $\epsilon_{\mu \mu} -\epsilon_{e e}= \epsilon_{\tau \tau}-\epsilon_{ee}=-0.3$ can be reproduced.
ii) In the second scenario, the  leptons are not charged under $U(1)^\prime$. A new Dirac fermion, denoted by $\Psi$, with a mass of 1~GeV which is singlet under SM gauge group but charged under $U(1)^\prime$ is introduced which mixes with neutrinos. As a result, neutrinos obtain coupling to $Z^\prime$ through mixing with the new fermion but  charged leptons do not couple to $Z^\prime$ at the tree level. If the new fermion mixes with more than one flavor, both LFV and LFC NSI will be induced.
Within this scenario, new fermions are needed to cancel the gauge anomalies. We have discussed different possibilities. To give masses to these new fermions, new scalars charged under $U(1)^\prime$ are required whose VEV also gives a significant contribution to the mass of $Z^\prime$ boson.

We have suggested two mechanisms for inducing a mixing between $\Psi$ and neutrinos: (1) Introducing a new Higgs doublet, $H^\prime$, with $U(1)^\prime$ charge equal to that of $\Psi$ which couples to left-handed lepton doublets and $\Psi$.
$H^\prime$ obtains a VEV of few MeV which induces mixing. (2) Introducing a sterile neutrino, $N$ (singlet both under SM gauge group and $U(1)^\prime$) and a new scalar singlet with a $U(1)^\prime$ charge equal to that of $\Psi$ which couples to $N$ and $\Psi$. Its VEV then induces the coupling.

Even though the mass of $Z^\prime$ particle is taken to be low ({\it i.e.,} of order of solar neutrino energies and much smaller than the typical energies of atmospheric neutrinos or the energies of the neutrinos of long baseline experiments), the effect of new interaction on propagation of neutrinos in matter can be described by an effective four-Fermi  Lagrangian integrating out $Z^\prime$ because at forward scattering of neutrinos off the background matter, the energy momentum transfer is zero. At high energy scattering experiments, such as NuTeV and CHARM, the energy momentum transfer, $q^2$, is much higher than $m_{Z^\prime}^2$ so  the  effective four-Fermi  coupling loses its viability. The amplitude of new effects will be suppressed by a factor of  $\epsilon m_{Z^\prime}^2/q^2 \ll 1 $ relative to SM amplitude and will be negligible. Thus, unlike the case that the intermediate state responsible for NSI is heavy, these experiments cannot constrain $\epsilon \sim 1$. However, by studying scattering of low energy neutrinos ($E_\nu \sim$few 10 MeV) off matter, these models can be tested. The current COHERENT experiment and the upcoming CONUS experiment\footnote{https://indico.cern.ch/event/606690/contributions/2591545/attachments/1499330/2336272/Taup2017\_CONUS\_talk\_JHakenmueller.pdf} are  ideal set-ups to eventually test this model. An alternative way to test such models is to search for a dip in the energy spectrum of high energy cosmic neutrinos around few hundred TeV.

%%%%%%%%%%%%%%%%%%%%%%%%%%%%%%%%

\vspace{-0.2cm}

\section* {Acknowledgements}

We thank J. Jeeck, I. Shoemaker, A. Yu Smirnov, M. Lindner, M.M. Sheikh-Jabbari and O. G. Miranda for useful discussions.
MT is  supported by a Ram\'{o}n y Cajal contract (MINECO) and the Spanish grants  FPA2014-58183-P and SEV-2014-0398 (MINECO) and PROMETEOII/2014/084 and GV2016-142 (Generalitat Valenciana).
YF thanks  MPIK in Heidelberg where a part of this work was done for their hospitality.
 This project has received funding from the European Union\'~\!s Horizon 2020 research and innovation programme under the Marie Sklodowska-Curie grant agreement No 674896 and No 690575.
YF is also grateful to ICTP associate office and Iran National Science Foundation (INSF) for partial financial support under contract 94/saad/43287.

\bibliographystyle{utcaps_mod}

\bibliography{refs}

\end{document}